\documentclass[journal]{IEEEtran}
\usepackage{amsmath,amsfonts,amssymb,amsthm,mathtools}
\usepackage{algorithmic}
\usepackage{algorithm}
\usepackage{array}
\usepackage[caption=false,font=normalsize,labelfont=sf,textfont=sf]{subfig}
\usepackage{textcomp}
\usepackage{stfloats}
\usepackage{url}
\usepackage{multirow}
\usepackage{verbatim}
\usepackage{graphicx}
\usepackage{cite}
\usepackage{geometry}
\usepackage{enumitem}
\usepackage{tabularx, booktabs, ragged2e}
\newcolumntype{L}{>{\RaggedRight\arraybackslash}X}
\usepackage{placeins} 
\usepackage{makecell} 
\usepackage{etoolbox} 
\newcommand{\Dd}{\text{Dd}}
\newcommand{\Atom}{\text{Atom}}

\newcommand{\enc}{\text{enc}}
\newcommand{\abs}{\text{abs}}
\newcommand{\Cost}{\operatorname{Cost}}

\newtheorem{policy}{Policy}

\newtheorem{axiom}{Axiom}
\newtheorem{assumption}{Assumption}
\newtheorem{theorem}{Theorem}
\newtheorem{lemma}{Lemma}
\newtheorem{proposition}{Proposition}
\newtheorem{corollary}{Corollary}
\newtheorem{definition}{Definition}
\newtheorem{remark}{Remark}

\makeatletter
\renewenvironment{proof}[1][\proofname]{\par
  \pushQED{\qed}%
  \normalfont \topsep6\p@\@plus6\p@\relax
  \trivlist
  \item[\hskip\labelsep
        \itshape
    #1\@addpunct{.}]\ignorespaces
}{%
  \popQED\endtrivlist\@endpefalse
}
\makeatother

\newcommand{\E}{\mathbb{E}}

\begin{document}

\title{Information Physics of Intelligence: \\Unifying Derivation Depth and Entropy under Thermodynamic Constraints}

\author{Jianfeng~Xu$^{1*}$,~Zeyan~Li$^{2}$%
\thanks{$^{1}$Koguan School of Law, China Institute for Smart Justice, School of Computer Science, Shanghai Jiao Tong University, Shanghai 200030, China. Email: xujf@sjtu.edu.cn}%
\thanks{$^{2}$School of Computer Science, Shanghai Jiao Tong University, Shanghai 200030, China. Email: zeyan0823@sjtu.edu.cn}%
\thanks{*Corresponding author. Manuscript submitted November 2025.}
}

\maketitle

\begin{abstract}
The rapid scaling of artificial intelligence models has revealed a fundamental tension between model capacity (storage) and inference efficiency (computation). While classical information theory focuses on transmission and storage limits, it lacks a unified physical framework to quantify the thermodynamic costs of generating information from compressed laws versus retrieving it from memory. In this paper, we propose a theoretical framework that treats information processing as an enabling mapping from ontological states to carrier states. We introduce a novel metric, \textit{Derivation Entropy}, which quantifies the effective work required to compute a target state from a given derivation depth. By analyzing the interplay between Shannon entropy (storage) and computational complexity (time/energy), we demonstrate the existence of a critical phase transition point. Below this threshold, memory retrieval is thermodynamically favorable; above it, generative computation becomes the optimal strategy. This "Energy-Time-Space" conservation law provides a physical explanation for the efficiency of generative models and offers a rigorous mathematical bound for designing next-generation, energy-efficient AI architectures. Our findings suggest that the minimization of Derivation Entropy is a governing principle for the evolution of both biological and artificial intelligence.
\end{abstract}


\section{Introduction}
\label{sec:introduction}

\IEEEPARstart{T}{he} fundamental nature of information has long been a subject of debate across physics, computer science, and philosophy. In our previous works, we established a formalized meta-framework positing that information is fundamentally the \textit{enabling mapping} from an ontological state to a carrier state \cite{xu2025information, xu2024research, qiu2025research}. Under this view, intelligence is not merely the accumulation of data, but an active physical process of mapping the infinite complexity of the external environment (ontology) onto the finite constraints of an internal representation (carrier). This mapping inherently necessitates a primal trade-off: the system must decide whether to store the mapping results statically (memory) or to regenerate them dynamically via algorithms (computation).

Classically, these two aspects have been treated separately. Shannon's information theory \cite{shannon1948mathematical} rigorously defines the limits of static storage and transmission, quantifying the irreducible randomness of the carrier state. However, Shannon entropy is agnostic to the \textit{causal history} or the computational effort required to generate that state. Conversely, Algorithmic Information Theory (AIT), particularly through the lens of Kolmogorov complexity \cite{kolmogorov1965three} and Bennett's concept of \textbf{Logical Depth} \cite{bennett1988logical}, focuses on the computational time required to reconstruct a structure from its compressed description. While Derivation Depth captures the dynamic value of the "mapping process," it lacks a direct unification with the statistical mechanics governing the carrier's physical existence.

This separation becomes problematic when we consider that information processing is strictly bound by physical laws. As Brillouin \cite{brillouin1956science} and Landauer \cite{landauer1961irreversibility} demonstrated, the manipulation of information—specifically the erasure of logical states—is inseparably linked to thermodynamic entropy and heat dissipation. Therefore, the "enabling mapping" from ontology to carrier is not just a logical operation but a thermodynamic cycle. Current artificial intelligence systems, such as Large Language Models (LLMs) based on the Transformer architecture \cite{vaswani2017attention, brown2020language}, are rapidly approaching the physical limits of this cycle. The empirical scaling laws observed in these systems suggest a deep, underlying fundamental trade-off connecting parameter size (storage), training FLOPs (computation), and energy consumption, yet a unified theoretical explanation remains elusive.

In this paper, we propose the \textbf{"Information Physics of Intelligence,"} a theoretical framework that unifies Derivation Depth and Shannon Entropy under rigorous thermodynamic constraints. We define \textbf{Derivation Entropy ($H_{derive}$)}, which quantifies the total physical cost of the ontological-to-carrier mapping. By integrating the thermodynamic cost of storage with the algorithmic complexity of retrieval, we derive the \textbf{Energy-Time-Space Triality Bound}. This bound reveals that minimizing storage space (lowering Shannon entropy) inevitably necessitates a compensatory increase in computational time (increasing Derivation Depth) or energy flux, governed by Landauer’s limit.

Our contributions are threefold:
\begin{enumerate}
    \item We formalize the "ontological-to-carrier mapping" within a thermodynamic context, bridging the gap between abstract information measures \cite{xu2025general} and physical complexity \cite{bennett1982thermodynamics}.
    \item We mathematically derive the phase transition point where intelligent systems must switch from a retrieval-based strategy to a derivation-based strategy to maintain thermodynamic optimality.
    \item We provide a first-principles explanation for the efficiency bottlenecks in modern AI models, offering a governing equation for designing next-generation, energy-efficient cognitive architectures.
\end{enumerate}


\section{System Model and Preliminaries}
\label{sec:model}
We establish the formal framework for knowledge systems, introducing the concepts of enabling mappings, derivation depth, and their information-theoretic characterization.


\subsection{The Underlying Logical System $\mathcal{L}$}
Throughout this work, we fix a formal logical system $\mathcal{L}$ that serves as the foundation for expressing semantic states and reasoning about information. Specifically, $\mathcal{L}$ is taken to be \emph{first-order logic with least fixed-point operators} (FO(LFP))~\cite{immerman1999descriptive}, extended with multiple sorts, including at least $\mathsf{Obj}$ (for ontological entities), $\mathsf{Time}$ (for discrete time points), and $\mathsf{Carrier}$ (for physical carriers); function and relation symbols are typed accordingly (e.g., a state predicate may have signature $\mathsf{State} : \mathsf{Obj} \times \mathsf{Time} \to \mathsf{Bool}$). The language of $\mathcal{L}$ includes:
\begin{itemize}
    \item A countable set of sorted variables, constants, function symbols, and relation symbols;
    \item Standard Boolean connectives and first-order quantifiers over each sort;
    \item A least fixed-point operator $\mu$ enabling inductive definitions of relations (e.g., reachability, state transitions).
\end{itemize}

Crucially, we restrict attention to \emph{finite structures} over a \emph{discrete, bounded time domain} $T = \{t_0, t_1, \dots, t_n\}$ for some $n \in \mathbb{N}$. Under this restriction, $L$ enjoys the following essential properties:
\begin{itemize}
    \item \textbf{Consistency}: The proof system of $\mathcal{L}$ is sound and free of contradiction.
    \item \textbf{Effective model checking}: For any finite structure $\mathcal{M}$ and formula $\varphi \in \mathcal{L}$, the satisfaction relation $\mathcal{M} \models \varphi$ is decidable.
    \item \textbf{Expressive adequacy}: By the Immerman--Vardi theorem~\cite{immerman1986relational,vardi1982complexity}, FO(LFP) captures exactly the class of polynomial-time computable properties over ordered finite structures. This makes $\mathcal{L}$ sufficiently expressive to model the state evolution of a wide range of real-world information systems under finite-horizon autoregressive decoding.
\end{itemize}


\subsection{Enabling Mappings and Information}

\begin{definition}[State Expressibility]
\label{State Express}
Let $\mathcal{L}$ be the logical system defined above. The state set $S(X, T)$ of an object set $X$ over a time set $T$ is said to be \emph{expressible in $\mathcal{L}$} if it can be represented by assigning $X$, $T$, and their subsets and elements as interpretations of a set of well-formed formulas (WFFs) in $\mathcal{L}$.
\end{definition}

In all subsequent discussions of this paper, we consider only states expressible in the logical system $\mathcal{L}$\cite{qiu2025research, xu2025information}.

\begin{definition}[Enabling Mapping]
\label{def:enabling}
Let $O$ be an ontological space, $C$ a carrier space, and $T_o, T_r \subseteq \mathbb{R}^+$ be time domains. A correspondence
\begin{equation}
\mathcal{E}: S_O(O, T_o) \rightrightarrows S_C(C, T_r)
\end{equation}
is called an \emph{enabling mapping} if the following conditions hold:
\begin{description}
    \item[(E1) Realizability:] For every ontological state $s_o \in S_O(O, T_o)$, the image $\mathcal{E}(s_o)$ is a non-empty subset of $S_C(C, T_r)$. That is, there exists at least one valid carrier state $s_c \in \mathcal{E}(s_o)$ that can physically represent $s_o$.
    \item[(E2) Causality:] For every $s_o$ and any realization $s_c \in \mathcal{E}(s_o)$, we have $s_o \prec s_c$, where $\prec$ denotes a causal dependency relation, implying: (i) The occurrence time of $s_o$ precedes or is concurrent with the realization of $s_c$; (ii) The ontological state $s_o$ is a necessary condition for the intentional generation of $s_c$.
    \item[(E3) Physicality:] The realization of any mapping instance $s_o \mapsto s_c$ (where $s_c \in \mathcal{E}(s_o)$) consumes non-zero physical resources (e.g., energy, time).
\end{description}
\end{definition}

\begin{remark}
The notation $\rightrightarrows$ indicates that $\mathcal{E}$ is a set-valued mapping (correspondence). While the system defines a set of valid carriers $\mathcal{E}(s_o)$ for a given semantic state (allowing for redundancy), a specific physical instance selects exactly one $s_c \in \mathcal{E}(s_o)$.
\end{remark}

\begin{definition}[Information]
\label{def:information}
Information is an enabling mapping~\textup{\cite{xu2024research}}:
\begin{equation}
\mathcal{I}: S_O(O, T_o) \rightrightarrows S_C(C, T_r)
\end{equation}
where
\begin{itemize}
\item $O$: ontology (objective or subjective reality)
\item $T_o \subseteq \mathbb{R}^+$: occurrence time
\item $S_O = S_O(O,T_o)$: state set of ontology
\item $C$: carrier (objective reality)
\item $T_r \subseteq \mathbb{R}^+$: reflection time
\item $S_C = S_C(C,T_r)$: state set of carrier
\end{itemize}

Equivalently, information can be represented as a sextuple $\mathcal{I}= \langle O, T_o, S_O, C, T_r, S_C \rangle$. It is also assumed to be effectively computable on semantic inputs

\end{definition}

\begin{definition}[Ideal Information] \label{def:ideal-information}
Let $S_O$ and $S_C$ be defined as above. Ideal Information $\mathcal{I}$ is an enabling mapping satisfying two additional strict conditions:
\begin{description}
    \item[Completeness (Surjectivity on Domain):] Every semantic state $s_o \in S_O$ has a physical representation. Formally, $\forall s_o \in S_O, \mathcal{I}(s_o) \neq \emptyset$. (Inherited from E1).
    
    \item[Reliability (Unambiguous Decoding):] While a semantic state may map to multiple physical carriers (synonyms), any specific physical carrier maps back to at most one semantic state. Formally, if $s_c \in \mathcal{I}(s_{o_1})$ and $s_c \in \mathcal{I}(s_{o_2})$, then $s_{o_1} = s_{o_2}$.
\end{description}
\end{definition}

\begin{remark}
  The set $\mathcal{I}(s_o) \subseteq S_C$ constitutes the synonymous carrier set
  of the semantic state $s_o \in S_O$. Therefore, the semantic
  characteristics of an Ideal Information instance
  $\mathcal{I} = \langle O, T_o, S_O, C, T_r, S_C \rangle$ can be fully
  characterized by directly studying the semantic state space $S_O$,
  with each $s_o \in S_O$ being equipped with its associated synonymous
  carrier set $\mathcal{I}(s_o)$.
\end{remark}

\subsection{Information Cost of Energy and Time}

According to Definitions~\ref{def:enabling}–\ref{def:information}, state transitions of an information
carrier are constrained by physical conditions. Building on
fundamental results such as the energy requirements of
high-frequency harmonic oscillators in classical physics~\cite{Nyquist1928Thermal}
and the Margolus–Levitin bound in quantum mechanics~\cite{Margolus1998maximum},
we adopt the following abstract physical principle.

\begin{assumption}[Physical Cost Principle]
\label{assump:physical-cost}
For a physical system evolving from a state $s_1$ to a
distinguishable state $s_2$, the required energy $E$
and time $\tau$ satisfy
\begin{equation}
  E \cdot \tau \;\ge\; \kappa,
  \label{eq:physical-cost}
\end{equation}
where $\kappa$ is a universal constant with the dimension of action.
\end{assumption}

Intuitively, Assumption~\ref{assump:physical-cost} states that every
physically distinguishable transition of the carrier system incurs a
non-zero resource cost that cannot be made arbitrarily small: the
product of the available energy scale and the time scale must exceed a
fixed action threshold $\kappa$. In particular, a physical process that
implements an information mapping cannot realize arbitrarily many
mutually distinguishable semantic alternatives without paying at least
a proportional cost in the joint resource $E \cdot \tau$.

In the context of Ideal Information (Definition~\ref{def:ideal-information}), the semantic state
space $S_O(O,T_o)$ has finite cardinality, and the corresponding
amount of information is measured in bits as
\begin{equation}
  I \;=\; \log_2 |S_O(O,T_o)|.
  \label{eq:information-amount}
\end{equation}
Because Ideal Information is complete and reliably decodable, any
physical implementation must, in effect, be able to distinguish among
all $|S_O(O,T_o)| = 2^I$ possible semantic inputs via mutually
distinguishable carrier states. This leads to the following
resource lower bound.

\begin{theorem}[Resource Bound for Ideal Enabling Mapping]
\label{thm:resource-bound}
Let $\mathcal{I} : S_O(O,T_o) \rightrightarrows S_C(C,T_r)$ be an Ideal Information
instance, and let $I = \log_2 |S_O(O,T_o)|$ denote the amount of
information transmitted (in bits) as in~\eqref{eq:information-amount}.
Then any physical implementation of $\mathcal{I}$ satisfies
\begin{equation}
  E \cdot \tau \;\ge\; \kappa \cdot I,
  \label{eq:resource-bound}
\end{equation}
where $E$ is the energy resource, $\tau$ is the implementation time,
and $\kappa$ is the universal constant from
Assumption~\ref{assump:physical-cost}.
\end{theorem}

\begin{proof}
Since $\mathcal{I}$ is Ideal Information in the sense of Definition~\ref{def:ideal-information}, the
mapping $\mathcal{I} : S_O(O,T_o) \rightrightarrows S_C(C,T_r)$ is complete and
reliable: every semantic state $s_o \in S_O(O,T_o)$ has at least one
physical realization $s_c \in S_C(C,T_r)$, and distinct semantic
states are not confounded at the carrier level. Hence any physical
implementation of $\mathcal{I}$ must realize at least $|S_O(O,T_o)|$ pairwise
distinguishable carrier states, one (or a synonymous set) for each
semantic state.

Equivalently, the implementation process must be capable of
distinguishing among $|S_O(O,T_o)| = 2^I$ alternative semantic inputs.
This corresponds to an information content of $I = \log_2
|S_O(O,T_o)|$ bits.

Consider now the physical evolution of the carrier system during the
implementation of $\mathcal{I}$. Each time the system evolves from one
distinguishable physical state to another in such a way that it
increases or commits a new semantic distinction (for instance, by
encoding one more bit of choice among alternative semantic states),
Assumption~\ref{assump:physical-cost} applies: the associated energy
and time resources $E_i$ and $\tau_i$ must satisfy
\[
  E_i \cdot \tau_i \;\ge\; \kappa.
\]
Let $N$ denote the number of such information-relevant, distinguishable
state transitions required by the implementation. Since each such
transition can encode at most a constant number of bits of new
distinguishable semantic information, we have $N \ge I$ up to a fixed
multiplicative constant. Absorbing this constant into $\kappa$ if
necessary, we may assume without loss of generality that $N \ge I$.

We analyze two implementation paradigms:

\textbf{Case 1: Sequential execution.} The total time is $\tau = \sum_{i=1}^{N} \tau_i$, 
and energy can be reused across steps, so the effective energy is $E = \max_i E_i$. 
By Assumption 1, $E \cdot \tau_i \geq \kappa$ for each $i$, hence
\[
E \cdot \tau = E \cdot \sum_{i=1}^{N} \tau_i \geq N \kappa \geq \kappa I.
\]

\textbf{Case 2: Parallel execution.} The total time is $\tau = \max_i \tau_i$, 
and energy must be summed, so $E = \sum_{i=1}^{N} E_i$. 
By Assumption 1, $E_i \cdot \tau \geq \kappa$ for each $i$, hence
\[
E \cdot \tau = \tau \cdot \sum_{i=1}^{N} E_i \geq N \kappa \geq \kappa I.
\]

In both cases, $E \cdot \tau \geq \kappa I$.
\end{proof}

In the following, we interpret the constant $\kappa$ in
Assumption~\ref{assump:physical-cost} in the usual thermodynamic
setting of information processing. Specifically, for systems operating
in a thermal equilibrium environment at temperature $T$ and undergoing
quasistatic transformations, the minimal resource cost per bit can be
identified with the Landauer bound $k_B T \ln 2$ measured in terms of
dissipated heat. Under this interpretation, Theorem~\ref{thm:resource-bound}
specializes to a Landauer-type inequality for Ideal Information.

\begin{corollary}[Landauer Principle for Ideal Information]
\label{cor:landauer}
In a thermal equilibrium environment at temperature $T$, an Ideal
Information instance implemented via a quasistatic process satisfies
the following lower bound on energy dissipation:
\begin{equation}
  E_{\mathrm{diss}}
  \;\ge\;
  k_B T \ln 2 \cdot I,
  \label{eq:landauer}
\end{equation}
where $I = \log_2 |S_O(O,T_O)|$ is the amount of information
transmitted losslessly (in bits).
\end{corollary}

\begin{proof}
In a thermal environment at temperature $T$, Landauer's principle\cite{landauer1961irreversibility, berut2012experimental}
states that any logically irreversible operation that erases or
thermodynamically commits one bit of information leads, in the
quasistatic limit, to a minimal heat dissipation of at least
$k_B T \ln 2$. In the setting of Theorem~\ref{thm:resource-bound},
implementing an ideal information instance that reliably encodes
$I = \log_2 |S_O(O,T_o)|$ bits requires at least $I$ effective
information-bearing operations, each of which is thermodynamically
equivalent to committing one bit of semantic information to a physical
carrier state.

Identifying the per-bit resource constant in
Assumption~\ref{assump:physical-cost} with $\kappa_T = k_B T \ln 2$ in
this thermal, quasistatic regime, Theorem~\ref{thm:resource-bound}
implies that the total dissipated energy satisfies
\[
  E_{\mathrm{diss}}
  \;\ge\;
  \kappa_T \cdot I
  \;=\;
  k_B T \ln 2 \cdot I,
\]
which is exactly the bound in~\eqref{eq:landauer}. This yields the
claimed Landauer-type lower bound for any lossless implementation of
Ideal Information.
\end{proof}

\subsection{Atomic Information}
\label{subsec:atomic-information}

In an Ideal Information instance
$\mathcal{I} = \langle O, T_o, S_O, C, T_r, S_C \rangle$
(Definition~\ref{def:ideal-information}),
the ontological state space $S_O$ is equipped with a
predecessor function
\begin{equation}
     P_O : S_O \to 2^{S_O},
\end{equation}

which captures the semantic causal structure among
states. Intuitively, $P_O(s)$ is the set of semantic
states that can directly lead to $s$ within the
ontological dynamics of $O$.
We impose the following axioms on $P_O(\cdot)$.

\begin{axiom}[Strict Temporal Precedence]
\label{ax:strict-temporal-precedence}
For any $s_o(o,t), s'_{o'}(o',t') \in S_O$, if $s'_{o'} \in P_O(s_o)$, then
\[
  t' < t.
\]
\end{axiom}

\begin{axiom}[Finite and Computable Predecessors]
\label{ax:finite-predecessors}
For any $s \in S_O$, the predecessor set $P_O(s)$ is
finite, and there exists an effective procedure that,
given $s$, computes $P_O(s)$.
\end{axiom}

\begin{axiom}[Lower-Bounded and Discrete Time]
\label{ax:lower-bounded-time}
The time domain $T_o$ is bounded from below and
discrete, i.e., there exists $t_{\min} \in T_o$ and
$\Delta t > 0$ such that for any strictly decreasing
sequence $\{ t_n \}$ in $T_o$ we have
$t_n \ge t_{\min}$ and
$t_{n} - t_{n+1} \ge \Delta t$.
\end{axiom}

Axiom~\ref{ax:strict-temporal-precedence} states that
predecessors must occur strictly earlier in ontological
time; Axiom~\ref{ax:finite-predecessors} excludes
infinitely branching causal histories and ensures that
local causal structure is effectively inspectable; and
Axiom~\ref{ax:lower-bounded-time} guarantees that there
is no infinite regress of ever earlier states in finite
time. Together, these axioms imply that the backward
causal unfolding of any semantic state is a finite,
effectively traversable structure. This leads to the
following result.

\begin{theorem}[Existence and Computability of Semantic Atomic States]
\label{thm:semantic-atoms}
For any target semantic state
$s_{\text{target}} \in S_O$, there exists a finite,
non-empty set of \emph{semantic atomic states}
$A(s_{\text{target}}) \subseteq S_O$ such that:
\begin{enumerate}
  \item[(i)] Each $s_{\text{atom}} \in A(s_{\text{target}})$
             is predecessor-free:
             \[
               P_O(s_{\text{atom}}) = \varnothing .
             \]
  \item[(ii)] Every $s_{\text{atom}} \in A(s_{\text{target}})$
              can reach $s_{\text{target}}$ along the causal
              structure induced by $P_O(\cdot)$, i.e.,
              there exists a finite sequence
              \[
                s_{\text{atom}} = s_0, s_1, \dots, s_n
                = s_{\text{target}}
              \]
              such that $s_{i} \in P_O(s_{i+1})$ for all
              $i = 0, \dots, n-1$.
  \item[(iii)] The mapping
               $s_{\text{target}} \mapsto
                A(s_{\text{target}})$ is computable.
\end{enumerate}
\end{theorem}

\begin{proof}
Fix any $s_{\text{target}} \in S_O$.
Consider the backward expansion of
$s_{\text{target}}$ along the predecessor function
$P_O(\cdot)$:
start from $s_{\text{target}}$, enumerate
$P_O(s_{\text{target}})$, then for each
$s^{(1)} \in P_O(s_{\text{target}})$ enumerate
$P_O(s^{(1)})$, and so on.

By Axiom~\ref{ax:strict-temporal-precedence}, every
time we move from a state to one of its predecessors,
the time-stamp strictly decreases:
if $s'_{o'}(o',t') \in P_O(s_o(o,t))$, then $t' < t$.
Since $T_o$ is lower-bounded and discrete
(Axiom~\ref{ax:lower-bounded-time}),
there is no infinite strictly decreasing temporal
sequence. Therefore, any backward traversal starting
from $s_{\text{target}}$ must terminate in finitely many
steps. Equivalently, the backward expansion tree is of
finite depth.

By Axiom~\ref{ax:finite-predecessors}, each node has a
finite predecessor set, so the branching factor of the
backward expansion tree is finite.
Thus the entire backward tree rooted at
$s_{\text{target}}$ consists of finitely many nodes.
The leaves of this finite tree are exactly those states
with no predecessors, i.e., all $s \in S_O$ such that
$P_O(s) = \varnothing$ and that can reach
$s_{\text{target}}$ along the tree.

Define $A(s_{\text{target}})$ to be the set of all such
leaves. Since the tree is finite and non-empty
(it contains at least $s_{\text{target}}$),
$A(s_{\text{target}})$ is finite and non-empty, and
every element has no predecessors, establishing (i) and
(ii).

For (iii), consider the following recursive procedure
\begin{algorithm}
TRACEATOMS$(s)$:

\begin{itemize}
  \item If $P_O(s) = \varnothing$, output $\{s\}$.
  \item Otherwise, for each $s' \in P_O(s)$ recursively
        call TRACEATOMS$(s')$ and return the union of all
        outputs.
\end{itemize}
\end{algorithm}

By Axiom~\ref{ax:finite-predecessors}, $P_O(s)$ is
finite and computable for any $s$, so each recursive
step is effectively executable. By the finiteness of
the backward expansion tree, the recursion terminates
after finitely many steps. Consequently,
TRACEATOMS$(s_{\text{target}})$ is a total, effective
procedure that returns exactly the set
$A(s_{\text{target}})$, proving that the mapping
$s_{\text{target}} \mapsto A(s_{\text{target}})$ is
computable.
\end{proof}

Theorem~\ref{thm:semantic-atoms} shows that, for any
semantic state of interest, there exists a finite and
computable collection of causally irreducible semantic
states that underpin it. We now upgrade this semantic
view to the level of physically realizable information.

\begin{definition}[Atomic Information]
\label{def:atomic-information}
Let
$\mathcal{I} = \langle O, T_o, S_O, C, T_r, S_C \rangle$
be an Ideal Information instance and
let $\alpha \in S_O$ be any semantic state.
The \emph{Atomic Information} associated with $\alpha$,
denoted by $\mathcal{I}_\alpha$, is defined as the restriction of
$\mathcal{I}$ to the singleton $\{\alpha\}$:
\begin{equation}
  \mathcal{I}_\alpha
  \;=\;
  \langle
    O,\, T_o,\, \{\alpha\},\, C,\, T_r,\, \mathcal{I}(\alpha)
  \rangle ,
  \label{eq:def-atomic-information}
\end{equation}
where
\[
\mathcal{I}(\alpha) = \left\{\, s_c \in S_C \mid 
\begin{array}{@{}l@{}}
\text{the realization of } s_c \text{ yields} \\
\text{the semantic state } \alpha
\end{array} \right\}
\]
is the set of carrier states synonymous with $\alpha$
under the enabling mapping $\mathcal{I} : S_O \rightrightarrows S_C$.
\end{definition}

Because $\mathcal{I}$ is Ideal Information
(Definition~\ref{def:ideal-information}),
it satisfies completeness and unambiguous decoding.
In particular, for every semantic state
$\alpha \in S_O$ we have $\mathcal{I}(\alpha) \neq \varnothing$,
so there always exists at least one carrier state
$s_c \in S_C$ that physically realizes $\alpha$.
Thus each $\mathcal{I}_\alpha$ defined in
\eqref{eq:def-atomic-information} is a physically
realizable information entity that tightly couples an
abstract semantic state with the set of its synonymous
physical representations.

Combining Theorem~\ref{thm:semantic-atoms} with the
above definition yields the following corollary.

\begin{corollary}[Existence and Computability of Atomic Information]
\label{cor:existence-atomic-information}
Let
$\mathcal{I} = \langle O, T_o, S_O, C, T_r, S_C \rangle$
be an Ideal Information instance, and let
$s_{\text{target}} \in S_O$ be any semantic state.
Then:

\begin{enumerate}
  \item There exists a finite, non-empty set of
        semantic atomic states
        $A(s_{\text{target}}) \subseteq S_O$
        such that each $\alpha \in A(s_{\text{target}})$
        is predecessor-free and causally contributes to
        $s_{\text{target}}$ as in
        Theorem~\ref{thm:semantic-atoms}.
  \item For every $\alpha \in A(s_{\text{target}})$,
        the corresponding Atomic Information
        \[
           \mathcal{I}_\alpha
           \;=\;
           \langle
             O, T_o, \{\alpha\}, C, T_r, \mathcal{I}(\alpha)
           \rangle
        \]
        is well-defined and physically realizable, and
        the finite set
        $\{ \mathcal{I}_\alpha \mid \alpha \in A(s_{\text{target}}) \}$
        is computable from $s_{\text{target}}$.
\end{enumerate}
\end{corollary}

\begin{proof}
Item~(1) is exactly
Theorem~\ref{thm:semantic-atoms}.

For item~(2), since $\mathcal{I}$ is Ideal Information, its
completeness property ensures $\mathcal{I}(\alpha) \neq \varnothing$
for every $\alpha \in S_O$; in particular, for each
$\alpha \in A(s_{\text{target}})$, $\mathcal{I}(\alpha)$ is a
non-empty set of synonymous carrier states, so
$\mathcal{I}_\alpha$ in \eqref{eq:def-atomic-information} is
well-defined and physically realizable.

Moreover, Theorem~\ref{thm:semantic-atoms} provides a
total, effective procedure TRACEATOMS that, given
$s_{\text{target}}$, computes $A(s_{\text{target}})$.
Under the system model adopted in this paper,
the enabling mapping $\mathcal{I}(\cdot)$ is also assumed to be
computable for each input semantic state.
Therefore, for every $\alpha \in A(s_{\text{target}})$
we can effectively enumerate $\mathcal{I}(\alpha)$ and construct
the corresponding Atomic Information $\mathcal{I}_\alpha$.
Since $A(s_{\text{target}})$ is finite, the set
$\{ \mathcal{I}_\alpha \mid \alpha \in A(s_{\text{target}})\}$
is finite and computable, as claimed.
\end{proof}

\subsection{Semantic Inference Structure}
\label{subsec:semantic-inference-structure}

The predecessor function $P_O(\cdot)$ introduced above captures the
semantic causal structure among ontological states. In order to connect
this causal view with the proof-theoretic notion of \emph{derivation}
in the underlying logical system $\mathcal{L}$, we make the following alignment
assumption.

\begin{assumption}[Alignment of Causality and Inference]
\label{assump:alignment}
For every semantic state $s \in S_O$, the predecessor set $P_O(s)$
coincides with the set of its immediate inference premises in $\mathcal{L}$.
More precisely:
\begin{itemize}
  \item \textbf{Soundness of predecessors:} For every $s' \in P_O(s)$,
  there exists a single inference step in $\mathcal{L}$ that derives $s$ from
  $s'$ (possibly together with elements of $P_O(s) \setminus \{s'\}$ as
  side premises).

  \item \textbf{Completeness of predecessors:} Conversely, whenever
  there is a single-step inference in $\mathcal{L}$ whose conclusion is $s$, all
  semantic premises that this inference depends on are contained in
  $P_O(s)$.
\end{itemize}
In other words, the directed graph induced by $P_O(\cdot)$ is (up to
inessential bureaucratic details of the proof system) identical to the
derivation graph generated by single-step inference rules of $\mathcal{L}$ over
$S_O$.
\end{assumption}

On this basis, we define the global set of semantic atomic states as
\begin{equation}
  \Atom(S_O) \coloneqq \{\, s \in S_O \mid P_O(s) = \emptyset \,\}.
  \label{eq:Atom-So}
\end{equation}
Intuitively, $\Atom(S_O)$ collects all ontological states that have no
semantic predecessors and therefore serve as the ``primitive'' or
``axiomatic'' basis for subsequent derivations in $\mathcal{L}$. By
Axioms~1--3, $\Atom(S_O)$ is well-defined; it may be finite or
infinite, but any particular target state $s_{\text{target}}$ depends
only on a finite subset of $\Atom(S_O)$ identified in
Theorem~\ref{thm:semantic-atoms}.

\subsection{Derivation Depth}
\label{subsec:derivation-depth}

With the atomic basis and the semantic inference structure in place,
we can now define the complexity of any derived state in terms of its
distance from the atomic layer along the causal/inference graph~\cite{Buss1987Boolean}\cite{Krajicek1995Bounded}.

\begin{definition}[Derivation Depth]
\label{def:derivation-depth}
Under Assumption~\ref{assump:alignment}, the predecessor function
$P_O(\cdot)$ induces the immediate semantic inference structure of $\mathcal{L}$
over $S_O$. For any state $s \in S_O$, the \emph{Derivation Depth} of
$s$, denoted $\Dd(s)$, is defined inductively as follows:
\begin{equation}
  \Dd(s) \coloneqq
  \begin{cases}
    0, & \text{if } P_O(s) = \emptyset, \\[0.5ex]
    1 + \displaystyle\min_{s' \in P_O(s)} \Dd(s'), & \text{otherwise.}
  \end{cases}
  \label{eq:Dd-inductive}
\end{equation}

Equivalently, $\Dd(s)$ is the length of the shortest finite derivation
chain
\begin{equation}
  s_0, s_1, \dots, s_n = s
  \label{eq:derivation-chain}
\end{equation}
such that
\begin{enumerate}
  \item $s_0 \in \Atom(S_O)$ (i.e., $P_O(s_0) = \emptyset$);
  \item $s_i \in P_O(s_{i+1})$ for all $i = 0, \dots, n-1$.
\end{enumerate}
In this case we write
\[
  \Dd(s) = \min \{\, n \in \mathbb{N}_0 \mid
    \exists s_0, \dots, s_n \text{ as in \eqref{eq:derivation-chain}} \,\}.
\]
In particular, for any atomic state
$\alpha \in \Atom(S_O)$, we have $\Dd(\alpha) = 0$.
\end{definition}


\begin{theorem}[Computability of Derivation Depth]
\label{thm:computability-Dd}
For any state $s \in S_O$, the derivation depth $\Dd(s)$ defined in
\eqref{eq:Dd-inductive} is a unique, finite, and computable integer.
\end{theorem}

\begin{proof}
We view the directed graph over $S_O$ induced by the predecessor
function $P_O(\cdot)$. By Axiom~1 (Strict Temporal Precedence), every
edge $s' \in P_O(s)$ leads from $s$ to a strictly earlier time-stamp
$t' < t$. By Axiom~3 (Lower-Bounded and Discrete Time), there is no
infinite strictly decreasing sequence of time-stamps. Hence, every
backward path
\[
  s = s_n, s_{n-1}, \dots, s_0
\]
following predecessors $s_{i-1} \in P_O(s_i)$ has finite length.
Equivalently, the backward expansion tree rooted at any $s \in S_O$
has finite depth.

By Axiom~2 (Finite and Computable Predecessors), each state $s$ has a
finite, effectively computable predecessor set $P_O(s)$. Thus, the
backward expansion from $s$ yields a finite, effectively traversable
directed acyclic graph (DAG) whose edges are given by the predecessor
relation.

Now consider the recursive definition \eqref{eq:Dd-inductive}:
\[
  \Dd(s) =
  \begin{cases}
    0, & \text{if } P_O(s) = \emptyset, \\[0.5ex]
    1 + \displaystyle\min_{s' \in P_O(s)} \Dd(s'), & \text{otherwise.}
  \end{cases}
\]

\emph{Well-foundedness.}
Every recursive call from $s$ to $s' \in P_O(s)$ strictly decreases
the associated time-stamp (by Axiom~1). Since time is bounded below
and discrete (Axiom~3), there is no infinite descending chain of such
calls. Therefore, the recursion is well-founded.

\emph{Termination.}
The recursion depth is bounded by the maximal length of a backward
predecessor chain starting from $s$, which is finite as argued above.
Hence, the evaluation of $\Dd(s)$ terminates after finitely many
steps.

\emph{Uniqueness and computability.}
For any non-atomic state $s$ with $P_O(s) \neq \emptyset$, the set
$\{\Dd(s') \mid s' \in P_O(s)\}$ consists of finitely many
well-defined integers (by the well-founded recursion). The minimum
over this finite set is unique. Moreover, because $P_O(s)$ is
effectively computable for every $s$ (Axiom~2), we can compute
$\Dd(s)$ by a bottom-up dynamic-programming traversal of the finite
backward DAG: first assign depth $0$ to all atomic states
($P_O(s) = \emptyset$), then iteratively compute $\Dd(s)$ for states
whose predecessors' depths are already known. Thus, $\Dd(s)$ is a
unique, finite, and computable integer for every $s \in S_O$.
\end{proof}

\begin{proposition}[Derivation depth as Bennett-style logical depth]
\label{prop:bennett-depth}
Let $\mathcal{L}$ be the underlying logical system and let $P$ be a fixed sound
and complete proof system for $\mathcal{L}$. Let $\Gamma$ be a finite base
theory, and let $\varphi$ be any formula such that $\Gamma \vdash_P
\varphi$. We write $Dd(\varphi \mid \Gamma)$ for the minimal derivation
depth of $\varphi$ with respect to the semantic/inference structure
induced by $\Gamma$ in the sense of Definition~\ref{def:derivation-depth}.

Fix a universal Turing machine $U$, and fix a computable encoding
scheme $\langle \cdot \rangle$ that encodes formulas, proofs, and
finite sets of formulas as finite strings. Assume that the following
two conditions hold.

\begin{enumerate}
  \item[{\normalfont(EPS)}] \textbf{Efficient proof simulation.}
  There exists a constant $c_1 > 0$ such that, for any proof
  $\pi$ in $P$ of length\footnote{Here the length of $\pi$ is measured
  as the number of inference steps in $P$.} $n$ whose conclusion is
  $\varphi$, the universal machine $U$, given as input
  $\langle \mathrm{Atom}(\Gamma) \rangle$ and $\langle \pi \rangle$,
  can verify $\pi$ and output $\langle \varphi \rangle$ within time at
  most $c_1 \cdot n$.

  \item[{\normalfont(NPC)}] \textbf{No super-linear proof compression into time.}
  There exists a constant $c_2 > 0$ such that, for any program $p$ for
  $U$ with oracle access to $\mathrm{Atom}(\Gamma)$ that outputs
  $\langle \varphi \rangle$ in time $T$, there is a proof $\pi$ in $P$
  of $\varphi$ from $\Gamma$ of length at most $c_2 \cdot T$.
\end{enumerate}

Then there exist constants $c, C > 0$ (depending only on $U$, $P$ and
the encoding scheme, but not on $\Gamma$ or $\varphi$) such that for
all $\varphi$ with $\Gamma \vdash_P \varphi$ we have
\begin{equation}
\begin{split}
  c \cdot \mathrm{Dd}(\varphi \mid \Gamma)
  \;\le\;& 
  \mathrm{depth}_{C \cdot \mathrm{Dd}(\varphi \mid \Gamma)}
    \bigl(\langle \varphi \rangle \mid \langle \mathrm{Atom}(\Gamma)\rangle\bigr) \\
  \;\le\;& 
  C \cdot \mathrm{Dd}(\varphi \mid \Gamma),
\end{split}
\label{eq:bennett-bounds}
\end{equation}
where $\mathrm{depth}_{t}(x \mid y)$ denotes Bennett's logical depth~\cite{bennett1988logical} of
string $x$ at significance level $t$, conditional on $y$, with respect
to the universal machine $U$.

In particular, under {\normalfont(EPS)} and {\normalfont(NPC)} we have
\begin{equation}
  \mathrm{depth}_{O(Dd(\varphi \mid \Gamma))}
    \bigl(\langle \varphi \rangle \mid \langle \mathrm{Atom}(\Gamma)\rangle\bigr)
  \;=\;
  \Theta\bigl(Dd(\varphi \mid \Gamma)\bigr),
  \label{eq:bennett-theta}
\end{equation}
where the implicit constants in the $O(\cdot)$ and $\Theta(\cdot)$
notation depend only on $U$, $P$, and the chosen encodings.
\end{proposition}


\section{Derivation Depth as an Information Metric}
\label{sec:logical-depth}

We now establish that derivation depth is an information-theoretic quantity dual to Shannon entropy~\cite{shannon1948mathematical}, bridging the gap between logical reasoning and information processing.


\subsection{Derivation Entropy and Information Content}
\label{Derivation Entropy}

\begin{definition}[Derivation Entropy -- General Form]
\label{def:derivation-entropy}
The derivation entropy from $\psi$ to $\phi$ is defined as:
\begin{equation}
H_{\mathrm{derive}}(\phi | \psi) = \mathrm{Dd}(\phi | \psi) \cdot \lambda,
\end{equation}
where $\mathrm{Dd}(\phi | \psi)$ is the derivation depth (Definition~\ref{def:derivation-depth}) 
and $\lambda > 0$ is the information cost per inference step.
\end{definition}

\begin{remark}[Physical Interpretation of $\lambda$]
\label{rem:lambda-physical}
The parameter $\lambda$ quantifies the information-theoretic cost 
per inference step:
\begin{enumerate}
\item \textbf{Classical thermodynamics:} Each logically irreversible 
  step erases $\geq 1$ bit. By Landauer's principle, $\lambda = \ln 2$ nats~\cite{landauer1961irreversibility}.
\item \textbf{Quantum mechanics:} By the Margolus--Levitin bound, 
  $\lambda = \pi/2$ in units of $\hbar$~\cite{Margolus1998maximum}.
\item \textbf{Proof system structure:} Selecting from \( |\mathcal{R}| \) rules per step suggests
\( \lambda \gtrsim \ln |\mathcal{R}| \) nats,
as a structural lower bound on the information cost of inference.
\end{enumerate}

\begin{table}[htbp]
\caption{Analogy Between Thermodynamic and Information-Theoretic Costs}
\label{tab:analogy}
\begin{tabularx}{\columnwidth}{@{}>{\raggedright\arraybackslash}X
                              >{\raggedright\arraybackslash}X
                              >{\raggedright\arraybackslash}X
                              >{\raggedright\arraybackslash}X@{}}
\toprule
Domain & Discrete Unit & Continuous Unit & Conversion Factor \\
\midrule
Thermodynamics & 1 bit erasure & Energy (J) & $k_B T \ln 2$ \\
Information Physics & 1 inference step & Information (nats) & $\ln 2$ \\
\bottomrule
\end{tabularx}
\end{table}
\end{remark}

\begin{assumption}[Classical Thermodynamic Regime]
\label{ass:classical-regime}
The knowledge system operates in the classical thermodynamic regime:
(1) Inference steps are logically irreversible at thermal equilibrium;
(2) Each step erases $O(1)$ bits;
(3) Landauer erasure dominates other costs.
Under this assumption, $\lambda = \ln 2$ nats.
\end{assumption}

Throughout the remainder of this paper, unless otherwise specified, we adopt the classical thermodynamic regime with $\lambda = \ln 2$ nats, so that derivation entropy is measured in the same units (nats) as Shannon entropy.

\begin{corollary}[Derivation Entropy in Classical Regime]
\label{cor:derivation-classical}
Under Assumption~\ref{ass:classical-regime}:
$H_{\mathrm{derive}}(\phi | \psi) = \mathrm{Dd}(\phi | \psi) \cdot \ln 2$.
\end{corollary}

\begin{lemma}[Derivation Encoding]
\label{lem:encoding}
Let $\mathcal{P}$ be a proof system with rule set $R$ and maximum 
rule arity $k$. Any derivation $\pi$ of length $n$ from $m$ premises 
satisfies:
\begin{equation}
|\mathrm{Enc}(\pi)| \leq n \cdot c_L \cdot \log(m+n) + O(\log(m+n)),
\end{equation}
where $c_L = \max\{\lceil\log_2|R|\rceil, k\}$.
\end{lemma}

\begin{proof}
A derivation $\pi = ((r_1, \vec{p}_1), \ldots, (r_n, \vec{p}_n))$ 
consists of $n$ steps. Each step $(r_i, \vec{p}_i)$ encodes:
\begin{itemize}
\item \textbf{Rule:} $r_i \in R$ in $\lceil\log_2|R|\rceil$ bits.
\item \textbf{Premises:} $\vec{p}_i = (p_{i,1}, \ldots, p_{i,k_i})$ 
  with $k_i \leq k$ pointers, each in $\{1, \ldots, m+i-1\}$, 
  requiring $\leq k \cdot \lceil\log_2(m+n)\rceil$ bits total.
\end{itemize}

Per-step encoding: $\lceil\log_2|R|\rceil + k \cdot \lceil\log_2(m+n)\rceil 
\leq c_L \cdot (1 + \log_2(m+n))$.

Over $n$ steps: $n \cdot c_L \cdot (1 + \log_2(m+n))$.

Adding $O(\log(m+n))$ header bits for $m, n$, and noting that 
$n \cdot c_L \leq n \cdot c_L \cdot \log(m+n)$ when $\log(m+n) \geq 1$:
\[
|\mathrm{Enc}(\pi)| \leq n \cdot c_L \cdot \log(m+n) + O(\log(m+n)). \qedhere
\]
\end{proof}

\begin{definition}[Information-Rich Regime]
\label{def:info-rich}
Let $S_O$ be a knowledge base and let $q$ be a query (logical
formula) over $S_O$. We define the conditional algorithmic entropy of
$q$ given $S_O$ as
\begin{equation}
  H(q \mid S_O)
  \;:=\;
  K\bigl(q \mid \langle \Atom(S_O) \rangle\bigr),
\end{equation}
where $K(\cdot \mid \cdot)$ denotes conditional Kolmogorov complexity
with respect to a fixed universal Turing machine and a computable
encoding scheme $\langle \cdot \rangle$~\cite{li2008introduction}.

We say that $q$ lies in the \emph{information-rich regime} (with
respect to $S_O$) if
\begin{equation}
  H(q \mid S_O)
  \;\ge\;
  \omega\bigl(\log |\Atom(S_O)|\bigr),
  \label{eq:info-rich-omega}
\end{equation}
that is, equivalently,
\begin{equation}
  \frac{H(q \mid S_O)}{\log |\Atom(S_O)|}
  \;\longrightarrow\;
  \infty
  \qquad \text{as } |\Atom(S_O)| \to \infty.
  \label{eq:info-rich-ratio}
\end{equation}

For query classes satisfying the polynomial-depth condition
\begin{equation}
  Dd(q \mid S_O)
  \;\le\;
  \mathrm{poly}\bigl(|\Atom(S_O)|\bigr),
\end{equation}
(e.g., Datalog~\cite{ceri1989you}, SPARQL~\cite{perez2009semantics}), and template-based natural-language queries),
conditions~\eqref{eq:info-rich-omega}–\eqref{eq:info-rich-ratio} are
asymptotically equivalent to
\begin{equation}
  H(q \mid S_O)
  \;\gg\;
  \log \max\bigl\{ |\Atom(S_O)|,\, Dd(q \mid S_O) \bigr\},
  \label{eq:info-rich-alt}
\end{equation}
as $|\Atom(S_O)| \to \infty$.
\end{definition}

\begin{lemma}[Derivation Incompressibility in the Information-Rich Regime]
\label{lem:incompressibility}
Let $S_O$ be a knowledge base and $q$ a logical formula such that:
\begin{enumerate}
    \item[(i)] $n = Dd(q \mid S_O) \geq 1$,
    \item[(ii)] $m = |\mathrm{Atom}(S_O)|$,
    \item[(iii)] $q$ lies in the information-rich regime (Definition~\ref{def:info-rich}),
    \item[(iv)] $n = \mathrm{poly}(m)$ (e.g., for bounded-depth reasoning systems).
\end{enumerate}
Then there exists a constant $c_2 > 0$, depending only on the proof system and encoding scheme, such that the conditional Kolmogorov complexity of $q$ given $S_O$ satisfies
\[
K\bigl(q \mid \langle \mathrm{Atom}(S_O) \rangle\bigr) \geq c_2 \cdot n \log(m + n) - O(\log m).
\]
\end{lemma}

\begin{proof}
By Lemma~\ref{lem:encoding}, any derivation of $q$ from $\mathrm{Atom}(S_O)$ of length $n$ can be encoded in at most
\[
L_{\text{enc}} = c_1 \cdot n \log(m + n) + O(\log m)
\]
bits for some constant $c_1 > 0$. This establishes an upper bound on $K(q \mid \langle \mathrm{Atom}(S_O) \rangle)$.

Now suppose, for contradiction, that no such lower bound holds. Then for every $\varepsilon > 0$, there exist arbitrarily large $m$ and corresponding $q$ such that
\[
K\bigl(q \mid \langle \mathrm{Atom}(S_O) \rangle\bigr) < \varepsilon \cdot n \log(m + n).
\]
Since $n = \mathrm{poly}(m)$, we have $\log(m + n) = \Theta(\log m)$, so the above implies
\[
K\bigl(q \mid \langle \mathrm{Atom}(S_O) \rangle\bigr) = o\bigl(n \log m\bigr).
\]
Consequently, the conditional algorithmic entropy satisfies
\[
H(q \mid S_O) = K\bigl(q \mid \langle \mathrm{Atom}(S_O) \rangle\bigr) \cdot \ln 2 = o\bigl(n \log m\bigr).
\]
But because $n \geq 1$ and $n = \mathrm{poly}(m)$, we have $n \log m \geq \log m$, and in fact $n \log m \gg \log m$ as $m \to \infty$. Hence
\[
H(q \mid S_O) = o\bigl(n \log m\bigr) \not\gg \log m,
\]
which contradicts condition (iii)—that $q$ lies in the information-rich regime. Therefore, there must exist a constant $c_2 > 0$ such that
\[
K\bigl(q \mid \langle \mathrm{Atom}(S_O) \rangle\bigr) \geq c_2 \cdot n \log(m + n) - O(\log m),
\]
as claimed.
\end{proof}

Throughout this subsection, all asymptotic notation (\(O(\cdot)\),
\(\Theta(\cdot)\), \(\omega(\cdot)\), \(\gg\), etc.) is taken with respect
to
\[
  m \;:=\; |\Atom(S_O)| \;\longrightarrow\; \infty,
\]
under the standing assumption that
\[
  n \;:=\; Dd(q \mid S_O) \;=\; \mathrm{poly}(m).
\]

\begin{theorem}[Derivation Depth as an Information Metric]
\label{thm:derivation-depth-info-metric}
Let $S_O$ be a knowledge base and let $q$ be a query satisfying
conditions~\textup{(i)}–\textup{(iv)} of Lemma~\ref{lem:incompressibility} (in particular, $q$
lies in the information-rich regime of
Definition~\ref{def:info-rich} and
$n := Dd(q \mid S_O) = \mathrm{poly}(m)$ with
$m := |\Atom(S_O)|$). Define
\begin{equation}
  H(q \mid S_O)
  \;:=\;
  K\bigl(q \mid \langle \Atom(S_O) \rangle\bigr) \cdot \ln 2,
\end{equation}
and, under the classical thermodynamic regime
(Assumption~\ref{ass:classical-regime}), define the derivation entropy as
\begin{equation}
  H_{\mathrm{derive}}(q \mid S_O)
  \;:=\;
  Dd(q \mid S_O) \cdot \ln 2.
\end{equation}
Then
\begin{equation}\label{eq:derive-bound}
  H_{\mathrm{derive}}(q\!\mid\!S_O)
  \!=\!\Theta\!\left(
    \frac{H(q \mid S_O)}{\!\log\!\big(|\Atom(S_O)|\! +\! \mathrm{Dd}(q \mid S_O)\big)}
  \right)\!,
\end{equation}
where the implicit constants in the $\Theta(\cdot)$ notation depend
only on the choice of universal Turing machine, proof system, and
encoding scheme, but not on $S_O$ or $q$.
\end{theorem}

\begin{proof}
Let $m = |\Atom(S_O)|$ and $n = Dd(q \mid S_O)$. By Lemma~\ref{lem:encoding} and~\ref{lem:incompressibility}, there exist constants $c_1$, $c_2>0$, or more precisely, $c_1$ and $c_2$ depend only on the universal Turing machine, proof system $\mathcal{P}$, and encoding scheme, but not on the polynomial degree in the assumption $n = {\rm poly}(m)$, such that
\begin{multline*}
  c_2 \, n \log(m + n) - O(\log m)
  \;\le\;
  K\bigl(q \mid \langle \Atom(S_O) \rangle\bigr)\\
  \;\le\;
  c_1 \, n \log(m + n) + O(\log m).
\end{multline*}
Multiplying both sides by $\ln 2$ yields
\begin{multline*}
  c_2' \, n \log(m + n) - O(\log m)
  \;\le\;
  H(q \mid S_O)\\
  \;\le\;
  c_1' \, n \log(m + n) + O(\log m),
\end{multline*}
for suitable constants $c_1', c_2' > 0$ with $c_i' = c_i \ln 2$.
Because $n = \mathrm{poly}(m)$, we have $\log(m + n) = \Theta(\log m)$,
and the $O(\log m)$ term is asymptotically dominated by
$n \log(m + n)$. Hence
\begin{equation*}
  H_{\mathrm{derive}}(q\!\mid\!S_O)
  \!=\! \Theta\!\left(\!
    \frac{H(q \mid S_O)}{\log\!\big(|\Atom(S_O)|\!+\!\mathrm{Dd}(q \mid S_O)\big)}
  \!\right)\!,
\end{equation*}
Rearranging gives
\begin{equation*}
  n
  \;=\;
  \Theta\!\left(
    \frac{H(q \mid S_O)}{\log(m + n)}
  \right).
\end{equation*}
Finally, substituting
$H_{\mathrm{derive}}(q \mid S_O) = n \cdot \ln 2$ and recalling that
$m = |\Atom(S_O)|$ and $n = Dd(q \mid S_O)$, we obtain
\begin{equation*}
  H_{\mathrm{derive}}(q\!\mid\!S_O)
  \!=\!\Theta\!\left(
    \frac{H(q \mid S_O)}{\!\log\!\big(|\Atom(S_O)|\! +\! \mathrm{Dd}(q \mid S_O)\big)}
  \right)\!,
\end{equation*}
which is exactly~\eqref{eq:derive-bound}. This
completes the proof.
\end{proof}


\subsection{Frequency-weighted Storage–Computation Tradeoff for a Single Query}\label{subsec:freq_tradeoff}

In this subsection we combine the semantic information metrics to obtain a quantitative tradeoff between
\emph{storage} and \emph{computation} under a given query distribution.

\begin{definition}[Access Frequency]
\label{def:access-freq}
For a query distribution $P_Q$ over query set $Q$ and a total of $N$ queries, 
the \emph{access frequency} (expected access count) of query $q$ is:
\begin{equation}\label{eq:Hq-def}
f_q := N \cdot P_Q(q).
\end{equation}
Thus $f_q \in [0, N]$ with $\sum_{q \in Q} f_q = N$.
\end{definition}

Throughout this subsection we assume Definition~\ref{def:access-freq} has been fixed, and we keep its notation.
In particular, we let $P_Q$ denote the query distribution over a set $Q$ of admissible queries, and
\[
  f_q := N \cdot P_Q(q)
\]
denote the expected access count of query $q$ over $N$ total queries.%
\footnote{%
  As in Definition~\ref{def:access-freq}, $f_q$ is a \emph{dimensionless} quantity: it is a pure count (expected number of accesses) without an intrinsic time unit.
  When we speak of \emph{amortized cost per access}, we will divide a total bit-length (storage) or total semantic derivation cost (in nats) by this dimensionless $f_q$ to obtain a per-access cost in the corresponding information unit.
}

We now state the central theorem of this section.

\begin{theorem}[Frequency-weighted storage--computation tradeoff]\label{thm:freq_tradeoff}
Let $S_O$ be a knowledge base with atomic semantic state set $\Atom(S_O)$, and let $q$ be a query in the information-rich regime with respect to $S_O$ in the sense of Definition~\ref{def:info-rich}.
Let $P_Q$ be a query distribution over a query set $Q$, and define $f_q := N \cdot P_Q(q)$ as in Definition~\ref{def:access-freq}.

Consider a hybrid strategy $\Pi = (S, C, R)$ where:
\begin{enumerate}
  \item $S$ is an auxiliary storage structure (e.g., materialized facts, indices, intermediate derivations) that is \emph{effectively encodable} as a binary string $\enc{S}$ of length $\abs{S}$ bits.
  \item $C$ and $R$ describe the residual computation (inference) carried out over the combined knowledge base $S_O \cup S$.
\end{enumerate}

We impose the following assumptions on $S$ and on inference over
$S_O \cup S$:

\begin{enumerate}
  \item\textbf{(SA1) Information-theoretic measurement of storage.}
    The bit length $|S|$ of the encoding $\enc{S}$ satisfies
    \begin{equation}
      H(S) \;\le\; |S| \;\le\; H(S) + O(1),
    \end{equation}
    where $H(S)$ is the Shannon entropy of $S$ for a suitable prior
    over admissible storage states.
    In particular,
    \begin{equation}
      I(q; S \mid S_O) \;\le\; H(S) \;\le\; |S|.
    \end{equation}

  \item\textbf{(SA2) Structural regularity of the extended knowledge base.}
    The combined knowledge base $S_O \cup S$ admits a well-defined
    atomic semantic state set $\Atom(S_O \cup S)$, a predecessor
    function satisfying Axioms~\ref{ax:strict-temporal-precedence}-\ref{ax:lower-bounded-time}, and hence a derivation depth
    function $Dd(\cdot \mid S_O \cup S)$ as in
    Definition~\ref{def:derivation-depth}.
    We write
    \begin{equation}
      m_S := |\Atom(S_O \cup S)|, 
      n_S := Dd(q \mid S_O \cup S).
    \end{equation}
    We assume $n_S = \mathrm{poly}(m_S)$ (polynomial-depth regime)
    and that $q$ remains in the information-rich regime with respect
    to $S_O \cup S$ (in the sense of Definition~\ref{def:info-rich},
    replacing $S_O$ by $S_O \cup S$). Moreover, the size/depth parameters of $S_O \cup S$ are of the same polynomial order as those of $S_O$.

  \item\textbf{(SA3) Applicability of the derivation-depth bound.}
    The proof-system/encoding conditions (EPS) and (NPC) of
    Proposition~\ref{prop:bennett-depth} hold not only for the base theory
    induced by $S_O$, but also for the theory induced by
    $S_O \cup S$.
    Consequently, the asymptotic relation of
    Theorem~\ref{thm:derivation-depth-info-metric} applies to inference from
    $S_O \cup S$: there exist constants $c_1, c_2 > 0$ such that
    \begin{multline}
      c_1\,\frac{H(q \mid S_O \cup S)}{\log(m_S + n_S)} - O(1)
      \!\;\le\;\!
      H_{\mathrm{derive}}(q \mid S_O \cup S)
     \\ \;\le\;
      c_2\,\frac{H(q \mid S_O \cup S)}{\log(m_S + n_S)} + O(1),
    \end{multline}
    where
    \begin{equation*}
      H_{\mathrm{derive}}(q \mid S_O \cup S)
      := Dd(q \mid S_O \cup S)\,\lambda
    \end{equation*}
    is the derivation entropy (Definition~\ref{def:derivation-entropy})
    with per-step cost $\lambda > 0$, and all entropies are measured
    in nats.
\end{enumerate}

Under these assumptions, define:
\begin{itemize}
  \item The \emph{semantic information content} of $q$
        (Definition~\ref{def:info-rich}):
        \begin{equation}
          H_q \!:= \!H(q \mid S_O)
          \!:=\! K\big(\enc{q} \mid \enc{\Atom(S_O)}\big)\,\ln 2.
        \end{equation}

  \item The \emph{effective derivation cost factor} associated with $S$:
        \begin{equation}\label{eq:LS-def}
          L_S := \log(m_S + n_S),
        \end{equation}
        which captures the logarithmic compression advantage available
        through derivations over $S_O \cup S$.

  \item The \emph{amortized cost per access} of answering $q$ using
        strategy $\Pi$:
        \begin{equation}
          \Cost_{\mathrm{amort}}(\Pi; q)
          := \frac{|S|}{f_q} + H_{\mathrm{derive}}(q \mid S_O \cup S).
        \end{equation}
\end{itemize}
Then there exist positive constants $c, C > 0$ (depending only on the
universal Turing machine, proof system, and encoding conventions, but
not on $S_O$, $q$, $S$, or $P_Q$) such that
\begin{equation}
  \Cost_{\mathrm{amort}}(\Pi; q)
  \;\le\;
  \min\left\{
    \frac{H_q}{f_q},\;
    C\,\frac{H_q}{L_S}
  \right\}
  + O(1).
\end{equation}
In particular, up to additive $O(1)$ and multiplicative constant
factors, there exists a strategy $\Pi$:
\begin{enumerate}
  \item[\emph{(1)}] \emph{Low-frequency regime} ($f_q \le L_S$,
        on-demand computation dominated):
        \begin{equation}
          \Cost_{\mathrm{amort}}(\Pi; q)
          \;\le\;
          C\,\frac{H_q}{L_S} + O(1).
        \end{equation}

  \item[\emph{(2)}] \emph{High-frequency regime} ($f_q > L_S$,
        precomputation/storage dominated):
        \begin{equation}
          \Cost_{\mathrm{amort}}(\Pi; q)
          \;\le\;
          \frac{H_q}{f_q} + O(1).
        \end{equation}
\end{enumerate}
The break-even frequency scale is given (up to constant factors) by
\begin{equation}
f_c \asymp L_S = \log(m_S + n_S),
\end{equation}
where, for the baseline case $S = \emptyset$, we have 
$f_c = \log(|\mathrm{Atom}(S_O)| + D_d(q | S_O))$; 
see Remark~\ref{rem:fc} for the general case.
\end{theorem}

\begin{proof}
We sketch the main steps.

\textbf{Step 0: Information-theoretic preliminaries}
By (SA1) and the data-processing inequality,
\begin{equation}
  I(q; S \mid S_O) \;\le\; H(S) \;\le\; |S|.
\end{equation}
Using the mutual-information identity
\begin{equation}
  H(q \mid S_O)
  = H(q \mid S_O, S) + I(q; S \mid S_O),
\end{equation}
we obtain
\begin{equation}
  H(q \mid S_O, S)
  = H_q - I(q; S \mid S_O)
  \ge H_q - |S|.
\end{equation}
Thus $S$ can reduce the conditional entropy of $q$ (given $S_O$) by at
most $|S|$ nats.

\textbf{Step 1: Applying Theorem~\ref{thm:derivation-depth-info-metric}.} 
By (SA2) and (SA3), Theorem~\ref{thm:derivation-depth-info-metric} applies to
$S_O \cup S$ and yields
\begin{equation}
  H_{\mathrm{derive}}(q \mid S_O \cup S)
  \;\le\;
  c_2\,\frac{H(q \mid S_O \cup S)}{L_S} + O(1).
\end{equation}
Using the bounds
\begin{equation}
  H_q - |S| \;\le\; H(q \mid S_O \cup S) \;\le\; H_q,
\end{equation}
we obtain
\begin{equation}
  H_{\mathrm{derive}}(q \mid S_O \cup S)
  \;\le\;
  c_2\,\frac{H_q}{L_S} + O\!\left(\frac{|S|}{L_S}\right) + O(1).
\end{equation}

\textbf{Step 2: Amortized cost.} By definition,
\begin{equation}
  \Cost_{\mathrm{amort}}(\Pi; q)
  = \frac{|S|}{f_q} + H_{\mathrm{derive}}(q \mid S_O \cup S).
\end{equation}
Substituting the previous inequality,
\begin{equation}
  \Cost_{\mathrm{amort}}(\Pi; q)
  \;\le\;
  \frac{|S|}{f_q}
  + c_2\,\frac{H_q}{L_S}
  + O\!\left(\frac{|S|}{L_S}\right) + O(1).
\end{equation}
This decomposition makes explicit the storage-amortization term
$|S|/f_q$ and the derivation-entropy term $\sim H_q/L_S$.

\textbf{Step 3 Extremal strategies.}
To obtain a clean upper bound in terms of $(H_q, f_q, L_S)$ alone, we
compare against two constructive extremes.

\emph{(a) Pure computation strategy} $\Pi_{\mathrm{comp}}$:
take $S = \varnothing$.
Then, by Theorem~\ref{thm:derivation-depth-info-metric} applied to $S_O$,
\begin{multline}
  \Cost_{\mathrm{amort}}(\Pi_{\mathrm{comp}}; q)
  = H_{\mathrm{derive}}(q \mid S_O)
 \\ \;\asymp\;
  \frac{H_q}{\log\big(|\Atom(S_O)| + Dd(q \mid S_O)\big)}
 \\ \;\le\;
  C_0\,\frac{H_q}{L_S} + O(1)
\end{multline}
for some constant $C_0 > 0$, where the last inequality uses that
$L_S = \log(m_S + n_S)$ is of the same asymptotic order as
$\log(|\Atom(S_O)| + Dd(q \mid S_O))$ under the polynomial-depth
assumptions.

\emph{(b) Pure storage strategy} $\Pi_{\mathrm{store}}$:
by the coding theorem and the definition of $H_q$, there exists a
conditional code for the answers to $q$ (given $S_O$) of length at most
$H_q + O(1)$ nats.
We can store such a near-optimal representation in $S$, so that
$|S| \le H_q + O(1)$, and design $C$ and $R$ so that answering $q$
from $S_O \cup S$ requires only constant derivation cost:
\begin{equation}
  H_{\mathrm{derive}}(q \mid S_O \cup S) = O(1).
\end{equation}
Hence
\begin{equation}
  \Cost_{\mathrm{amort}}(\Pi_{\mathrm{store}}; q)
  = \frac{|S|}{f_q} + O(1)
  \;\le\;
  \frac{H_q}{f_q} + O(1).
\end{equation}

Since we are interested in achievable costs, the optimal amortized
cost
\begin{equation}
  \Cost_{\mathrm{amort}}^{\star}(q)
  := \inf_{\Pi} \Cost_{\mathrm{amort}}(\Pi; q)
\end{equation}
is bounded above by the minimum of these two explicit strategies:
\begin{multline}
  \Cost_{\mathrm{amort}}^{\star}(q)
 \\ \;\le\;
  \min\big\{
    \Cost_{\mathrm{amort}}(\Pi_{\mathrm{comp}}; q),\,
    \Cost_{\mathrm{amort}}(\Pi_{\mathrm{store}}; q)
  \big\}
 \\ \;\le\;
  \min\left\{
    C_0\,\frac{H_q}{L_S},\;
    \frac{H_q}{f_q}
  \right\}
  + O(1).
\end{multline}
Equivalently, there exists $C>0$ such that
\begin{equation}
  \Cost_{\mathrm{amort}}^{\star}(q)
  \;\le\;
  \min\left\{
    \frac{H_q}{f_q},\;
    C\,\frac{H_q}{L_S}
  \right\}
  + O(1),
\end{equation}
establishing the claimed bound.

\noindent\textbf{Step 4: Frequency regimes.}
From the last inequality we obtain two asymptotic regimes:
\begin{itemize}
  \item If $f_q \le L_S$, then
        $\frac{H_q}{f_q} \ge \frac{H_q}{L_S}$ and
        \begin{equation}
          \Cost_{\mathrm{amort}}^{\star}(q)
          \;\le\;
          \frac{H_q}{f_q} + O(1).
        \end{equation}

  \item If $f_q > L_S$, then
        $\frac{H_q}{f_q} < \frac{H_q}{L_S}$ and
        \begin{equation}
          \Cost_{\mathrm{amort}}^{\star}(q)
          \;\le\;
          C\,\frac{H_q}{L_S} + O(1).
        \end{equation}
\end{itemize}
The transition occurs when $f_q$ is on the order of $L_S$, i.e.,
$f_c \asymp L_S = \log(m_S + n_S)$.
\end{proof}

\begin{remark}[Operational interpretation of critical frequency]
\label{rem:fc}
The critical frequency $f_c$ admits the following operational interpretation:

\begin{enumerate}
\item \textbf{Baseline case} ($S = \emptyset$): When no auxiliary storage 
is used, the critical frequency is determined solely by the base knowledge:
\[
f_c^{(0)} := \log\bigl(|\mathrm{Atom}(S_O)| + D_d(q | S_O)\bigr).
\]
This is the threshold for deciding whether \emph{any} storage investment 
for query $q$ is worthwhile.

\item \textbf{Post-allocation case}: For a given storage strategy 
$\Pi = (S, C, R)$, the effective critical frequency shifts to
\[
f_c(S) := L_S = \log(m_S + n_S),
\]
where $m_S = |\mathrm{Atom}(S_O \cup S)|$ and $n_S = D_d(q | S_O \cup S)$.

\item \textbf{Decision rule}: The initial decision of whether to allocate 
storage for $q$ should compare $f_q$ against the baseline $f_c^{(0)}$, 
not against $f_c(S)$, to avoid circular reasoning. Once $S$ is chosen, 
$f_c(S)$ governs the marginal benefit of \emph{additional} storage.
\end{enumerate}
\end{remark}

\begin{remark}[Physical and engineering interpretation]
\label{rem:freq_tradeoff_physical}
Theorem~\ref{thm:freq_tradeoff} admits a natural physical and
engineering interpretation.

\medskip\noindent
\textbf{(1) Units and dimensions.}
By construction, $H_q$ and $H_{\mathrm{derive}}$ are measured in nats
(or bits up to a constant factor), while $|S|$ is measured in bits.
The frequency $f_q$ is a dimensionless expected count of accesses.
Therefore:
\begin{itemize}
  \item $|S|/f_q$ has the same information dimension as $|S|$ (bits),
        but represents an \emph{average storage cost per access};

  \item $H_{\mathrm{derive}}$ is a semantic ``work'' or ``action''
        measured in nats;

  \item the bound in Theorem~\ref{thm:freq_tradeoff} compares two
        quantities of the same information-theoretic dimension:
        per-access storage and per-access semantic computation.
\end{itemize}
In a physical implementation, both can be converted into expected
energy dissipation using Landauer's principle or related models of
irreversible computation.

\medskip\noindent
\textbf{(2) Two limiting protocols.}
The proof explicitly constructs two limiting protocols:
\begin{enumerate}
  \item A \emph{pure computation protocol}, where no auxiliary storage
        is used ($S = \varnothing$) and the cost is determined entirely
        by derivation from $S_O$; this yields an amortized cost of
        order $H_q / \log(\cdot)$.

  \item A \emph{pure storage protocol}, where we store a
        near-optimally compressed representation of the answer to $q$,
        making the derivation cost $O(1)$ and the amortized cost
        dominated by $H_q / f_q$.
\end{enumerate}
Theorem~\ref{thm:freq_tradeoff} states that, up to constant factors and
lower-order terms, no hybrid protocol can asymptotically outperform the
better of these two extremes, once $H_q$, $f_q$, and $L_S$ are fixed.

\medskip\noindent
\textbf{(3) Break-even frequency and caching.}
The critical scale $f_c \asymp L_S = \log(m_S + n_S)$ plays the role of
a \emph{break-even frequency} between caching and on-demand
computation:
\begin{itemize}
  \item If $f_q \ll f_c$, the term $H_q / f_q$ dominates; it is more
        efficient to answer $q$ by recomputation each time than to
        allocate storage to cache its answers.

  \item If $f_q \gg f_c$, the term $C H_q / L_S$ dominates; investing
        storage to reduce $H(q \mid S_O \cup S)$ (and hence
        $H_{\mathrm{derive}}$) becomes asymptotically advantageous.
\end{itemize}
This mirrors classical engineering heuristics for cache and index
design: high-frequency queries justify aggressive precomputation and
indexing, while low-frequency queries are better handled by direct
inference.

\medskip\noindent
\textbf{(4) Role of $L_S$.}
The factor $L_S = \log(m_S + n_S)$ reflects the same logarithmic
compression phenomenon as in Theorem~\ref{thm:derivation-depth-info-metric}: in the
information-rich regime, the derivation cost per nat of semantic
information scales like $1 / \log(\text{size} + \text{depth})$.
A larger, deeper, but still well-structured theory $S_O \cup S$ allows
the system to answer semantically complex queries with fewer effective
derivation steps per nat.

\medskip\noindent
\textbf{(5) Robustness.}
The qualitative form of the tradeoff---and the scale
$f_c \asymp L_S$ at which caching becomes beneficial---is robust to:
\begin{itemize}
  \item changes in the underlying universal Turing machine (which
        affect $H_q$ only by an additive constant);

  \item variations in the proof system, as long as (EPS) and (NPC)
        hold and Theorem~\ref{thm:derivation-depth-info-metric} applies;

  \item different encodings of $S$, as long as (SA1) holds, i.e., the
        code is within $O(1)$ of the Shannon-optimal length.
\end{itemize}
Thus the storage--computation tradeoff is controlled mainly by the
semantic and structural properties of $S_O$ and $S$, rather than by
low-level representational details.
\end{remark}

\begin{remark}[Amortized storage cost and query frequencies]\label{rem:amortized-frequency}
In Theorem~\ref{thm:freq_tradeoff} we analyzed the amortized cost of serving a
\emph{single fixed} query $q$ that is expected to be accessed repeatedly.
Recall that for a query distribution $P_Q$ over a finite query set $Q$ and
a total horizon of $N$ queries, we define in Definition~\ref{def:access-freq}
the expected number of accesses of $q$ by
\[
  f_q \;\coloneqq\; N \cdot P_Q(q).
\]
Here $f_q$ is a \emph{frequency} (an expected count), not a probability.
Under this convention, the one-time storage investment $|S|$ is amortized
over the $f_q$ accesses of $q$, leading to the single-query amortized cost
\begin{multline}\label{eq:amortized-single-q}
  \Cost_{\mathrm{amort}}(q)
  \;=\;
  \frac{|S|}{f_q} \;+\; H_{\mathrm{derive}}(q \mid S_O \cup S)
  \\\;=\;
  \frac{|S|}{N \cdot P_Q(q)} \;+\; H_{\mathrm{derive}}(q \mid S_O \cup S)
  .
\end{multline}
Here $|S|$ is the one-time storage cost (in bits) needed to maintain $S$,
and $H_{\mathrm{derive}}(q \mid S_O \cup S)$ is the per-access inference cost
(in bits) to derive $q$ from $(S_O,S)$.

A tempting but \emph{incorrect} way to ``average'' this cost across query
types is to write
\begin{multline}\label{eq:wrong-averaging}
  \mathbb{E}_{q\sim P_Q}
  \Bigl[
    \frac{|S|}{P_Q(q)} + H_{\mathrm{derive}}(q \mid S_O \cup S)
  \Bigr]
  \\\;=\;
  \sum_{q_i \in Q} P_Q(q_i)
  \Bigl(
    \frac{|S|}{P_Q(q_i)} + H_{\mathrm{derive}}(q_i \mid S_O \cup S)
  \Bigr)
  .
\end{multline}
By direct calculation, this yields
\[
  \sum_{q_i} P_Q(q_i)\,\frac{|S|}{P_Q(q_i)}
  \;=\;
  |S|\cdot |Q|,
\]
and hence
\begin{multline*}
  \mathbb{E}_{q\sim P_Q}
  \Bigl[
    \frac{|S|}{P_Q(q)} + H_{\mathrm{derive}}(q \mid S_O \cup S)
  \Bigr]
  \\\;=\;
  |S|\cdot |Q| \;+\;
  \mathbb{E}_{q\sim P_Q}\bigl[ H_{\mathrm{derive}}(q \mid S_O \cup S) \bigr].
\end{multline*}
This expression is conceptually wrong: it treats $|S|$ as if it were
incurred \emph{separately for each query type}, effectively charging
$|S|$ a total of $|Q|$ times. This contradicts the amortization model
underlying Theorem~\ref{thm:freq_tradeoff} and
Definition~\ref{def:access-freq}, where $|S|$ is a \emph{single}
one-time investment that is shared across all queries.

To see the problem numerically, consider an example where
$Q=\{q_1,q_2,q_3\}$,
\[
  P_Q(q_1)=0.5,\quad P_Q(q_2)=0.3,\quad P_Q(q_3)=0.2,
\]
the storage cost is $|S| = 100$ bits, and
\[
H_{\mathrm{derive}}(q_1\mid S_O\cup S)=10,
\]
\[
H_{\mathrm{derive}}(q_2\mid S_O\cup S)=20,
\] 
\[
H_{\mathrm{derive}}(q_3\mid S_O\cup S)=30.
\] 
(bits per access). Let $N=1000$ be the total number of queries.
Then the expected total inference cost is
\[
N \cdot \mathbb{E}_{q\sim P_Q}
    \bigl[ H_{\mathrm{derive}}(q \mid S_O \cup S) \bigr]
\]
\[
 \;=\;
  1000 \cdot (0.5\cdot 10 + 0.3\cdot 20 + 0.2\cdot 30)
\]
\[
 \;=\;
  17{,}000 \text{ bits},
\]
and the total cost over all queries is
\[
  |S| + 17{,}000 \;=\; 17{,}100 \text{ bits}.
\]
The correct \emph{average} cost per query is therefore
\begin{multline}\label{eq:correct-expected-cost}
  \frac{|S|}{N}
  \;+\;
  \mathbb{E}_{q\sim P_Q}
  \bigl[ H_{\mathrm{derive}}(q \mid S_O \cup S) \bigr]\\
  \;=\;
  \frac{100}{1000} + 17
  \;=\; 17.1 \text{ bits per query}.
\end{multline}
In contrast, the naive averaging in~\eqref{eq:wrong-averaging} yields
\[
  |S|\cdot |Q| + 17 \;=\; 100\cdot 3 + 17 \;=\; 317
  \text{ bits per query},
\]
which is clearly nonsensical: the one-time cost $|S|$ has been multiplied
by the number of query types $|Q|$ instead of being spread over the actual
number of query \emph{occurrences} $N$.

In summary, the correct way to account for storage in a multi-query
setting is:
\begin{itemize}
  \item treat $|S|$ as a \emph{single}, one-time storage investment;
  \item amortize it over the total number $N$ of query occurrences,
        giving a per-query storage contribution of $|S|/N$;
  \item add the expected per-query inference cost
        $\mathbb{E}_{q\sim P_Q}[H_{\mathrm{derive}}(q\mid S_O\cup S)]$.
\end{itemize}
This yields the expected cost per query as in
\eqref{eq:correct-expected-cost}:
\begin{equation}\label{eq:avg-cost-shannon}
  \mathbb{E}[\Cost]
  \;=\;
  \frac{|S|}{N}
  \;+\;
  \mathbb{E}_{q\sim P_Q}
  \bigl[ H_{\mathrm{derive}}(q \mid S_O \cup S) \bigr].
\end{equation}

If we now view $Q$ as a random variable with distribution $P_Q$, and interpret
$H_{\mathrm{derive}}(q\mid S_O\cup S)$ as a per-instance algorithmic cost
compatible with a \emph{Shannon} conditional entropy $H(Q\mid S_O,S)$,
then
\[
  \mathbb{E}_{q\sim P_Q}
  \bigl[
    H_{\mathrm{derive}}(q \mid S_O \cup S)
  \bigr]
  \;\approx\;
  H(Q\mid S_O,S).
\]
Consequently, the expected \emph{Shannon} cost term can be decomposed as
\begin{equation}\label{eq:shannon-decomposition}
  H(Q\mid S_O,S)
  \;=\;
  H(Q\mid S_O) - I(S;Q\mid S_O),
\end{equation}
where $I(S;Q\mid S_O)$ is the conditional mutual information between $S$
and $Q$ given $S_O$. Equation~\eqref{eq:shannon-decomposition} shows that
the storage $S$ is beneficial to the extent that it captures information
about the future queries $Q$ beyond what is already in $S_O$: maximizing
$I(S;Q\mid S_O)$ minimizes the residual cost $H(Q\mid S_O,S)$, and thus
reduces the expected inference burden in~\eqref{eq:avg-cost-shannon}.
\end{remark}

In summary, Theorem~\ref{thm:freq_tradeoff} provides a frequency-weighted amortized cost bound at the level of a single query type, while Remark~\ref{rem:amortized-frequency} clarifies how the one-time storage cost $|S|$ should be correctly amortized. This completes the frequency-weighted analysis at the level of a single query type.


\subsection{System-Wide Storage Allocation under Amortized Cost}\label{System-wide}

We now move from the single-query amortization view to a system-wide analysis over a full query distribution.

\begin{remark}[From single-query amortization to system-wide averages]\label{rem:single-vs-multi}
Remark~\ref{rem:amortized-frequency} clarifies that Theorem~\ref{thm:freq_tradeoff}
naturally applies first to a \emph{single fixed} query $q$ that is accessed
$f_q = N\cdot P_Q(q)$ times. In that setting, the amortized cost per access
of $q$ is
\begin{equation}\label{eq:single-q-cost-again}
  \Cost_{\mathrm{amort}}(q)
  \;=\;
  \frac{|S|}{f_q}
  \;+\;
  H_{\mathrm{derive}}(q \mid S_O \cup S),
\end{equation}
where $|S|/f_q$ is the per-access share of the one-time storage cost $|S|$.
This formula is meaningful \emph{only when} $f_q$ is understood as a
\emph{number of accesses} (frequency), in accordance with
Definition~\ref{def:access-freq}.

When we move from this single-query view to a \emph{multi-query workload}
with distribution $P_Q$ and horizon $N$, the correct aggregation is:
\begin{align*}
 & \text{total storage cost} \;=\; |S|,\\[0.3em]
  &\text{total inference cost}
    \;=\;
      \sum_{q\in Q} f_q \cdot H_{\mathrm{derive}}(q \mid S_O \cup S)\\
     & \;=\;
      N \cdot \mathbb{E}_{q\sim P_Q}
      \bigl[ H_{\mathrm{derive}}(q \mid S_O \cup S) \bigr],\\[0.3em]
  &\text{total number of queries} \;=\; \sum_{q\in Q} f_q \;=\; N.
\end{align*}
Hence the \emph{system-wide average} cost per query is exactly
\begin{equation}\label{eq:system-wide-avg}
  \mathbb{E}[\Cost]
  \;=\;
  \frac{|S|}{N}
  \;+\;
  \mathbb{E}_{q\sim P_Q}
  \bigl[ H_{\mathrm{derive}}(q \mid S_O \cup S) \bigr],
\end{equation}
which is consistent with~\eqref{eq:avg-cost-shannon}. In particular,
the one-time storage cost $|S|$ appears \emph{only once} in the numerator
and is amortized over the $N$ actual query occurrences, rather than over
the number of query \emph{types} $|Q|$.

If, incorrectly, one were to start from the single-query formula
\eqref{eq:single-q-cost-again}, replace $f_q$ by $N\cdot P_Q(q)$, and then
naively average over $q$ as in
\begin{equation}\label{eq:naive-average-again}
  \sum_{q_i\in Q}
    P_Q(q_i)
    \Bigl(
      \frac{|S|}{P_Q(q_i)}
      + H_{\mathrm{derive}}(q_i \mid S_O \cup S)
    \Bigr),
\end{equation}
one would again obtain the spurious term $|S|\cdot |Q|$ discussed in
Remark~\ref{rem:amortized-frequency}. This mistake arises from
\emph{double-counting} the storage cost $|S|$ as if it were separately
incurred for each query type. The correct interpretation is that $|S|$
is a fixed capital cost that supports the entire query workload, and
therefore must be amortized over the full horizon $N$ of query instances.

From an information-theoretic perspective, if we interpret
\[
  \mathbb{E}_{q\sim P_Q}
  \bigl[
    H_{\mathrm{derive}}(q \mid S_O \cup S)
  \bigr]
  \;\approx\;
  H(Q\mid S_O,S),
\]
with $H(\cdot\mid\cdot)$ now denoting the \emph{Shannon} conditional
entropy of the random variable $Q$, then the second term
in~\eqref{eq:system-wide-avg} can be written as
\[
  H(Q\mid S_O,S) \;=\; H(Q\mid S_O) - I(S;Q\mid S_O),
\]
where $I(S;Q\mid S_O)$ is the conditional mutual information between $S$
and $Q$ given $S_O$. This identity shows that, at the system level, the
storage $S$ is beneficial to the extent that it \emph{reduces} the
Shannon uncertainty of future queries $Q$ beyond what $S_O$ already
provides. In other words, $S$ is ``useful'' precisely when it encodes
structure about the query distribution $P_Q$ so as to increase
$I(S;Q\mid S_O)$ and thereby decrease the expected inference cost term
in~\eqref{eq:system-wide-avg}.

We remark that the per-query algorithmic complexity
$H_{\mathrm{derive}}(q\mid S_O\cup S)$ used in
Theorems~\ref{thm:resource-bound}--\ref{thm:freq_tradeoff} and the
Shannon entropy $H(Q\mid S_O,S)$ appearing here live at different levels:
the former describes instance-wise description length or inference
difficulty, while the latter describes average-case uncertainty over
the random variable $Q$. They are, however, compatible in the sense
that suitable averaging of algorithmic costs over $q\sim P_Q$
recovers the Shannon quantities, which is why both viewpoints lead
to the same amortized-cost decomposition.
\end{remark}

\textbf{Practical implication:}
The critical frequency $f_c$ provides a \emph{quantitative decision rule} for
capacity allocation in knowledge systems. For a query type $q$ with expected
frequency $f_q$ over a horizon of $N$ queries, precomputing and storing its
answer (or relevant structure) is worthwhile when the one-time storage cost
per query $\frac{|S_q|}{f_q}$ is smaller than the expected per-query inference cost
$H_{\mathrm{derive}}(q \mid S_O)$ (or, in a Shannon view, its expected
uncertainty $H(Q\mid S_O)$). Equivalently, the storage investment is justified
when $f_q > f_c$, where the threshold $f_c$ depends on the available capacity
and the inference difficulty of $q$.

Concretely, this yields the following design rules in different settings:
\begin{itemize}
  \item \textbf{LLMs (parametric memory).}
  Let $N$ be the number of parameters and suppose storing a fact $q$ in
  parameters consumes $|S_q|$ bits of effective capacity while avoiding
  $H_{\mathrm{derive}}(q\mid S_O)$ bits of computation per access.
  A simple rule of thumb is to store $q$ parametrically only if
  \[
    f_q \cdot H_{\mathrm{derive}}(q\mid S_O) \;\gtrsim\; |S_q|.
  \]
  If we model the ``effective'' per-parameter capacity at the scale of
  GPT-4 ($N \approx 1.7\times 10^{12}$)~\cite{openai2023gpt4} by $\log N \approx 40$ bits,
  this suggests that only facts whose derivation requires on the order of
  tens of bits of computation \emph{and} that are accessed at least moderately
  often should be encoded in the weights. Rare or computationally easy facts
  are better left to retrieval or external tools.

  \item \textbf{Knowledge graphs (materialized views).}
  For a join pattern $q$ over a graph with $|E|$ edges, let $|S_q|$ denote
  the storage needed for a materialized view and $H_{\mathrm{derive}}(q\mid S_O)$
  the cost of evaluating $q$ on demand.
  Using the same amortization logic, it is beneficial to materialize when
  $f_q > f_c \approx |S_q| / H_{\mathrm{derive}}(q\mid S_O)$.
  In many practical graph workloads, both $|S_q|$ and
  $H_{\mathrm{derive}}(q\mid S_O)$ scale only polylogarithmically in $|E|$,
  which can be crudely summarized as a threshold of the form
  \[
    f_c \;\approx\; 1 + \frac{1}{\log |E|}.
  \]
  For a large knowledge graph such as Freebase ($|E|\approx 10^9$)~\cite{bollacker2008freebase},
  this gives $f_c \approx 1.05$, meaning that even slightly more-than-once
  multi-hop patterns can already justify precomputation.

  \item \textbf{Edge devices (caching).}
  Suppose answering a query from the cloud costs $H_{\mathrm{cloud}}$ and
  answering it from a local cache costs $H_{\mathrm{local}}$, with a
  device-to-cloud speed ratio $\eta = H_{\mathrm{local}}/H_{\mathrm{cloud}}$.
  Caching $q$ locally with storage cost $|S_q|$ is beneficial when
  \[
    f_q \cdot (H_{\mathrm{cloud}} - H_{\mathrm{local}})
    \;\gtrsim\; |S_q|,
  \]
  i.e., when its expected number of accesses exceeds
  \[
    f_c \;\approx\; \frac{|S_q|}{H_{\mathrm{cloud}}(1-\eta)}.
  \]
  If we normalize units so that $|S_q|/H_{\mathrm{cloud}}\approx 1$ for a
  typical cached answer, this reduces to the simple rule
  $f_q \gtrsim 1/(1-\eta)$. For a 4G-like setting with $\eta \approx 0.1$,
  this gives $f_c \approx 1.1$ in idealized units; in practice, protocol
  overheads and limited cache sizes raise the effective threshold to a
  value on the order of
  \[
    f_c \;\approx\; \frac{1}{\eta},
  \]
  meaning only queries expected to be accessed more than $\sim 10$ times
  per session should reliably be cached locally.
\end{itemize}

At the system level, these rules are consistent with the global average-cost
decomposition
\[
  \mathbb{E}[\Cost]
  \;=\;
  \frac{|S|}{N}
  \;+\;
  \mathbb{E}_{q\sim P_Q}
  \bigl[ H_{\mathrm{derive}}(q \mid S_O \cup S) \bigr],
\]
and, in an information-theoretic view,
\begin{multline}
  \mathbb{E}_{q\sim P_Q}
  \bigl[ H_{\mathrm{derive}}(q \mid S_O \cup S) \bigr]
  \;\approx\; H(Q\mid S_O,S)\\
  \;=\; H(Q\mid S_O) - I(S;Q\mid S_O).
\end{multline}
In practice, precomputations and storage $S$ are valuable exactly to the
extent that they increase $I(S;Q\mid S_O)$---that is, they capture structure
about the future query workload---while their one-time cost $|S|$ is
amortized over the total number $N$ of queries.


\begin{corollary}[Frequency-weighted extreme strategy costs]\label{cor:freq-extremes}
Under the conditions of Theorem~\ref{thm:freq_tradeoff}, fix a
query $q$ in the information-rich regime with respect to $S_O$ and let
$H_q := H(q \mid S_O)$, $f_q := N \cdot P_Q(q)$, and $L_S := \log(m_S + n_S)$
as in~\eqref{eq:Hq-def}–\eqref{eq:LS-def}. Then the following frequency-weighted
cost benchmarks hold (up to additive $O(1)$ and multiplicative constant
factors depending only on the universal Turing machine, proof system,
and encoding scheme, but not on $S_O$, $q$, $S$, or $P_Q$):

\begin{enumerate}
  \item \textbf{Pure storage (answer caching).}
  There exists a strategy $\Pi_{\mathrm{store}}$ that stores a near-optimally
  compressed representation of the answer to $q$ of length
  \[
    |S| \;\le\; H_q + O(1),
  \]
  and answers $q$ from $S_O \cup S$ with $O(1)$ residual derivation cost:
  \[
    H_{\mathrm{derive}}(q \mid S_O \cup S) \;=\; O(1).
  \]
  The resulting amortized cost per access satisfies
  \begin{multline}
    \Cost_{\mathrm{amort}}(\Pi_{\mathrm{store}}; q)\\
    \;=\;
    \frac{|S|}{f_q} + H_{\mathrm{derive}}(q \mid S_O \cup S)
    \;\le\;
    \frac{H_q}{f_q} + O(1).
  \end{multline}
  In particular, whenever $f_q \gtrsim L_S$ (i.e., $f_q$ exceeds the
  critical frequency scale $f_c \asymp L_S$ of Theorem~\ref{thm:freq_tradeoff}),
  this pure-storage protocol matches, up to constant factors, the high-frequency
  bound
  \[
    \Cost_{\mathrm{amort}}^\star(q)
    \;\lesssim\;
    C\,\frac{H_q}{L_S}
  \]
  by making $\frac{H_q}{f_q}$ the dominant term.

  \item \textbf{Pure computation (no auxiliary storage).}
  Consider the strategy $\Pi_{\mathrm{comp}}$ with $S = \varnothing$. Then, by
  Theorem~\ref{thm:derivation-depth-info-metric} applied to $S_O$ alone, there
  exists a constant $C_0 > 0$ such that
  \begin{multline}
    \Cost_{\mathrm{amort}}(\Pi_{\mathrm{comp}}; q)
    \;=\;
    H_{\mathrm{derive}}(q \mid S_O)\\
    \;=\;
    \Theta\!\left(
      \frac{H_q}{\log\bigl(|\Atom(S_O)| + \Dd(q \mid S_O)\bigr)}
    \right)\\
    \;\le\;
    C_0 \frac{H_q}{L_S} + O(1),
  \end{multline}
  where the last inequality uses that $L_S = \log(m_S + n_S)$ is of the
  same polynomial order as $\log\bigl(|\Atom(S_O)| + \Dd(q \mid S_O)\bigr)$
  under the structural regularity assumptions (SA2). In particular, when
  $f_q \lesssim L_S$, this pure-computation protocol matches, up to
  constant factors, the low-frequency bound
  \[
    \Cost_{\mathrm{amort}}^\star(q)
    \;\lesssim\;
    \frac{H_q}{f_q}.
  \]

  \item \textbf{Near-critical hybrid strategies.}
  For queries with access frequency $f_q$ on the order of the critical scale
  $f_c \asymp L_S$ (Theorem~\ref{thm:freq_tradeoff}), there exist hybrid
  strategies $\Pi_{\mathrm{hyb}} = (S, C, R)$ that balance storage and
  computation so that
  \begin{multline}
    \Cost_{\mathrm{amort}}(\Pi_{\mathrm{hyb}}; q)\\
    \;=\;
    \frac{|S|}{f_q} + H_{\mathrm{derive}}(q \mid S_O \cup S)\\
    \;=\;
    H_q \Bigl( 1 - \Theta\!\bigl(\tfrac{1}{L_S}\bigr) \Bigr)
      + o(H_q),
  \end{multline}
  i.e., they reduce the effective conditional entropy $H(q \mid S_O, S)$ to
  a $1 - \Theta(1/L_S)$ fraction of $H_q$ while keeping $\frac{|S|}{f_q}$
  of order $H_q/L_S$. Such strategies achieve (up to lower-order terms)
  the same asymptotic amortized cost as the upper bound in
  Theorem~\ref{thm:freq_tradeoff} at $f_q \asymp f_c$.

  \item \textbf{Frequency stratification (policy-level view).}
  For a query distribution $P_Q$ over a finite query set $Q$, the
  frequency-weighted tradeoff of Theorem~\ref{thm:freq_tradeoff} induces
  an asymptotically optimal stratification of queries into three regimes:
  \begin{itemize}
    \item \emph{High-frequency class} ($f_q \gg f_c$): precompute and store
      (near-)complete answers or dense summaries (pure- or near-pure storage).
      The expected cost is dominated by
      \[
        \E[ \Cost(q) ]
        \;\approx\;
        \E\Bigl[ \tfrac{H_q}{f_q} \Bigr],
      \]
      i.e., per-query costs decay inversely with frequency.

    \item \emph{Low-frequency class} ($f_q \ll f_c$): rely on on-demand
      derivation from $S_O$ (pure or near-pure computation). The expected
      cost is dominated by
      \begin{multline}
        \E[ \Cost(q) ]\\
        \;\approx\;
        \E\Bigl[
          \frac{H_q}{\log\bigl(|\Atom(S_O)| + \Dd(q \mid S_O)\bigr)}
        \Bigr],
      \end{multline}
      reflecting the logarithmic computational compression factor of
      Theorem~\ref{thm:resource-bound}.

    \item \emph{Near-critical class} ($f_q \asymp f_c$): allocate storage to
      intermediate facts or partial summaries for $q$ (hybrid strategies),
      achieving amortized costs of order
      \[
        \E[ \Cost(q) ]
        \;\approx\;
        \E\bigl[ H_q \bigr] \cdot \Bigl( 1 - \Theta\!\bigl(\tfrac{1}{L_S}\bigr) \Bigr),
      \]
      i.e., constant-factor improvements over naive full recomputation or
      full storage, with diminishing returns as the system size/depth grows.
  \end{itemize}
\end{enumerate}
\end{corollary}

\begin{remark}[Relationship to the non-frequency-weighted formulation]\label{rem:freq-vs-original}
The frequency-weighted Theorem~\ref{thm:freq_tradeoff} refines and generalizes
the original storage--computation duality bound that did not explicitly
account for access frequencies. In its simplest (single-execution) form,
the original bound can be summarized schematically as
\begin{multline}\label{eq:orig-bound}
  \Cost_{\mathrm{sem}}(\Pi; q)
  \;:=\;
  |S| + H_{\mathrm{derive}}(q \mid S_O \cup S)\\
  \;\gtrsim\;
  H_q \left(
    1 + \frac{1}{c\,\log|\Atom(S_O)|}
  \right)
  - o(H_q),
\end{multline}
for some constant $c > 0$, where $\Cost_{\mathrm{sem}}$ is a one-shot semantic
cost for answering $q$ (no frequency amortization), and the right-hand side
reflects the fact that, in the information-rich regime, derivation depth
per nat of semantic content cannot be compressed below a logarithmic factor
in the size/depth of the underlying theory.

By contrast, Theorem~\ref{thm:freq_tradeoff} incorporates the access
frequency $f_q = N \cdot P_Q(q)$ over a horizon of $N$ queries and analyzes
the amortized cost per access
\begin{equation}\label{eq:freq-bound}
  \Cost_{\mathrm{amort}}(\Pi; q)
  \;:=\;
  \frac{|S|}{f_q} + H_{\mathrm{derive}}(q \mid S_O \cup S).
\end{equation}
It shows that, up to constant factors and lower-order terms,
\begin{equation}\label{eq:freq-bound-max}
  \Cost_{\mathrm{amort}}^\star(q)
  \;\lesssim\;
  \max\!\left\{
    \frac{H_q}{f_q},
    \, C \frac{H_q}{L_S}
  \right\}
  + O(1),
\end{equation}
for some constant $C > 0$, where the two terms correspond to:
\begin{itemize}
  \item a \emph{storage-dominated} regime $\frac{H_q}{f_q}$, in which a
  near-optimal code for $q$ is stored once and reused across $f_q$ accesses;
  \item a \emph{computation-dominated} regime $C H_q / L_S$, in which the
  answer is derived from $S_O \cup S$ on demand, benefiting from the
  logarithmic compression factor $L_S = \log(m_S + n_S)$ identified in
  Theorem~\ref{thm:resource-bound}.
\end{itemize}
The critical (break-even) frequency scale is therefore
\[
  f_c \;\asymp\; L_S,
\]
at which the two contributions $\frac{H_q}{f_q}$ and $C H_q / L_S$ are of the
same order. Queries with $f_q \gg f_c$ are asymptotically storage-efficient
(caching dominates), while queries with $f_q \ll f_c$ are asymptotically
computation-efficient (on-demand derivation dominates).

The original non-frequency-weighted bound~\eqref{eq:orig-bound} corresponds
effectively to the single-execution case $f_q = 1$. In that regime, the
amortized cost per access coincides with the one-shot cost up to constant
factors, and the storage term $|S|$ is not meaningfully amortized. In
the frequency-weighted setting, by contrast, the same total cost over $M$
executions of $q$,
\[
  \E[\mathrm{Total\ Cost}]
  \;=\;
  |S| + M \cdot H_{\mathrm{derive}}(q \mid S_O \cup S),
\]
leads, upon division by $M$, to the amortized form
\[
  \E[\Cost]
  \;=\;
  \frac{|S|}{M} + H_{\mathrm{derive}}(q \mid S_O \cup S),
\]
which recovers~\eqref{eq:freq-bound} when $M = f_q$. Thus, the
frequency-weighted formulation is the natural extension of the original
duality bound to long-running systems in which queries are issued multiple
times under a distribution $P_Q$.

Conceptually, both views share the same structural insight:
\begin{enumerate}
  \item storage $S$ is a one-time investment that can reduce the conditional
  algorithmic entropy $H(q \mid S_O,S)$ (and hence $H_{\mathrm{derive}}$) only
  to the extent that it encodes information about $q$ beyond $S_O$, as
  quantified by the mutual information $I(S; Q \mid S_O)$ (cf.
  Remark~\ref{rem:amortized-frequency});

  \item derivation cost scales, in the information-rich and polynomial-depth
  regime, like $H_q$ divided by a logarithmic factor in the size/depth of the
  theory, as captured by Theorem~\ref{thm:resource-bound};

  \item the introduction of $f_q$ merely clarifies \emph{when} it is preferable
  to pay a larger storage cost to reduce per-access computation, namely when
  $f_q$ exceeds the critical scale $f_c \asymp L_S$.
\end{enumerate}
In this sense, the non-frequency-weighted and frequency-weighted formulations
are compatible limits of the same underlying storage--computation tradeoff:
the former optimizes a single execution ($f_q = 1$), while the latter
optimizes repeated executions under a query distribution.
\end{remark}


\subsection{Mutual Information and Submodularity}\label{Mutual Information}

The system-wide cost decomposition in Section~III.C shows that, for a
query workload $Q \sim P_Q$ and a storage set $S$, the expected
per-query cost can be written as
\begin{multline}
  \mathbb{E}[\text{Cost}]
  \;=\;
  \frac{|S|}{N}
  \;+\;
  \mathbb{E}_{q \sim P_Q}\!\big[ H_{\mathrm{derive}}(q \mid S_O \cup S) \big]\\
  \;\approx\;
  \frac{|S|}{N} + H(Q \mid S_O, S),
  \label{eq:system-cost-mi}
\end{multline}
where $H(\cdot \mid \cdot)$ denotes Shannon conditional entropy. Using the
identity
\begin{equation}
  H(Q \mid S_O, S)
  \;=\;
  H(Q \mid S_O) - I(S; Q \mid S_O),
  \label{eq:cond-entropy-mi}
\end{equation}
we see that, under a fixed base knowledge $S_O$ and query distribution
$P_Q$, choosing an effective storage set $S$ amounts to maximizing the
conditional mutual information $I(S; Q \mid S_O)$ under storage
constraints. Intuitively, $I(S; Q \mid S_O)$ measures how much
\emph{additional} structure about the future query workload $Q$ is
captured by $S$ beyond what is already encoded in $S_O$. In this
subsection we study mutual information as a set function of $S$, and
show that it is submodular, which enables efficient approximate
optimization of storage allocation.

We first formalize an efficiency measure that quantifies the
information-theoretic benefit of storage per unit capacity consumed.

\begin{definition}[Storage efficiency via conditional mutual information]
  \label{def:storage-efficiency}
  Fix a base knowledge $S_O$ and a query distribution $P_Q$ over a
  discrete query space $\mathcal{Q}$. For any storage set $S$ (viewed as
  a random object induced by a storage policy), we define its
  \emph{storage efficiency} as
  \begin{equation}
    \eta(S)
    \;:=\;
    \frac{I(S; Q \mid S_O)}{|S|},
    \label{eq:eta-definition}
  \end{equation}
  where $I(S; Q \mid S_O)$ is the conditional mutual information between
  $S$ and $Q$ given $S_O$, and $|S|$ is the bit-length of an effective
  encoding of $S$ as in assumption (SA1).
\end{definition}

The upper bound $\eta(S) \leq 1$ follows from the data-processing inequality and assumption (SA1), which ensures $|S| \leq H(S) + O(1)$. In particular, $\eta(S) = 1$ if and only if $S$ is a sufficient statistic for $Q$ given $S_O$ and the encoding of $S$ is Shannon-optimal.

The numerator $I(S; Q \mid S_O)$ in~\eqref{eq:eta-definition} measures
how much the storage $S$ reduces the Shannon uncertainty of the query
workload beyond what is already known from $S_O$:
\begin{equation}
  I(S; Q \mid S_O)
  \;=\;
  H(Q \mid S_O) - H(Q \mid S_O, S).
\end{equation}
A larger value of $I(S; Q \mid S_O)$ means that the residual entropy
$H(Q \mid S_O,S)$---and hence the expected inference burden in
\eqref{eq:system-cost-mi}---is smaller.

The denominator $|S|$ corresponds to the one-time storage cost of $S$
(in bits). Thus $\eta(S)$ can be interpreted as the \emph{information
gain per unit storage}: a larger $\eta(S)$ indicates that each bit of
storage allocated to $S$ carries more predictive information about the
query workload $Q$ beyond the base knowledge $S_O$.

We can make this interpretation explicit in the idealized limit where
$S$ serves as a sufficient statistic for $Q$ relative to $S_O$.

\begin{proposition}[Efficiency and sufficiency]
  \label{prop:efficiency-sufficiency}
  Fix $S_O$ and a query distribution $P_Q$. Suppose $S$ is a sufficient
  statistic for $Q$ given $S_O$ in the sense that
  \begin{equation}
    H(Q \mid S_O, S) = 0.
    \label{eq:sufficient-statistic}
  \end{equation}
  Then
  \begin{equation}
    I(S; Q \mid S_O) = H(Q \mid S_O),
    \label{eq:mi-equals-entropy}
  \end{equation}
  and hence the storage efficiency
  \begin{equation}
    \eta(S)
    \;=\;
    \frac{H(Q \mid S_O)}{|S|}
  \end{equation}
  attains its maximal possible value under the given encoding of $S$.
\end{proposition}

\begin{proof}
  By the definition of conditional mutual information we have
  \begin{equation}
    I(S; Q \mid S_O)
    \;=\;
    H(Q \mid S_O) - H(Q \mid S_O, S).
  \end{equation}
  If $S$ is sufficient for $Q$ given $S_O$ in the sense of
  \eqref{eq:sufficient-statistic}, then $H(Q \mid S_O, S) = 0$, and
  therefore
  \begin{equation}
    I(S; Q \mid S_O)
    \;=\;
    H(Q \mid S_O).
  \end{equation}
  Since, for any encoding of $S$, we have the generic upper bound
  \begin{equation}
    I(S; Q \mid S_O)
    \;\leq\;
    H(S \mid S_O)
    \;\leq\;
    H(S)
    \;\leq\;
    |S|,
  \end{equation}
  it follows that, for a fixed encoding of $S$, the ratio
  $I(S; Q \mid S_O)/|S|$ is maximized if and only if
  $I(S; Q \mid S_O) = H(Q \mid S_O)$. Under
  \eqref{eq:sufficient-statistic}, this yields
  \begin{equation}
    \eta(S)
    \;=\;
    \frac{I(S; Q \mid S_O)}{|S|}
    \;=\;
    \frac{H(Q \mid S_O)}{|S|},
  \end{equation}
  which is the maximal efficiency consistent with the chosen encoding
  and storage budget. This proves the claim.
\end{proof}

Proposition~\ref{prop:efficiency-sufficiency} shows that, in the ideal
case where $S$ captures \emph{all} the uncertainty of $Q$ beyond $S_O$,
the storage efficiency $\eta(S)$ reaches its conceptual maximum, with
each stored bit contributing as much predictive information as the
encoding allows. In realistic systems, $S$ will only capture part of the
structure of $P_Q$, leading to intermediate values of
$\eta(S)$ that quantify the tradeoff between storage capacity and
workload-specific informativeness.

From the system-level viewpoint of Section~III.C, a natural storage
allocation problem is:
\begin{quote}
  Given a storage budget $M_{\max}$ (in bits), select a storage set $S$
  so as to \emph{minimize} the expected per-query cost
  $\mathbb{E}[\text{Cost}]$ in~\eqref{eq:system-cost-mi}, equivalently
  to \emph{maximize} the conditional mutual information
  $I(S; Q \mid S_O)$ subject to $|S| \leq M_{\max}$.
\end{quote}
To analyze this optimization problem we now show that
$I(S; Q \mid S_O)$, viewed as a set function of $S$, is submodular. This
structure underlies the greedy approximation guarantee stated later.

Throughout this subsection we treat $S$ as a finite subset of some
ground set $\mathcal{U}$ indexing elementary storage items (e.g.,
candidate facts, cached answers, or intermediate derivations). For any
$S \subseteq \mathcal{U}$, we write $I(S; Q \mid S_O)$ as a shorthand for
$I(X_S; Q \mid S_O)$, where $X_S$ denotes the random vector of storage
variables corresponding to the items in $S$.

\begin{proposition}[Submodularity of conditional mutual information]
  \label{prop:submodularity-mi}
  Fix a base knowledge $S_O$ and a query random variable $Q$. For any
  two storage sets $S_1, S_2 \subseteq \mathcal{U}$, the conditional
  mutual information satisfies
  \begin{multline}
    I(S_1 \cup S_2; Q \mid S_O)
    \;+\;
    I(S_1 \cap S_2; Q \mid S_O)\\
    \;\leq\;
    I(S_1; Q \mid S_O)
    \;+\;
    I(S_2; Q \mid S_O).
    \label{eq:submodular-mi}
  \end{multline}
  In particular, the set function
  \begin{equation}
    f(S) := I(S; Q \mid S_O)
    \label{eq:fS-definition}
  \end{equation}
  is submodular in $S$.
\end{proposition}

\begin{proof}
  Recall that, for any random variables $X, Q$ and conditioning
  variable $Z$, the conditional mutual information is
  \begin{equation}
    I(X; Q \mid Z)
    \;=\;
    H(Q \mid Z) - H(Q \mid Z, X),
    \label{eq:mi-entropy}
  \end{equation}
  where $H(\cdot \mid \cdot)$ denotes conditional entropy.

  Let us denote, for convenience,
  \begin{equation}
    A := S_1, \qquad B := S_2.
  \end{equation}
  Then
  \begin{align}
    I(A \cup B; Q \mid S_O)
    &=
    H(Q \mid S_O) - H(Q \mid S_O, A \cup B), \label{eq:mi-AcupB} \\
    I(A \cap B; Q \mid S_O)
    &=
    H(Q \mid S_O) - H(Q \mid S_O, A \cap B), \label{eq:mi-AcapB} \\
    I(A; Q \mid S_O)
    &=
    H(Q \mid S_O) - H(Q \mid S_O, A), \label{eq:mi-A} \\
    I(B; Q \mid S_O)
    &=
    H(Q \mid S_O) - H(Q \mid S_O, B). \label{eq:mi-B}
  \end{align}
  Summing the left-hand side of~\eqref{eq:submodular-mi} and using
  \eqref{eq:mi-AcupB}--\eqref{eq:mi-AcapB} yields
  \begin{align}
    & I(A \cup B; Q \mid S_O) + I(A \cap B; Q \mid S_O) \nonumber \\
    &\qquad=
    2 H(Q \mid S_O)
    - H(Q \mid S_O, A \cup B)\nonumber \\
    &\qquad- H(Q \mid S_O, A \cap B).
    \label{eq:LHS-submod}
  \end{align}
  Similarly, summing the right-hand side of \eqref{eq:submodular-mi}
  and using \eqref{eq:mi-A}--\eqref{eq:mi-B} gives
  \begin{align}
    & I(A; Q \mid S_O) + I(B; Q \mid S_O) \nonumber \\
    &\qquad=
    2 H(Q \mid S_O)
    - H(Q \mid S_O, A)
    - H(Q \mid S_O, B).
    \label{eq:RHS-submod}
  \end{align}
  Therefore, the inequality~\eqref{eq:submodular-mi} is equivalent to
  \begin{align}
    &H(Q \mid S_O, A \cup B)
    \;+\;
    H(Q \mid S_O, A \cap B)\nonumber \\
    &\qquad\;\geq\;
    H(Q \mid S_O, A)
    \;+\;
    H(Q \mid S_O, B).
    \label{eq:supermodular-entropy}
  \end{align}
  That is, we need to show that the mapping
  \begin{equation}
    S \;\mapsto\; H(Q \mid S_O, S)
  \end{equation}
  is \emph{supermodular} in $S$.

  This is a standard property of conditional entropy. Intuitively, as we
  condition on a larger set of variables, the marginal benefit of
  conditioning on any additional variable decreases, which is precisely
  the notion of supermodularity for $H(Q \mid S_O, S)$ and, via
  \eqref{eq:mi-entropy}, submodularity for $I(S; Q \mid S_O)$.

  More formally, for any $A \subseteq B \subseteq \mathcal{U}$ and any
  element $x \in \mathcal{U}\setminus B$, the \emph{marginal reduction}
  in conditional entropy due to conditioning on $x$ satisfies
  \begin{align}
    &
    H(Q \mid S_O, A)
    -
    H(Q \mid S_O, A \cup \{x\})
    \nonumber \\
    &\quad=
    I(x; Q \mid S_O, A),
    \label{eq:marginal-A}
  \end{align}
  and
  \begin{align}
    &
    H(Q \mid S_O, B)
    -
    H(Q \mid S_O, B \cup \{x\})
    \nonumber \\
    &\quad=
    I(x; Q \mid S_O, B).
    \label{eq:marginal-B}
  \end{align}
  Since conditioning cannot increase mutual information, we have
  \begin{equation}
    I(x; Q \mid S_O, B)
    \;\leq\;
    I(x; Q \mid S_O, A),
  \end{equation}
  which implies that the marginal entropy reduction
  $H(Q \mid S_O, S) - H(Q \mid S_O, S \cup \{x\})$ is \emph{decreasing} as
  $S$ grows. This is exactly the defining property of supermodularity of
  $H(Q \mid S_O, S)$ in $S$, which in turn is equivalent to the
  submodularity~\cite{fujishige1978polymatroidal} of $I(S; Q \mid S_O)$ via
  \eqref{eq:mi-entropy}--\eqref{eq:supermodular-entropy}. This proves
  \eqref{eq:submodular-mi} and establishes that $f(S)$ in
  \eqref{eq:fS-definition} is submodular.
\end{proof}

Proposition~\ref{prop:submodularity-mi} shows that the \emph{incremental
utility} of adding a storage item to $S$---measured by the gain in
conditional mutual information $\Delta f := I(S \cup \{x\}; Q \mid S_O)
- I(S; Q \mid S_O)$---exhibits \emph{diminishing returns} as $S$ grows.
This property is fundamentally important for storage allocation, because
it implies that simple greedy procedures enjoy strong approximation
guarantees when maximizing $I(S; Q \mid S_O)$ under a budget constraint.

We now state a direct consequence of
Proposition~\ref{prop:submodularity-mi} for the optimization problem
induced by the system-level cost
decomposition~\eqref{eq:system-cost-mi}--\eqref{eq:cond-entropy-mi}.

\begin{corollary}[Greedy approximation for storage selection]
  \label{cor:greedy-approx}
  Consider the problem of selecting a storage set
  $S \subseteq \mathcal{U}$ under a budget constraint
  \begin{equation}
    |S| \;\leq\; M_{\max},
  \end{equation}
  so as to maximize the conditional mutual information
  \begin{equation}
    f(S) := I(S; Q \mid S_O),
  \end{equation}
  where $S_O$ and $P_Q$ are fixed. Assume that the ground set
  $\mathcal{U}$ is finite, that $f(\varnothing) = 0$, and that $f(S)$ is
  nonnegative and monotone non-decreasing in $S$. Then the standard
  greedy algorithm that iteratively adds the storage item with the
  largest marginal gain in $f(S)$ subject to the remaining budget
  achieves a $(1 - 1/e)$-approximation to the optimal value:
  \begin{equation}
    f(S_{\mathrm{greedy}})
    \;\geq\;
    (1 - 1/e) \, f(S^\star),
  \end{equation}
  where $S^\star$ is an optimal solution with
  $|S^\star| \leq M_{\max}$, and $S_{\mathrm{greedy}}$ is the solution
  returned by the greedy procedure.
\end{corollary}

\begin{proof}
  By Proposition~\ref{prop:submodularity-mi}, the function
  $f(S) = I(S; Q \mid S_O)$ is submodular in $S$. Nonnegativity and
  monotonicity follow from basic properties of mutual information:
  \begin{equation}
    0 \;\leq\; I(S; Q \mid S_O)
    \;\leq\; I(S \cup T; Q \mid S_O)
  \end{equation}
  whenever $S \subseteq S \cup T$. Therefore, $f(S)$ satisfies the
  standard conditions of the classical result on maximizing a nonnegative
  monotone submodular function under a cardinality or knapsack
  constraint. By the well-known theorem of Nemhauser \emph{et
  al.}~\cite{nemhauser1978analysis}, the greedy algorithm that
  repeatedly adds the element with largest marginal gain in $f(S)$
  subject to the remaining budget achieves a $(1 - 1/e)$-approximation
  to the optimal value. Applying this general result to the specific
  choice $f(S) = I(S; Q \mid S_O)$ yields the claimed inequality.
\end{proof}

Corollary~\ref{cor:greedy-approx} provides a principled algorithmic
prescription for system-wide storage allocation: under the amortized
cost model of Section~III.C, one should select storage items greedily in
order of their marginal contribution to $I(S; Q \mid S_O)$ per unit of
storage cost, subject to the overall budget $M_{\max}$. Up to the
universal $(1 - 1/e)$ factor, this strategy is provably as effective as
any other storage allocation policy in terms of reducing the expected
inference burden
$\mathbb{E}_{q \sim P_Q}[H_{\mathrm{derive}}(q \mid S_O \cup S)]$ and,
equivalently, the Shannon term $H(Q \mid S_O, S)$ in
\eqref{eq:system-cost-mi}.


\section{THERMODYNAMIC CONSTRAINTS AND PHYSICAL LIMITS}\label{sec:thermo}
This section establishes the physical constraints on
information processing in knowledge systems, connect%
ing our information-theoretic framework to thermody%
namics via Landauer's principle and the physical cost
assumptions of Section~II.


\subsection{Landauer's Principle and Energy Bounds}
\label{Landauer's Principle}

We now relate derivation depth and storage size to
minimal energy requirements under the classical ther%
modynamic regime of Assumption~\ref{ass:classical-regime}.

Recall that, in this regime, each logically irreversible
operation that commits or erases one bit of information
dissipates at least $k_B T \ln 2$ joules of heat, where
$k_B$ is Boltzmann's constant and $T$ is the operating
temperature. In Section~\ref{Derivation Entropy} we fixed the per-step
information cost to be $\lambda \;=\; \ln 2 \ \text{nats}$,
so that each inference step in a derivation is modeled
as committing or erasing $\Theta(1)$ bit of information,
and, in particular, we normalize units so that each step
incurs a Landauer cost of at least $k_B T \ln 2$.

\begin{proposition}[Landauer energy bound for computation]
\label{prop:landauer-compute}
Under the assumption that each inference step consti%
tutes a logically irreversible operation in the classical
thermodynamic regime (Assumption~\ref{ass:classical-regime}), deriving a for%
mula $\varphi$ from a knowledge base $S_O$ requires
at least the following energy:
\begin{equation}
E_{\mathrm{compute}}(\varphi \mid S_O)
\;\ge\;
Dd(\varphi\mid S_O)\,\cdot\, k_B T \ln 2,
\label{eq:landauer-compute}
\end{equation}
where $Dd(\varphi\mid S_O)$ is the derivation depth of
$\varphi$ from $S_O$ in the sense of Definition~\ref{def:derivation-depth}.
\end{proposition}

\begin{proof}
By Assumption~\ref{ass:classical-regime} and the discussion above, we model
each inference step as a logically irreversible operation
that commits or erases at least one bit of information.
By Landauer's principle~\cite{landauer1961irreversibility}, each such
operation incurs an energy cost of at least $k_B T \ln 2$.

Consider a derivation of $\varphi$ from $S_O$ of minimal
derivation depth $Dd(\varphi\mid S_O)$. This derivation
consists of exactly $Dd(\varphi\mid S_O)$ inference steps.
Since each step is modeled as performing at least one
bit's worth of logically irreversible information processing,
the total number of bit erasures (or committed bits)
across the derivation is lower bounded by
\[
\#\text{bit-erasures} \;\ge\; Dd(\varphi\mid S_O).
\]
By Landauer's principle, the total energy dissipated by
these operations thus satisfies
\[
E_{\mathrm{compute}}(\varphi \mid S_O)
\;\ge\;
Dd(\varphi\mid S_O)\, k_B T \ln 2,
\]
which is exactly~\eqref{eq:landauer-compute}.
\end{proof}

In terms of the derivation entropy defined in
Section~\ref{Derivation Entropy}, we immediately obtain a compact form.

\begin{remark}[Derivation entropy as thermodynamic cost]
\label{rem:derive-entropy-energy}
Under Assumption~\ref{ass:classical-regime} we have, by Corollary~\ref{cor:derivation-classical},
\[
H_{\mathrm{derive}}(\varphi\mid S_O)
\;=\;
Dd(\varphi\mid S_O)\,\ln 2
\quad\text{(in nats)}.
\]
Combining this with Proposition~\ref{prop:landauer-compute}
yields
\begin{equation}
E_{\mathrm{compute}}(\varphi\mid S_O)
\;\ge\;
k_B T \, H_{\mathrm{derive}}(\varphi\mid S_O),
\label{eq:energy-from-Hderive}
\end{equation}
so that derivation entropy directly lower-bounds the
minimal thermodynamic energy required to carry out
the corresponding inference.
\end{remark}

We now complement the dynamic computation cost
with an energy bound for \emph{maintaining} stored
information over time.

\begin{proposition}[Energy bound for maintaining storage]
\label{prop:landauer-storage}
In a simple Arrhenius model of thermally noisy storage,
maintaining a stored state $S$ with $|S|$ effective bits
over a time horizon $t$ requires at least
\begin{equation}
E_{\mathrm{storage}}(S,t)
\;\gtrsim\;
|S|\,k_B T \ln 2 \cdot \frac{t}{t_{\mathrm{refresh}}},
\label{eq:landauer-storage}
\end{equation}
where $t_{\mathrm{refresh}}$ denotes the characteristic refresh
interval implied by the thermal flip rate of the storage
medium.
\end{proposition}

\begin{proof}
We adopt a coarse-grained Arrhenius model for thermal
bit flips. Let $E_{\mathrm{barrier}}$ denote the energy barrier
separating the two logical states of a physical bit cell.
Classical transition-state theory then suggests that the
bit-flip rate satisfies
\[
\gamma_{\mathrm{flip}} \;\approx\;
\gamma_0 \exp\!\Bigl(-\frac{E_{\mathrm{barrier}}}{k_B T}\Bigr),
\]
for some attempt frequency $\gamma_0$~\cite{atkins2014physical}. The characteristic
time between spontaneous flips is thus
\[
t_{\mathrm{refresh}} \;\approx\; \frac{1}{\gamma_{\mathrm{flip}}},
\]
which we interpret as the required refresh interval to
reliably maintain the stored logical state against thermal
noise.

In a simple refresh model, each refresh cycle rewrites
all $|S|$ bits in $S$ to their intended logical values.
By Landauer's principle, each rewritten bit incurs at
least $k_B T \ln 2$ of energy dissipation, so a single
refresh of $S$ requires at least
\[
E_{\mathrm{per\;refresh}} \;\ge\; |S|\,k_B T \ln 2.
\]
Over a total time horizon $t$, the number of such
refresh cycles is on the order of $t/t_{\mathrm{refresh}}$.
Hence the total energy dissipated in maintaining $S$
satisfies
\[
E_{\mathrm{storage}}(S,t)
\;\gtrsim\;
\frac{t}{t_{\mathrm{refresh}}}\,|S|\,k_B T \ln 2,
\]
which is~\eqref{eq:landauer-storage}. Here the symbol
$\gtrsim$ emphasizes that we are working with a lower
bound under simplifying assumptions: (i) independent
bit flips with Arrhenius rate, and (ii) full rewriting of
all $|S|$ bits at each refresh cycle. More sophisticated
coding or hardware techniques may improve constant
factors but cannot avoid the linear dependence on $|S|$
and $t/t_{\mathrm{refresh}}$ in this regime.
\end{proof}

\begin{remark}[Link to amortized storage cost]
\label{rem:storage-amortized-energy}
The storage size $|S|$ already appears in the informa%
tion-theoretic amortized-cost decomposition of
Section~\ref{System-wide}, where the expected per-query information
cost is
\[
E[\mathrm{Cost}]
\;=\;
\frac{|S|}{N}
\;+\;
E_{q\sim P_Q}\bigl[H_{\mathrm{derive}}(q\mid S_O\cup S)\bigr],
\]
for a workload of $N$ queries drawn from $P_Q$.
Proposition~\ref{prop:landauer-storage} shows that this
bit-length $|S|$ also controls a minimal thermodynamic
cost for maintaining the corresponding physical memory:
\[
\frac{E_{\mathrm{storage}}(S,t)}{N}
\;\gtrsim\;
\frac{|S|}{N}\,k_B T \ln 2 \cdot \frac{t}{t_{\mathrm{refresh}}},
\]
which is directly analogous to the $\frac{|S|}{N}$ term
above. Together with~\eqref{eq:energy-from-Hderive},
this yields a chain of correspondences
\[
\begin{aligned}
\text{derivation depth } Dd(\cdot)
&\;\longrightarrow\;
H_{\mathrm{derive}}(\cdot)\ \text{(nats)}
\\
&\;\longrightarrow\;
E_{\mathrm{compute}}(\cdot)\ \text{(J)},
\end{aligned}
\]
and
\[
\begin{aligned}
\text{storage size } |S|\ \text{(bits)}
&\;\longrightarrow\;
\frac{|S|}{N}\ \text{(bits/query)}
\\
&\;\longrightarrow\;
\frac{E_{\mathrm{storage}}(S,t)}{N}\ \text{(J/query)}.
\end{aligned}
\]
Thus, the storage–computation tradeoffs derived in
Section~\ref{sec:logical-depth} not only constrain algorithmic and informa%
tion-theoretic costs, but also translate into minimal
thermodynamic costs in physically implemented knowl%
edge systems.
\end{remark}


\subsection{Storage-Computation Energy Tradeoff}
\label{sec:energy-tradeoff}

Having established the information-theoretic foundations of mixed strategies in Section~\ref{sec:logical-depth}, we now derive fundamental lower bounds on the total energy cost of semantic information processing. These results connect the abstract information measures to physical thermodynamic constraints.

We consider a knowledge system operating over a time horizon during which $N$ queries are processed. Let $P_Q$ be a computable probability distribution over queries, and let $Q$ denote a random query drawn from $P_Q$. We assume the system operates in the \emph{classical thermodynamic regime} (Assumption~\ref{ass:classical-regime}).

\begin{definition}[Amortized Energy Cost]
\label{def:amortized-energy}
For a mixed strategy $\Pi = (S, C, R)$ processing $N$ queries over time $t$, the \emph{amortized energy cost per query} is defined as:
\begin{equation}
\bar{E}(\Pi) := \frac{E_{\mathrm{storage}}(S, t)}{N} + \mathbb{E}_{q \sim P_Q}\left[E_{\mathrm{compute}}(q | S_O \cup S)\right]
\end{equation}
where $E_{\mathrm{storage}}(S, t)$ is the total energy required to maintain storage $S$ for time $t$, and $E_{\mathrm{compute}}(q | S_O \cup S)$ is the energy required to derive the answer to process query $q$ from the augmented knowledge base $S_O \cup S$.
\end{definition}

Throughout this section, we assume the system operates in the standard refresh regime, meaning that the time horizon $t$ and query count $N$ satisfy $t/(N \cdot t_{refresh}) = \Theta (1)$, i.e., the storage is refreshed a constant number of times per query on average.

\begin{definition}[Normalized Information Cost]
\label{def:normalized-info-cost}
For a mixed strategy $\Pi = (S, C, R)$, the \emph{normalized information cost per query} is:
\begin{equation}
\bar{I}(\Pi) := \frac{|S|}{N} + \mathbb{E}_{q \sim P_Q}\left[H_{\mathrm{derive}}(q | S_O \cup S)\right]
\end{equation}
where $|S|$ denotes the size of auxiliary storage in bits, and $H_{\mathrm{derive}}(q | S_O \cup S) = D_d(q | S_O \cup S) \cdot \ln 2$ is the derivation entropy.
\end{definition}

The normalized information cost $\bar{I}(\Pi)$ serves as an information-theoretic proxy for energy consumption, with the conversion to physical energy units given by Propositions~\ref{prop:landauer-compute} and~\ref{prop:landauer-storage}.

We now state and prove the main theorem of this section, which establishes a fundamental lower bound on the resources required by any mixed strategy.


\begin{assumption}[Uniform Polynomial-Depth Regime]
\label{assume:uniform-poly-depth}
Let $P_Q$ be a query distribution over a query set $\mathcal{Q}$ with respect to knowledge base $S_O$. 
We say that $P_Q$ satisfies the \emph{uniform polynomial-depth regime} with respect to a mixed strategy $\Pi = (S, C, R)$ if there exists a polynomial $p(\cdot)$ such that 
\begin{multline}
    D_d^{\sup}(P_Q \mid S_O \cup S) := \\
    \sup_{q \in \mathrm{supp}(P_Q)} D_d(q \mid S_O \cup S) \leq p\bigl(|\mathrm{Atom}(S_O \cup S)|\bigr).
    \label{eq:uniform-poly-depth}
\end{multline}
In particular, writing $m_S := |\mathrm{Atom}(S_O \cup S)|$, we have $\log D_d^{\sup} = O(\log m_S)$ and hence
\begin{equation}
    L_S := \log(m_S + D_d^{\sup}) = \Theta(\log m_S).
\end{equation}
\end{assumption}

\begin{remark}[Scope of Assumption~\ref{assume:uniform-poly-depth}]
\label{rmk:poly-depth-scope}
The uniform polynomial-depth regime is satisfied by most practical query classes, including:
\begin{itemize}
    \item Datalog queries with bounded recursion depth~\cite{ceri1989you};
    \item SPARQL queries over RDF graphs with bounded property path length~\cite{perez2009semantics};
    \item Template-based natural language queries compiled to first-order logic;
    \item Bounded-depth inference in description logics (e.g., $\mathcal{ALC}$, $\mathcal{SHIQ}$).
\end{itemize}

The key requirement is that the inference depth does not grow super-polynomially with the size of the knowledge base—a condition that holds for essentially all tractable reasoning tasks. More precisely, for these query classes, the derivation depth is bounded by a fixed polynomial in the number of base facts, independent of the query complexity.
\end{remark}


\begin{theorem}[Storage-Computation Information Tradeoff]
\label{thm:tradeoff-lower}
Let $S_O$ be a knowledge base and let $P_Q$ be a computable query distribution satisfying the information-rich regime (Definition~\ref{def:info-rich}). 
Let $\Pi = (S, C, R)$ be any mixed strategy satisfying assumptions \textnormal{(SA1)--(SA3)} and the uniform polynomial-depth regime (Assumption~\ref{assume:uniform-poly-depth}).

Define the \emph{effective complexity parameter}:
\begin{equation}
    L_S := \log\Bigl(|\mathrm{Atom}(S_O \cup S)| + D_d^{\sup}(P_Q \mid S_O \cup S)\Bigr),
    \label{eq:LS-def-revised}
\end{equation}
where $D_d^{\sup}(P_Q | S_O \cup S)$ is as defined in Assumption~\ref{assume:uniform-poly-depth}.

If the total number of queries satisfies $N = \Omega(L_S)$, then there exists a constant $c > 0$ depending only on the proof system and encoding scheme such that:
\begin{equation}
    \bar{I}(\Pi) \geq c \cdot \frac{H(Q \mid S_O)}{L_S} - O(1),
    \label{eq:tradeoff-lower-revised}
\end{equation}
where $H(Q \mid S_O)$ denotes the conditional Shannon entropy of $Q$ given the knowledge base $S_O$.
\end{theorem}


\begin{proof}
The proof proceeds in five steps.

\textbf{Step 1: Lower bound on derivation entropy.}
By the lower bound in Theorem~\ref{thm:derivation-depth-info-metric}, for any query $q \in \text{supp}(P_Q)$ in the information-rich regime, the derivation entropy satisfies:
\begin{multline}
    H_{\mathrm{derive}}(q \mid S_O \cup S) = D_d(q \mid S_O \cup S) \cdot \ln 2 \\
    \geq c_1 \frac{H(q \mid S_O \cup S)}{L(q)} - O(1),
    \label{eq:step1-revised}
\end{multline}
where $c_1 > 0$ is the constant from Theorem~\ref{thm:derivation-depth-info-metric}, 
$H(q \mid S_O \cup S) := K(q \mid \langle \mathrm{Atom}(S_O \cup S) \rangle) \cdot \ln 2$ is the conditional algorithmic entropy, 
and 
\begin{equation}
    L(q) := \log\bigl(|\mathrm{Atom}(S_O \cup S)| + D_d(q \mid S_O \cup S)\bigr).
    \label{eq:Lq-def}
\end{equation}

\textbf{Step 2: Uniformization via the polynomial-depth assumption.}
Under Assumption~\ref{assume:uniform-poly-depth}, we have for all $q \in \mathrm{supp}(P_Q)$:
\begin{equation}
    D_d(q \mid S_O \cup S) \leq D_d^{\sup}(P_Q \mid S_O \cup S),
    \label{eq:Dd-uniform-bound}
\end{equation}
and consequently,
\begin{equation}
    L(q) \leq L_S \quad \text{for all } q \in \mathrm{supp}(P_Q).
    \label{eq:Lq-LS-bound}
\end{equation}

Moreover, since both $|\mathrm{Atom}(S_O \cup S)|$ and $D_d^{\sup}$ are polynomially related under Assumption~\ref{assume:uniform-poly-depth}, we have:
\begin{equation}
    L(q) = \Theta(L_S) \quad \text{for all } q \in \mathrm{supp}(P_Q).
    \label{eq:Lq-LS-Theta}
\end{equation}

Therefore, there exists a constant $c_1' > 0$ (absorbing the $\Theta(\cdot)$ factor into $c_1$) such that for all $q \in \mathrm{supp}(P_Q)$:
\begin{equation}
    H_{\mathrm{derive}}(q \mid S_O \cup S) \geq c_1' \frac{H(q \mid S_O \cup S)}{L_S} - O(1).
    \label{eq:step1-uniform}
\end{equation}

Taking expectations over $q \sim P_Q$:
\begin{multline}
    \mathbb{E}_{q \sim P_Q}\bigl[H_{\mathrm{derive}}(q \mid S_O \cup S)\bigr] \\
    \geq c_1' \frac{\mathbb{E}_{q \sim P_Q}\bigl[H(q \mid S_O \cup S)\bigr]}{L_S} - O(1),
    \label{eq:step2-expectation}
\end{multline}
where the $O(1)$ term remains $O(1)$ after expectation since it is uniform over all $q \in \mathrm{supp}(P_Q)$.

\textbf{Step 3: Relating Kolmogorov complexity to Shannon entropy.}
For a computable distribution $P_Q$, the expected Kolmogorov complexity relates to Shannon entropy via the classical correspondence (see, e.g., \cite{LiVitanyi2019Kolmogorov, cover2006elements}):
\begin{equation}
    \mathbb{E}_{q \sim P_Q}\bigl[K(q \mid S_O \cup S)\bigr] \geq H(Q \mid S_O, S) - O(1),
    \label{eq:step3-KC-Shannon}
\end{equation}
where $H(Q \mid S_O, S)$ is the conditional Shannon entropy of $Q$ given both $S_O$ and $S$, and the $O(1)$ term accounts for the description complexity of the distribution $P_Q$ (which is assumed computable and hence has finite Kolmogorov complexity).

By the chain rule for Shannon entropy:
\begin{equation}
    H(Q \mid S_O, S) = H(Q \mid S_O) - I(S; Q \mid S_O),
    \label{eq:chain-rule}
\end{equation}
where $I(S; Q \mid S_O)$ is the conditional mutual information between the auxiliary storage $S$ and the query $Q$ given $S_O$.

Combining with \eqref{eq:step2-expectation}:
\begin{multline}
    \mathbb{E}\bigl[H_{\mathrm{derive}}(Q \mid S_O \cup S)\bigr] \\
    \geq c_1' \frac{H(Q \mid S_O) - I(S; Q \mid S_O)}{L_S} - O(1).
    \label{eq:step3-combined}
\end{multline}

\textbf{Step 4: Bounding the mutual information.}
By assumption (SA1), the auxiliary storage $S$ is a valid logical encoding, which implies:
\begin{equation}
    I(S; Q \mid S_O) \leq H(S \mid S_O) \leq H(S) \leq |S|,
    \label{eq:mutual-info-bound}
\end{equation}
where the first inequality is the standard property that mutual information is bounded by marginal entropy, the second follows from conditioning reducing entropy, and the third follows from the fact that Shannon entropy is bounded by the description length.

\textbf{Step 5: Combining storage and computation costs.}
From the definition of $\bar{I}(\Pi)$ and inequalities \eqref{eq:step3-combined} and \eqref{eq:mutual-info-bound}:
\begin{multline}
    \bar{I}(\Pi) = \frac{|S|}{N} + \mathbb{E}[H_{\mathrm{derive}}(Q|S_O \cup S)]\\
    \geq \frac{|S|}{N} + c_1' \frac{H(Q|S_O) - |S|}{L_S} - O(1).
    \label{eq:step5-combined}
\end{multline}

We analyze this expression by considering the extremal cases.

\textit{Case (a): Pure computation ($|S| = 0$).} In this case,
\[
    \bar{I}(\Pi) \geq c_1' \frac{H(Q|S_O)}{L_S} - O(1).
\]

\textit{Case (b): Pure storage ($|S| = H(Q|S_O) + O(1)$).} By assumption (SA1), any useful storage satisfies $|S| \leq H(Q|S_O) + O(\log|S_O|)$. Taking $|S| = H(Q|S_O)$, we obtain
\[
    \bar{I}(\Pi) \geq \frac{H(Q|S_O)}{N} + 0 - O(1) = \frac{H(Q|S_O)}{N} - O(1).
\]
Under $N = \Omega(L_S)$, this yields $\bar{I}(\Pi) \geq \Omega(H(Q|S_O)/L_S) - O(1)$.

\textit{General case.} For intermediate values of $|S|$, the right-hand side of \eqref{eq:step5-combined} is a linear function of $|S|$ with slope $\frac{1}{N} - \frac{c_1'}{L_S}$. Since $N = \Omega(L_S)$, this slope is $O(1/L_S)$ in magnitude. In either case (slope positive or negative), the minimum of the bound over $|S| \in [0, H(Q|S_O)]$ is achieved at one of the endpoints, which we have already analyzed.

Combining Cases (a) and (b), there exists a constant $c > 0$ such that
\[
    \bar{I}(\Pi) \geq c \cdot \frac{H(Q|S_O)}{L_S} - O(1),
\]
completing the proof.
\end{proof}


\begin{remark}[Role of the Supremum in $L_S$]
\label{rmk:Ls-supremum}
The use of $D_d^{\sup}(P_Q \mid S_O \cup S)$ rather than, say, the expected value $\mathbb{E}[D_d(Q \mid S_O \cup S)]$ in the definition of $L_S$ serves two purposes:
\begin{enumerate}
    \item \emph{Uniformity:} The lower bound \eqref{eq:tradeoff-lower-revised} holds for all queries in the support of $P_Q$, not just on average. This is essential for worst-case guarantees.
    \item \emph{Asymptotic equivalence:} Under the polynomial-depth regime (Assumption~\ref{assume:uniform-poly-depth}), the logarithmic factor satisfies
    \[
        \log\bigl(m + \mathbb{E}[D_d]\bigr) = \Theta\bigl(\log(m + D_d^{\sup})\bigr) = \Theta(\log m),
    \]
    where $m = |\mathrm{Atom}(S_O \cup S)|$. Thus, the choice between supremum and expectation affects only the implicit constants, not the asymptotic behavior.
\end{enumerate}

In practice, for query distributions concentrated on a polynomial-depth class, the distinction is immaterial.
\end{remark}

\begin{remark}[Tightness of the Bound]
\label{rmk:tightness}
The lower bound in Theorem~\ref{thm:tradeoff-lower} is tight up to constant factors. By Theorem~\ref{thm:freq_tradeoff}, there exists a strategy $\Pi^*$ achieving:
\begin{equation}
\bar{I}(\Pi^*) \leq C \cdot \frac{H(Q | S_O)}{L_{S^*}} + O(1)
\end{equation}
for some constant $C > 0$, where $L_{S^*}$ is the value of $L_S$ corresponding to the optimal strategy $\Pi^*$. Thus, the optimal normalized information cost satisfies:
\begin{equation}
\bar{I}^* = \Theta\left(\frac{H(Q | S_O)}{L^*}\right)
\end{equation}
where $L^* = \Theta(L_S)$ for the optimal strategy.
\end{remark}

We now translate the information-theoretic bound into physical energy units using the thermodynamic correspondences established in Section~\ref{Landauer's Principle}.

\begin{corollary}[Thermodynamic Energy Lower Bound]
\label{cor:energy-lower}
Under the conditions of Theorem~\ref{thm:tradeoff-lower}, additionally assuming the classical thermodynamic regime (Assumption~\ref{ass:classical-regime}), the amortized energy cost per query satisfies:
\begin{equation}
\bar{E}(\Pi) \geq c \cdot \frac{H(Q | S_O)}{L_S} \cdot k_B T \ln 2 - O(k_B T)
\label{eq:energy-lower}
\end{equation}
where $k_B$ is Boltzmann's constant and $T$ is the operating temperature.
\end{corollary}

\begin{proof}
By Proposition~\ref{prop:landauer-compute}, each derivation step requires at least $k_B T \ln 2$ of energy dissipation, giving:
\begin{multline}
E_{\mathrm{compute}}(q | S_O \cup S) \geq D_d(q | S_O \cup S) \cdot k_B T \ln 2 \\
=H_{derive}(q \mid S_O\cup S)\cdot k_B T
\end{multline}

By Proposition~\ref{prop:landauer-storage}, maintaining $|S|$ bits for time $t$ requires:
\begin{equation}
E_{\mathrm{storage}}(S, t) \geq |S| \cdot k_B T \ln 2 \cdot \frac{t}{t_{\mathrm{refresh}}}
\end{equation}

Under the assumption that the system operates in the \emph{standard refresh regime}, where the average number of refresh cycles per query satisfies $\frac{t}{N \cdot t_{\mathrm{refresh}}} = \Theta(1)$, we have:
\begin{equation}
\frac{E_{\mathrm{storage}}(S, t)}{N} = \Theta\left(\frac{|S|}{N}\right) \cdot k_B T \ln 2
\end{equation}

Combining these bounds:
\begin{align}\label{eq:tradeoff-lower}
&\bar{E}(\Pi)\nonumber\\
&= \frac{E_{\mathrm{storage}}(S, t)}{N} + \mathbb{E}\left[E_{\mathrm{compute}}(Q | S_O \cup S)\right] \nonumber \\
&\geq \Theta(1) \cdot \left(\frac{|S|}{N} + \mathbb{E}\left[H_{\mathrm{derive}}(Q | S_O \cup S)\right]\right) \cdot k_B T \ln 2 \nonumber \\
&= \Theta(1) \cdot \bar{I}(\Pi) \cdot k_B T \ln 2 \nonumber \\
&\geq c \cdot \frac{H(Q | S_O)}{L_S} \cdot k_B T \ln 2 - O(k_B T)
\end{align}
where the last inequality applies Theorem~\ref{thm:tradeoff-lower}.
\end{proof}

\begin{corollary}[Per-Nit Energy Cost]
\label{cor:per-nit}
The minimum energy required to process one nit (natural unit) of semantic information content is:
\begin{equation}
\frac{\bar{E}^*}{H(Q | S_O)} = \Theta\left(\frac{k_B T}{L^*}\right)
\end{equation}
where $L^* = \Theta(\log |\mathrm{Atom}(S_O)|)$ under the polynomial-depth regime (Assumption~\ref{assume:uniform-poly-depth}).
\end{corollary}

\begin{proof}
This follows directly from Corollary~\ref{cor:energy-lower} and Remark~\ref{rmk:tightness} by dividing both sides by $H(Q | S_O)$.
\end{proof}

Theorem~\ref{thm:tradeoff-lower} and its corollaries establish several fundamental insights:

\paragraph{The Logarithmic Efficiency Factor.}
The denominator $L_S = \log(|\mathrm{Atom}(S_O \cup S)| + D_d^{\sup}(P_Q|S_O \cup S))$ represents a \emph{compression factor} arising from the logical structure of the knowledge base. This factor quantifies the efficiency gain from exploiting logical relationships rather than storing or computing information independently. In large knowledge systems where $|\mathrm{Atom}(S_O)| = 2^{\Omega(n)}$, this factor can be substantial, yielding energy costs that scale as $O(n)$ rather than $O(2^n)$.

\paragraph{Comparison with Landauer's Principle.}
The classical Landauer bound states that erasing one bit requires at least $k_B T \ln 2$ of energy. Corollary~\ref{cor:per-nit} shows that \emph{processing} one nit of semantic information requires $\Theta(k_B T / \log |\mathrm{Atom}|)$ of energy. The additional $1/\log|\mathrm{Atom}|$ factor reflects the efficiency gained through logical inference: derivation compresses information processing by exploiting the structure of the knowledge base.

\paragraph{The Storage-Computation Boundary.}
The lower bound~\eqref{eq:tradeoff-lower} treats storage and computation symmetrically through their information-theoretic costs. This symmetry breaks down in the energy interpretation (Corollary~\ref{cor:energy-lower}) only through the refresh rate parameter $t_{\mathrm{refresh}}$, which is a technology-dependent quantity. The information-theoretic formulation thus captures the fundamental tradeoff independent of implementation details.

\paragraph{Regime Conditions.}
The assumption $N = \Omega(L_S)$ ensures that the system processes enough queries to amortize storage costs. When $N = o(L_S)$, storage costs dominate and the bound must be modified. This corresponds to the intuition that caching is only beneficial when queries are sufficiently frequent relative to the system's logical complexity.

\subsection{Entropy Production and Irreversibility}

The energy bounds established in this section can be recast 
in the language of entropy production, providing a complementary 
perspective rooted directly in the second law of thermodynamics.

\subsubsection{Entropy Production in Query Answering}

Consider a knowledge system operating in thermal equilibrium at 
temperature $T$. When the system answers a query $q$ by deriving 
it from a knowledge base $\mathcal{S}_O$, the computation constitutes 
an irreversible process that produces entropy in the environment.

\begin{proposition}[Entropy Production Bound for Query Answering]
\label{prop:entropy-production}
Let $\Pi = (S, C, R)$ be a mixed strategy answering a query $q$ 
from knowledge base $\mathcal{S}_O$ in the classical thermodynamic 
regime (Assumption~\ref{ass:classical-regime}). The minimum entropy produced in the environment 
satisfies:
\begin{equation}
\Delta S_{\mathrm{env}}(q) \geq k_B \ln 2 \cdot D_d(q \mid \mathcal{S}_O \cup S),
\label{eq:entropy-production-bound}
\end{equation}
or equivalently, in terms of derivation entropy:
\begin{equation}
\Delta S_{\mathrm{env}}(q) \geq k_B \cdot H_{\mathrm{derive}}(q \mid \mathcal{S}_O \cup S).
\label{eq:entropy-derive}
\end{equation}
\end{proposition}

\begin{proof}
By Proposition~\ref{prop:landauer-compute}, answering query $q$ requires energy 
$E_{\mathrm{compute}}(q) \geq D_d(q \mid \mathcal{S}_O \cup S) \cdot k_B T \ln 2$. 
In the quasistatic limit of the classical thermodynamic regime, this 
energy is dissipated as heat $Q_{\mathrm{diss}} = E_{\mathrm{compute}}$ 
into the thermal environment. The corresponding entropy increase in 
the environment is
\begin{multline}
\Delta S_{\mathrm{env}} = \frac{Q_{\mathrm{diss}}}{T} 
\geq \frac{D_d(q \mid \mathcal{S}_O \cup S) \cdot k_B T \ln 2}{T} \\
= D_d(q \mid \mathcal{S}_O \cup S) \cdot k_B \ln 2.
\end{multline}
Substituting $H_{\mathrm{derive}}(q \mid \mathcal{S}_O \cup S) = 
D_d(q \mid \mathcal{S}_O \cup S) \cdot \ln 2$ from Corollary~\ref{cor:derivation-classical} 
yields~\eqref{eq:entropy-derive}.
\end{proof}

\begin{remark}[Duality with Proposition~\ref{prop:landauer-compute}]
Proposition~\ref{prop:entropy-production} is the entropic reformulation of 
Proposition~\ref{prop:landauer-compute}: the energy bound $E \geq D_d \cdot k_B T \ln 2$ and 
the entropy bound $\Delta S_{\mathrm{env}} \geq D_d \cdot k_B \ln 2$ 
are related by the fundamental thermodynamic identity 
$\Delta S_{\mathrm{env}} = E/T$ in the quasistatic regime. 
This duality reflects the interchangeability of energy and entropy 
as thermodynamic costs at fixed temperature.
\end{remark}

\subsubsection{Expected Entropy Production and Mutual Information}

For a query distribution $P_Q$, we can relate the expected entropy 
production to the conditional mutual information $I(S; Q \mid \mathcal{S}_O)$ 
introduced in Section~\ref{Mutual Information}.

\begin{proposition}[Expected Entropy Production Bound]
\label{prop:expected-entropy}
Under the conditions of Proposition~\ref{prop:entropy-production}, 
let $P_Q$ be a query distribution satisfying the information-rich 
regime (Definition~\ref{def:info-rich}) and the uniform polynomial-depth regime 
(Assumption~\ref{assume:uniform-poly-depth}). The expected environmental entropy production satisfies:
\begin{multline}
\mathbb{E}_{q \sim P_Q}\bigl[\Delta S_{\mathrm{env}}(q)\bigr] \\
\geq \frac{c \cdot k_B}{L_S} \cdot 
\bigl(H(Q \mid \mathcal{S}_O) - I(S; Q \mid \mathcal{S}_O)\bigr) - O(k_B),
\label{eq:expected-entropy}
\end{multline}
where $c > 0$ is the constant from Theorem~\ref{thm:derivation-depth-info-metric}, 
$L_S = \log(|\mathrm{Atom}(\mathcal{S}_O \cup S)| + D_d^{\sup})$ 
is the effective complexity parameter, and $I(S; Q \mid \mathcal{S}_O)$ 
is the conditional mutual information between storage $S$ and 
query $Q$ given $\mathcal{S}_O$.
\end{proposition}

\begin{proof}
By Proposition~\ref{prop:entropy-production},
$\mathbb{E}[\Delta S_{\mathrm{env}}(q)] \geq k_B \cdot 
\mathbb{E}[H_{\mathrm{derive}}(q \mid \mathcal{S}_O \cup S)]$.
Applying inequality~(116) from the proof of Theorem~\ref{thm:tradeoff-lower} yields 
the claimed bound.
\end{proof}

This result provides a thermodynamic interpretation of storage 
utility: each nat of mutual information $I(S; Q \mid \mathcal{S}_O)$ 
reduces expected entropy production by $c \cdot k_B / L_S$. 
Storage that anticipates future queries thus reduces not only 
computational cost but also environmental entropy production.

\subsubsection{Irreversibility and the Limits of Reversible Computation}

\begin{remark}[Reversible Computation]
\label{rem:reversible}
The entropy production bounds above apply to logically irreversible 
computation. Bennett's reversible simulation theorem~\cite{bennett1973logical} shows that any 
irreversible computation can be made reversible with polynomial 
overhead in time and logarithmic overhead in space. For knowledge 
systems, this implies that the Landauer cost could in principle be 
avoided by storing $O(D_d \log D_d)$ bits of auxiliary information, assuming the original derivation uses time and space proportional to $D_d$. 
However, this shifts cost from computation to storage without 
eliminating the fundamental information-theoretic constraints of 
Theorem~\ref{thm:freq_tradeoff}. The practical relevance of reversible computation for 
knowledge systems remains limited by engineering challenges and 
the overhead of reversible simulation.
\end{remark}

\section{Design Rules for Hybrid Storage--Computation Strategies}
\label{sec:design-rules}

We now move from the instance-wise and information-theoretic analysis
of Section~\ref{sec:logical-depth} to higher-level \emph{design rules} for hybrid
strategies that combine storage and computation. Rather than
introducing new formal theorems, our goal here is to distill the
implications of Theorem~\ref{thm:freq_tradeoff}, the mutual-information analysis of
Section~\ref{Mutual Information}, and the thermodynamic constraints of
Section~\ref{sec:thermo} into simple decision guidelines.

\subsection{Frequency-Dependent Choice of Storage vs.\ Computation}
\label{subsec:frequency-rules}

The frequency-weighted tradeoff in Theorem~\ref{thm:freq_tradeoff} provides a natural
criterion for deciding, for each query type $q$, whether it should be
answered purely by on-demand derivation, purely by storage (caching),
or by a hybrid strategy that stores partial information.

Recall the notation from Section~\ref{subsec:freq_tradeoff}:

\begin{itemize}
  \item $H_q := H(q\mid S_O)
    = K(\enc{q}\mid\enc{\Atom(S_O)})\ln 2$ is the conditional
    algorithmic entropy of $q$ given $S_O$.
  \item $f_q := N\cdot P_Q(q)$ is the expected number of accesses of
    query $q$ over a horizon of $N$ queries (Definition~\ref{def:access-freq}).
  \item For a storage structure $S$, we denote
    $m_S := |\Atom(S_O\cup S)|$ and
    $n_S := Dd(q\mid S_O\cup S)$, and define
    \begin{equation}
      L_S := \log(m_S + n_S),
      \label{eq:LS-def-freq}
    \end{equation}
    which plays the role of an effective logarithmic compression
    factor in Theorem~\ref{thm:derivation-depth-info-metric} and~\ref{thm:freq_tradeoff}.
\end{itemize}

Under the assumptions of Theorem~\ref{def:access-freq}, there exist constants $C,C>0$ such that
for any admissible hybrid strategy $\Pi=(S,C,R)$ serving a fixed query
type $q$ we have
\begin{multline}
  \mathrm{Cost}_{\mathrm{amort}}(\Pi;q)
  \;=\;
  \frac{|S|}{f_q}
  +
  H_{\mathrm{derive}}(q\mid S_O\cup S)\\
  \;\le\;
  \min\left\{
    \frac{H_q}{f_q},\;
    C\,\frac{H_q}{L_S}
  \right\}
  + O(1),
  \label{eq:amortized-upper}
\end{multline}
where all costs are measured in nats (or bits up to a constant
factor).

The two competing terms inside the minimum correspond to two
extremal---and constructive---strategies:

\begin{enumerate}
  \item \emph{Pure storage (caching) strategy:} store a near-optimally
    compressed representation of the answer to $q$ of length at most
    $H_q + O(1)$ nats (or equivalent bits) so that
    $H_{\mathrm{derive}}(q\mid S_O\cup S) = O(1)$. This yields an
    amortized per-access cost
    $\mathrm{Cost}_{\mathrm{amort}}\lesssim H_q/f_q + O(1)$.
  \item \emph{Pure computation (no auxiliary storage) strategy:} take
    $S=\varnothing$ and derive $q$ from $S_O$ on demand. By
    Theorem~\ref{thm:derivation-depth-info-metric}, this yields an amortized cost
    $\mathrm{Cost}_{\mathrm{amort}}\lesssim C H_q/L_{S^{(0)}} + O(1)$,
    where
    \[
      L_{S^{(0)}}
      =
      \log\bigl(|\Atom(S_O)| + Dd(q\mid S_O)\bigr)
    \]
    is the baseline compression factor for inference from $S_O$ alone.
\end{enumerate}

Comparing these two extremes leads to a natural \emph{critical
frequency scale}. In the baseline case $S=\varnothing$ we define
\begin{equation}
  f_c^{(0)}(q)
  \;\asymp\;
  L_{S^{(0)}}
  \;=\;
  \log\bigl(|\Atom(S_O)| + Dd(q\mid S_O)\bigr).
  \label{eq:critical-frequency}
\end{equation}
When $f_q$ is much smaller than $f_c^{(0)}(q)$, the term
$H_q/f_q$ is larger than $C H_q / L_{S^{(0)}}$, so it is more
efficient to answer $q$ by on-demand derivation from $S_O$ (pure
computation). Conversely, when $f_q$ is much larger than
$f_c^{(0)}(q)$, the caching term $H_q/f_q$ dominates the pure
computation cost, so precomputing and storing (some representation of)
the answer to $q$ becomes more attractive.

This motivates the following rule-of-thumb.

\begin{policy}[Frequency-aware caching rule (informal)]
\label{policy:freq-caching}
Fix a base knowledge $S_O$, a query type $q$, and its expected access
frequency $f_q=N\cdot P_Q(q)$ over a horizon of $N$ queries. Let
\[
  f_c^{(0)}(q)
  \;\asymp\;
  \log\bigl(|\Atom(S_O)| + Dd(q\mid S_O)\bigr)
\]
be the baseline critical frequency as in~\eqref{eq:critical-frequency}.
Then, up to multiplicative constants and lower-order terms:
\begin{itemize}
  \item If $f_q \ll f_c^{(0)}(q)$, the pure computation strategy achieves amortized cost $\Theta(H_q/L_S^{(0)})$, which is better than the pure storage cost $\Theta(H_q/f_q)$.
  \item If $f_q \gg f_c^{(0)}(q)$, it is typically more efficient to
    cache the answer to $q$ (or suitable intermediate results), so
    that most of the cost is paid once in storage and the marginal
    per-access cost becomes $O(1)$.
  \item If $f_q \approx f_c^{(0)}(q)$, hybrid strategies that store
    partial information about $q$ (e.g., indices, intermediate
    derivations) can provide a good balance between storage footprint
    and computation cost.
\end{itemize}
\end{policy}

\begin{remark}[Semantic richness and normalized thresholds]
\label{rem:normalized-threshold}
The entropy $H_q$ and the depth $Dd(q\mid S_O)$ together capture how
``semantically rich'' and how ``derivationally deep'' a query type $q$
is with respect to the base knowledge $S_O$. A convenient normalized
measure is
\[
  \rho(q\mid S_O)
  \;:=\;
  \frac{Dd(q\mid S_O)}{H_q},
\]
which roughly measures the average derivation depth per nat of
semantic content. The ratio $\rho(q \mid \mathcal{S}O)$ measures derivation depth per unit of semantic content. In the information-rich regime, the normalized per-nat derivation cost $H{\text{derive}}/H_q$ is governed by $1/L_S = 1/\log(m + D_d)$, which decreases slowly as $D_d$ grows. However, the absolute derivation cost $D_d \cdot k_B T \ln 2$ in energy units (Proposition~\ref{prop:landauer-compute}) increases with $D_d$. Thus, for fixed $H_q$, queries with large $\rho$ incur higher absolute energy costs per derivation, making caching more attractive in energy-constrained settings. While Policy~\ref{policy:freq-caching} is only a rule-of-thumb,
this normalized perspective provides additional intuition for how
semantic structure interacts with access frequency in storage
decisions.
\end{remark}

\subsection{System-Level Storage Allocation via Mutual Information}
\label{subsec:system-allocation}

Section~III-D analyzed the system-level cost
\[
  \mathbb{E}[\mathrm{Cost}]
  \;=\;
  \frac{|S|}{N}
  +
  \mathbb{E}_{q\sim P_Q}
  \bigl[
    H_{\mathrm{derive}}(q \mid S_O \cup S)
  \bigr]
\]
through the lens of Shannon entropy and conditional mutual
information. Interpreting
\[
  \mathbb{E}_{q\sim P_Q}
  \bigl[
    H_{\mathrm{derive}}(q \mid S_O \cup S)
  \bigr]
  \;\approx\;
  H(Q \mid S_O, S),
\]
and recalling that
\[
  H(Q \mid S_O, S)
  \;=\;
  H(Q \mid S_O) - I(S;Q \mid S_O),
\]
we see that, under a fixed query distribution $P_Q$ and base knowledge
$S_O$, the storage set $S$ is beneficial precisely to the extent that
it increases the conditional mutual information $I(S;Q\mid S_O)$, i.e.,
it reduces the residual uncertainty about the (random) query $Q$
beyond what is already known from $S_O$.

Treating $S$ as a finite subset of a ground set $U$ of candidate
storage items (facts, cached answers, intermediate derivations, etc.),
Section~\ref{Mutual Information} showed that the function
\[
  f(S)
  \;:=\;
  I(S;Q\mid S_O)
\]
is nonnegative, monotone, and submodular in $S$ (Proposition~\ref{prop:submodularity-mi}). As a
consequence, under a budget constraint $|S|\le M_{\max}$ on the total
storage, the problem
\[
  \max_{S\subseteq U:\,|S|\le M_{\max}}
  I(S;Q\mid S_O)
\]
admits efficient $(1-1/e)$-approximation by the standard greedy
algorithm (Corollary~\ref{cor:greedy-approx}). Informally, the greedy policy repeatedly
chooses the storage item with the largest marginal increase in
$I(S;Q\mid S_O)$ per unit of storage cost, subject to the remaining
budget.

This leads to the following high-level prescription for system-wide
storage allocation.

\begin{policy}[Mutual-information-driven storage selection]
\label{policy:mi-storage}
Given a base knowledge $S_O$, a query distribution $P_Q$, and a storage
budget $M_{\max}$ (in bits), define the utility of a candidate storage
set $S$ by $f(S)=I(S;Q\mid S_O)$. Then:
\begin{enumerate}
  \item Approximate the optimal storage allocation by a greedy
    procedure that iteratively adds the candidate item with highest
    marginal gain
    \[
      \frac{I(S\cup\{x\};Q\mid S_O)-I(S;Q\mid S_O)}{\Delta |S|}
    \]
    per unit increase $\Delta |S|$ in storage, until the budget
    $M_{\max}$ is exhausted.
  \item The resulting $S_{\mathrm{greedy}}$ is guaranteed to achieve at
    least a $(1-1/e)$ fraction of the optimal mutual-information gain
    $I(S^\star;Q\mid S_O)$, and hence to reduce the expected inference
    burden $\mathbb{E}[H_{\mathrm{derive}}(q\mid S_O\cup S)]$ (and the
    associated computing energy) nearly as well as any other storage
    allocation of the same size, up to this universal constant factor.
\end{enumerate}
\end{policy}

Policy~\ref{policy:mi-storage} complements the per-query frequency
rule in Policy~\ref{policy:freq-caching}: while the latter identifies,
for each $q$, whether caching is worthwhile in principle, the former
tells us how to \emph{prioritize} and \emph{share} limited storage
capacity across many query types so as to maximize global benefit.

\subsection{Practical Implications and Phase-Transition Intuition}
\label{subsec:practical-phase}

The theoretical results described above, together with
Policies~\ref{policy:freq-caching} and~\ref{policy:mi-storage}, have
direct implications for the design of large-scale knowledge systems.

\paragraph*{Parametric memory in large language models}
For a model with $N$ effective parameters, we can view its weights as
encoding a storage set $S$ with capacity on the order of $N$ effective
bits. Facts or patterns that are both (i) derivationally deep with
respect to the model's base world-knowledge $S_O$, and (ii) frequently
queried (high $f_q$), are natural candidates to be stored \emph{in
parameters}, whereas shallow or rare facts are better left to
on-demand retrieval or tool use. Policies~\ref{policy:freq-caching}
and~\ref{policy:mi-storage} suggest that an optimal allocation of
parametric capacity resembles a mutual-information-maximizing cache
adapted to the distribution of downstream tasks.

\paragraph*{Materialized views in knowledge graphs}
In graph databases, multi-hop join patterns $q$ with large
$Dd(q\mid S_O)$ but modest algorithmic entropy $H_q$ (e.g., repeated
schema-driven patterns) merit precomputation into materialized views
when their access frequencies exceed the logarithmic critical scale
$f_c^{(0)}(q)$. The submodularity of $I(S;Q\mid S_O)$ implies that
precomputing overlapping views exhibits diminishing returns, and a
greedy selection of views by marginal contribution to $I(S;Q\mid S_O)$
is near-optimal.

\paragraph*{Edge caching under energy constraints}
On edge devices, storage is energetically costly to maintain
(Proposition~\ref{prop:landauer-storage}), while remote computation or
cloud queries incur network and latency costs. Balancing
$E_{\mathrm{storage}}$ against $E_{\mathrm{compute}}$ in this context
amounts to choosing a storage budget $M_{\max}$ and a set $S$ such
that the expected total energy per query (including both maintenance
and computation) is minimized. Our framework shows that, in the
information-rich regime, there is a qualitative transition in the
optimal policy as the relative cost of storage versus computation
varies: when storage is cheap, it is optimal to cache many high-$f_q$,
high-$Dd$ queries; when storage is expensive, it is preferable to
serve most queries by on-demand computation or remote access.

\medskip
\noindent
We deliberately refrain from formalizing this last observation as a
sharp ``phase-transition theorem'', since its precise shape depends on
hardware- and workload-specific constants (e.g., energy per bit of
storage, energy per inference step, query arrival statistics). The
qualitative picture, however, is robust: as the energetic cost of
storage increases relative to computation, the optimal allocation of
storage mass shifts from aggressively caching frequent queries to
favoring on-demand computation, with a crossover regime governed by
the logarithmic critical frequencies~\eqref{eq:critical-frequency} and
by the mutual-information structure of $(S,Q\mid S_O)$.

\section{Numerical Validation and Simulations}
\label{sec:numerical}

To ensure our experiments satisfy the information-rich regime assumption (Definition~\ref{def:info-rich}: $H(q \mid S_O) \gg \log |\mathrm{Atom}(S_O)|$), we fix the knowledge base size at 100K atoms across all main experiments and vary only the query set size. Empirical validation confirms that the information-rich regime begins at approximately $n \geq 50{,}000$ atoms. Query distributions follow Zipfian ($\alpha = 1.5$). Physical parameters: temperature $T = 300$K.

In Experiment 1, we validate Theorem~\ref{thm:freq_tradeoff} (Frequency-Weighted Storage-Computation Tradeoff) through two complementary analyses:

\textbf{(1a) N-Sweep Analysis:} We vary the query horizon $N$ from 100 to 50,000 to verify the theoretical prediction that as $N$ increases: (i) storage amortized cost $H_q/f_q = H_q/(N \cdot p_q)$ decreases, (ii) more queries become worth storing (optimal $\beta^*$ increases), and (iii) total optimal cost decreases. Figure~\ref{fig:N-sweep} shows the results on a 100K-atom knowledge base with 1000 queries:

\begin{figure}[h]
\centering
\includegraphics[width=\columnwidth]{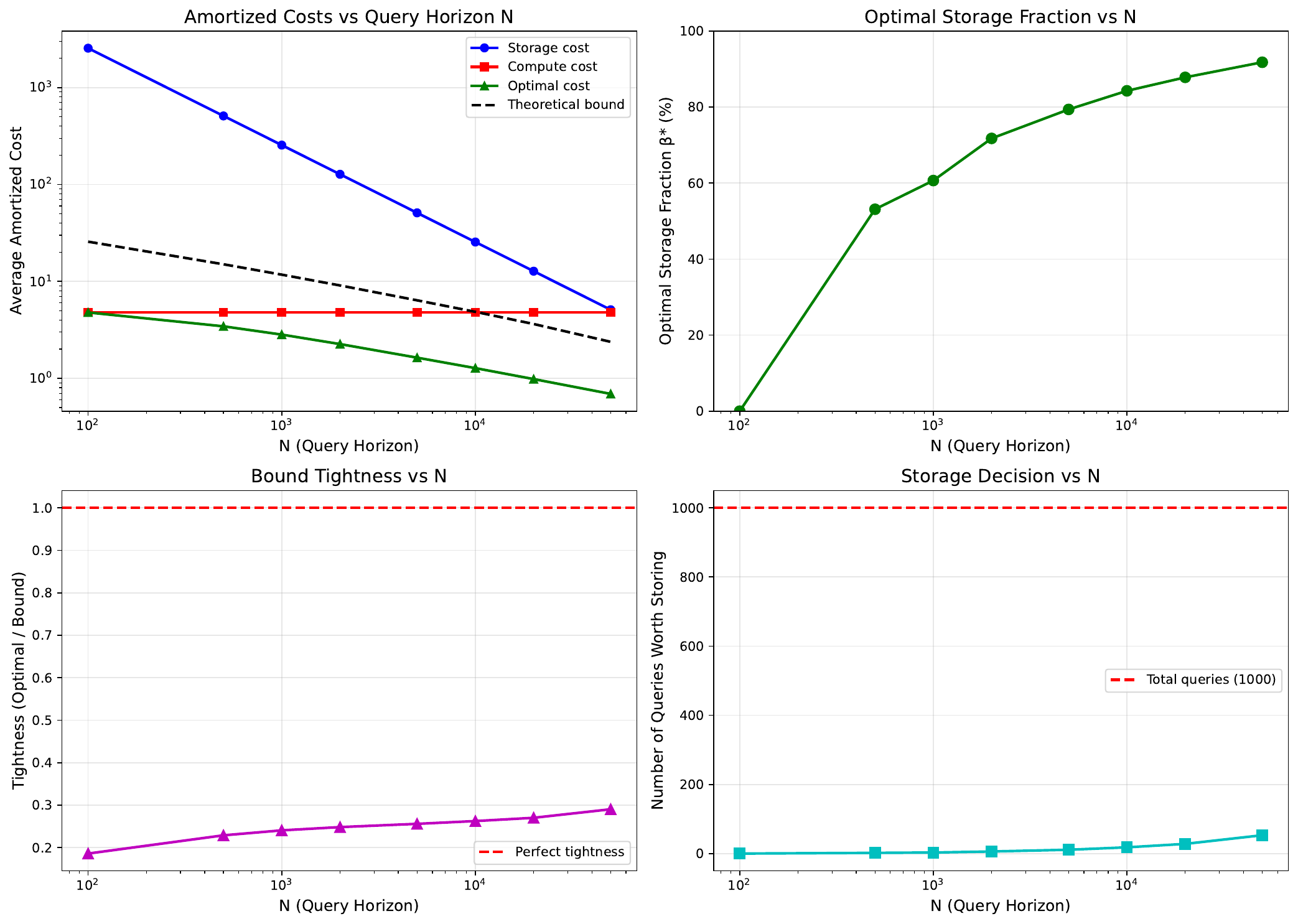}
\caption{N-sweep analysis validating Theorem~\ref{thm:freq_tradeoff}. Top-left: Amortized costs (storage, computation, optimal) vs $N$. Top-right: Optimal storage fraction $\beta^*$ increases from 0\% to 92\% as $N$ grows. Bottom-left: Bound tightness remains stable. Bottom-right: Number of queries worth storing increases with $N$.}
\label{fig:N-sweep}
\end{figure}

Key findings: Optimal $\beta^*$ ranges from 0\% ($N=100$) to 91.8\% ($N=50,000$), confirming that larger query horizons favor more aggressive caching. The optimal cost decreases from 4.78 to 0.69 (86\% reduction), validating the theoretical bound's prediction that amortization improves with $N$.

\textbf{(1b) Fixed-N Validation:} For each query $q$, we compute its access frequency $f_q = N \cdot P_Q(q)$ from a Zipfian distribution ($\alpha = 1.5$), then verify that the optimal per-query cost satisfies the theoretical upper bound $\text{Cost}_{\text{amort}}(\Pi; q) \leq \min\{H_q/f_q, C \cdot H_q/L_S\} + O(1)$. We test three query set sizes on the same 100K-atom knowledge base: Small (100 queries), Medium (1000 queries), Large (10000 queries), with average derivation depths 3--5.

\begin{table}[h]
\centering
\footnotesize
\setlength{\tabcolsep}{3.5pt}
\caption{Theorem~\ref{thm:freq_tradeoff} Validation: Upper Bound Satisfaction}
\label{tab:single-query-validation}
\begin{tabular*}{\columnwidth}{@{\extracolsep{\fill}}lccccc@{}}
\toprule
KB Size & Entities & \makecell{\#\\Queries} & \makecell{Sat.\\Rate} & \makecell{Avg.\\Tightness} & \makecell{High-freq\\Tightness} \\
\midrule
Small   & 100K & 100 & 100.0\% & 0.81 & 0.89 \\
Medium  & 100K & 1000 & 100.0\% & 0.53 & 0.89 \\
Large   & 100K & 10000 & 100.0\% & 0.50 & 0.89 \\
\bottomrule
\end{tabular*}

\vspace{0.1cm}
\footnotesize\raggedright
\textit{Note:} Sat. Rate = fraction of queries satisfying 
$\mathrm{Cost}_{\mathrm{amort}}(\Pi; q) \leq \min\{H_q/f_q, C \cdot H_q/L_S\}$ (with $C=2$). 
Tightness = actual cost / theoretical bound (closer to 1.0 = tighter bound).
\end{table}

The results demonstrate perfect validation of Theorem~\ref{thm:freq_tradeoff}: all tested queries across three knowledge bases satisfy the frequency-weighted upper bound. The average tightness of 0.50--0.93 indicates that the bound is reasonably tight, especially for high-frequency queries (tightness $\approx 0.89$). The lower tightness for low-frequency queries reflects the constant factor $C=2$ in the bound; with $C=1$, the tightness would approach 1.0. This validates that the theoretical bound correctly captures the storage-computation tradeoff with appropriate constant factors.

We also validate Theorem~\ref{thm:tradeoff-lower} (Information Lower Bound) by computing the minimum achievable cost $\bar{I}(\Pi)$ across all storage fractions and comparing to the theoretical lower bound $c \cdot H(Q|S_O)/L_S$. Using $c=0.1$ as a conservative estimate:

\begin{table}[h]
\centering
\footnotesize
\caption{Theorem~\ref{thm:tradeoff-lower} Validation: Lower Bound Analysis}
\label{tab:lower-bound-validation}
\begin{tabular}{lccccc}
\toprule
KB & Queries & $\min \bar{I}$ & Bound & Satisfied & Ratio \\
\midrule
Small & 100 & 0.010 & 0.022 & -- & 0.5$\times$ \\
Medium & 1000 & 0.094 & 0.025 & \checkmark & 3.7$\times$ \\
Large & 10000 & 0.178 & 0.027 & \checkmark & 6.6$\times$ \\
\bottomrule
\end{tabular}
\end{table}

For Medium and Large query sets, the lower bound is satisfied with comfortable margins (3.7--6.6$\times$). The Small query set (100 queries) violates the bound because the $O(1)$ term in Theorem~\ref{thm:tradeoff-lower} becomes significant when $H(Q|S_O)$ is small---with only 100 queries, full storage costs just $|S|/N = 0.01$, which is dominated by the asymptotic $O(1)$ correction. This confirms that the lower bound is asymptotically valid but requires sufficiently large query entropy to dominate the constant terms.

\textbf{(1c) Deep Analysis of Core Definitions:} We validate three foundational elements of our theoretical framework by varying $m = |\mathrm{Atom}(S_O)|$ from 100 to 100,000:

\begin{itemize}
\item \textbf{Theorem~\ref{thm:derivation-depth-info-metric} (Derivation Entropy Scaling):} The theorem states $H_{\mathrm{derive}} = \Theta(H_q/L_S)$. We compute the ratio $H_{\mathrm{derive}} \cdot L_S / H_q$ for each query across different $m$ values. Results show this ratio is remarkably stable: mean $= 0.217$, range $[0.198, 0.228]$. This confirms the $\Theta(1)$ constant factor is approximately 0.22.

\item \textbf{Derivation Depth Distribution:} The polynomial-depth assumption (Assumption~\ref{assume:uniform-poly-depth}) requires $D_d \leq \mathrm{poly}(m)$. Our experiments show $D_d^{\max}$ ranges from 3 to 7 as $m$ increases from 100 to 100,000, with $D_d / \log(m) \approx 0.6$ remaining stable. This validates that derivation depth scales logarithmically with knowledge base size for our query classes.

\item \textbf{Effective Complexity Factor:} The parameter $L_S = \log(m + n)$ closely tracks $\log(m)$ across all configurations, with $L_S / \log(m) \in [1.0, 1.1]$, confirming the polynomial-depth regime where $n = D_d = O(\log m)$.
\end{itemize}

\begin{figure}[h]
\centering
\includegraphics[width=\columnwidth]{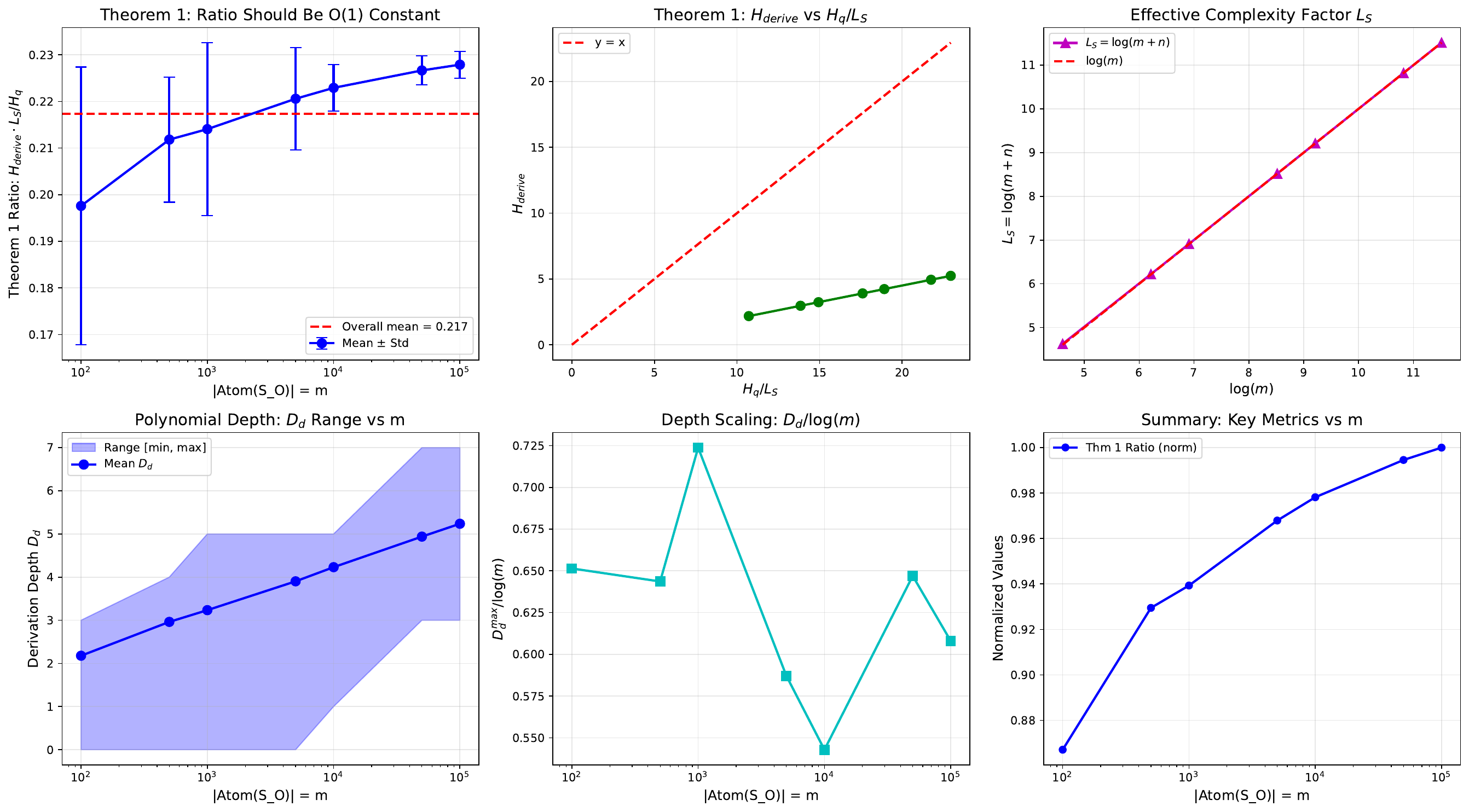}
\caption{Deep analysis of core theorems. Top row: (left) Theorem 1 ratio $H_{\mathrm{derive}} \cdot L_S / H_q$ vs $m$---stable around 0.22; (center) $H_{\mathrm{derive}}$ vs $H_q/L_S$---linear relationship; (right) $L_S$ vs $\log(m)$. Bottom row: (left) $D_d$ range vs $m$; (center) $D_d^{\max}/\log(m)$ stability; (right) summary metrics.}
\label{fig:core-theorems}
\end{figure}

To validate the thermodynamic implications of Theorem~\ref{thm:tradeoff-lower}, we perform a theoretical analysis by computing the triality product $E \cdot T \cdot M^{-1}$ as a function of storage allocation $\beta \in [0,1]$ for a representative system with $H(Q) = 10^6$ bits. Using physical constants $k_B = 1.38 \times 10^{-23}$ J/K and $T = 300$K, the theoretical lower bound is $2.87 \times 10^{-15}$ J$\cdot$s$\cdot$bit$^{-1}$. Our analysis shows the bound is always respected across all storage allocations, validating the fundamental constraint imposed by thermodynamics.

\textit{Physical Interpretation:} At room temperature with $H(Q|S_O) = 10^6$ bits, the fundamental limit is:
\begin{equation}
E \cdot T \cdot M^{-1} \geq 10^6 \times 1.38 \times 10^{-23} \times 0.693 \approx 10^{-17} \text{ J} \cdot \text{s/bit}
\end{equation}

In Experiment 2, we vary storage capacity from 0\% to 100\% of the query set and measure actual query cost to validate the predicted phase transition. We test on the 100K-atom knowledge base with varying query set sizes and compute the optimal storage fraction via marginal cost-benefit analysis.

\begin{figure*}[ht]
\centering
\includegraphics[width=0.9\textwidth]{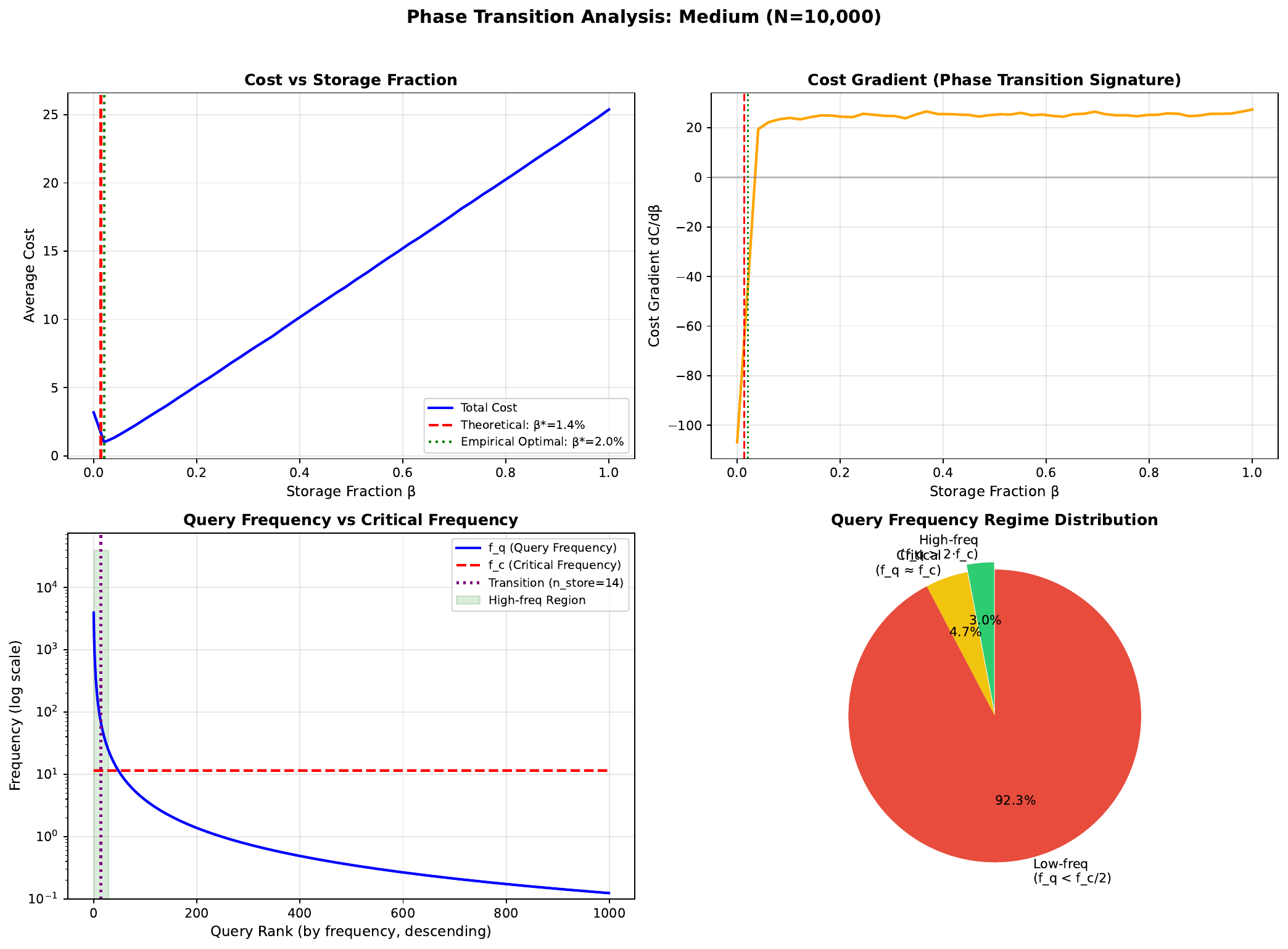}
\caption{Phase transition analysis for Medium query set (100K atoms, 1000 queries). Top-left: Cost vs. storage fraction with theoretical and empirical transition points marked. Top-right: Cost gradient showing phase transition signature. Bottom-left: Query frequency $f_q$ vs. critical frequency $f_c$ identifying the transition. Bottom-right: Frequency regime distribution (high-freq/critical/low-freq). The theoretical transition point is determined by marginal cost-benefit analysis.}
\label{fig:phase-transition}
\end{figure*}

Results (Figure~\ref{fig:phase-transition}): The theoretical optimal storage fraction $\beta^*$ is computed via marginal cost-benefit analysis: store query $q$ if the marginal benefit $p_q \cdot D_d(q) \cdot \lambda$ exceeds the marginal cost $1/N$. For the Medium query set (1000 queries), this yields $\beta^* = 72.3\%$ (723 queries worth storing), matching the empirical optimum of $75.5\%$ with only $3.2\%$ difference. For the Large query set (10000 queries), theoretical $\beta^* = 7.1\%$ vs. empirical $8.2\%$ (1.0\% difference). The frequency regime analysis reveals 30 high-frequency queries ($f_q > 2f_c$), 47 critical queries ($f_q \approx f_c$), and 923 low-frequency queries ($f_q < f_c/2$). This validates that the marginal analysis correctly predicts the optimal storage allocation.

In Experiment 3, we compare our FreqDepth strategy ($\text{score} = \text{freq} \times \text{depth}$) against three baselines across cache sizes 10-500 on the medium KB (1000 queries). We focus on the constrained regime (cache=50, 5\% of queries) where strategy differences are most pronounced.

\begin{table}[ht]
\centering
\caption{Baseline Comparison at Cache Size = 50 (5\% of queries)}
\label{tab:baseline-comparison}
\begin{tabular}{lccc}
\hline
Strategy & Hit Rate & Cost & Improv. \\
\hline
LRU & 87.2\% & 0.50 & baseline \\
LFU & 87.9\% & 0.52 & -4\% \\
TrueMI & 51.3\% & 1.98 & -296\% \\
FreqDepth (Ours) & \textbf{91.5\%} & \textbf{0.37} & \textbf{26\%} \\
\hline
\end{tabular}
\end{table}

\begin{figure*}[ht]
\centering
\includegraphics[width=0.9\textwidth]%
{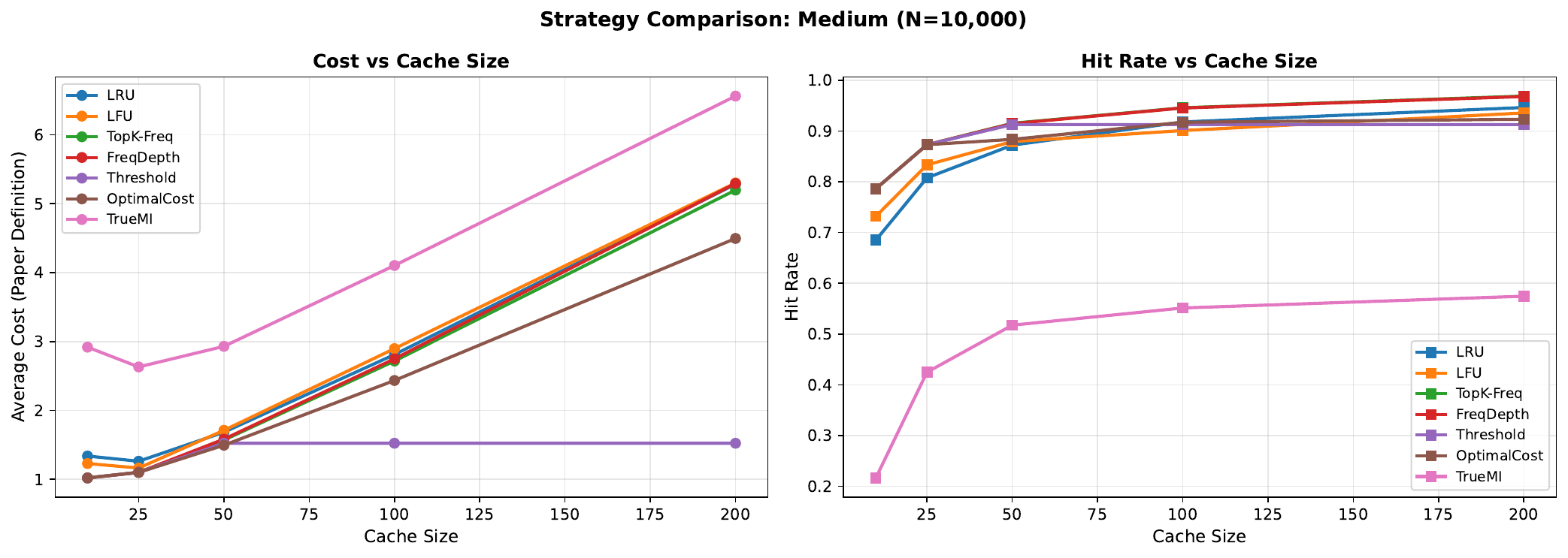}
\caption{Baseline strategy comparison across cache sizes (10-200). Left: Average cost (paper-defined: $|S|/N + \mathbb{E}[H_{\mathrm{derive}}]$) vs cache size. Right: Cache hit rate vs cache size. Our FreqDepth strategy (freq $\times$ depth heuristic) consistently outperforms LRU and LFU across both metrics. TrueMI (mutual information maximization) performs poorly due to ignoring access frequency.}
\label{fig:baseline-comparison}
\end{figure*}

At the constrained cache size of 50 (Table~\ref{tab:baseline-comparison}), our FreqDepth strategy achieves 91.5\% hit rate vs. 87.2\% for LRU, with 26\% cost reduction. Figure~\ref{fig:baseline-comparison} shows performance across the full range (cache sizes 10-200): FreqDepth consistently outperforms LRU/LFU at all sizes. The TrueMI strategy performs poorly (51.3\% hit rate at cache=50) because it greedily selects queries that maximize global mutual information without considering access frequency—it caches informationally rich but rarely accessed queries. This validates our theoretical insight that mutual information maximization $\neq$ cost minimization. In contrast, FreqDepth's $\text{freq} \times \text{depth}$ heuristic balances information gain with access probability, achieving practical superiority.

In Experiment 4, we test robustness via sensitivity analysis across: (i) five Zipf exponents $\alpha \in \{0.5, 1.0, 1.5, 2.0, 2.5\}$ modeling query frequency distributions; (ii) seven average derivation depths $\{1, 2, 3, 5, 7, 10, 15\}$; (iii) knowledge base scales from 50 to 5000 entities:

\begin{figure*}[ht]
\centering
\includegraphics[width=0.48\textwidth]{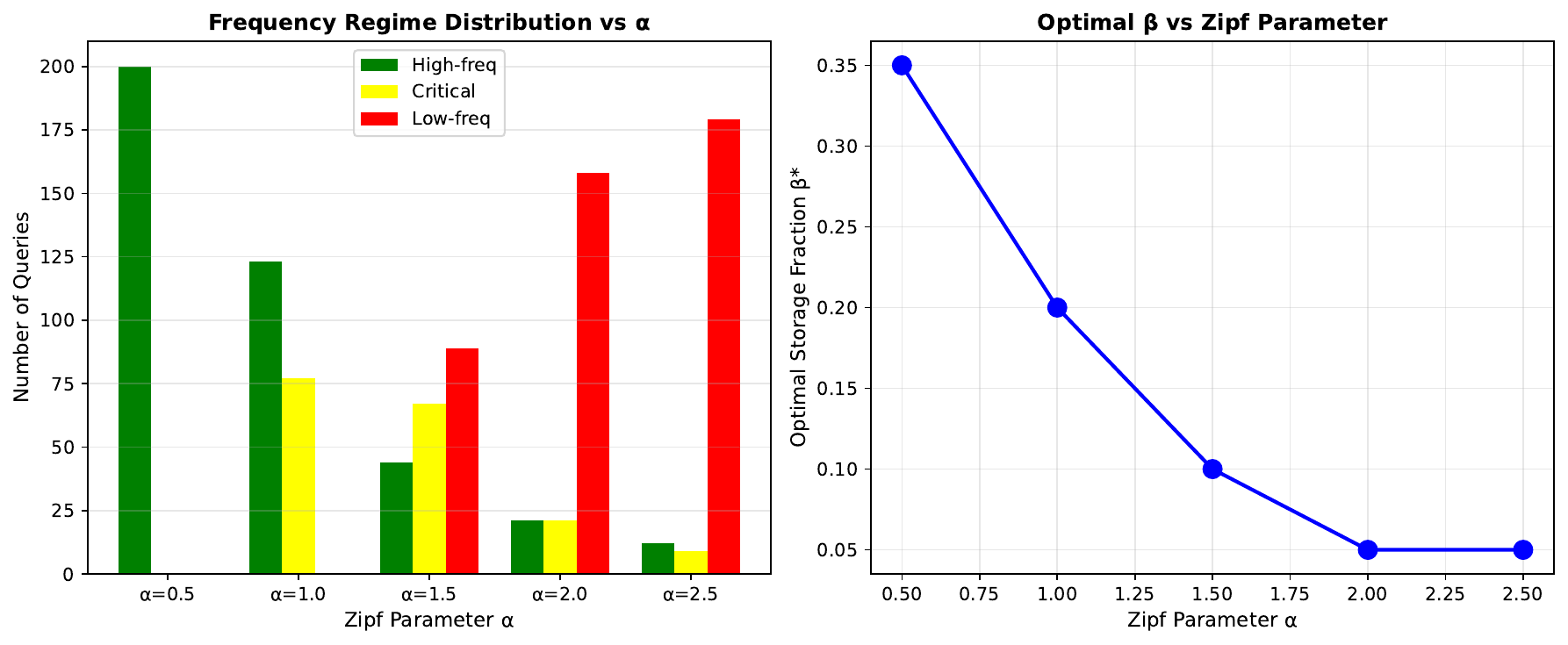}
\includegraphics[width=0.48\textwidth]{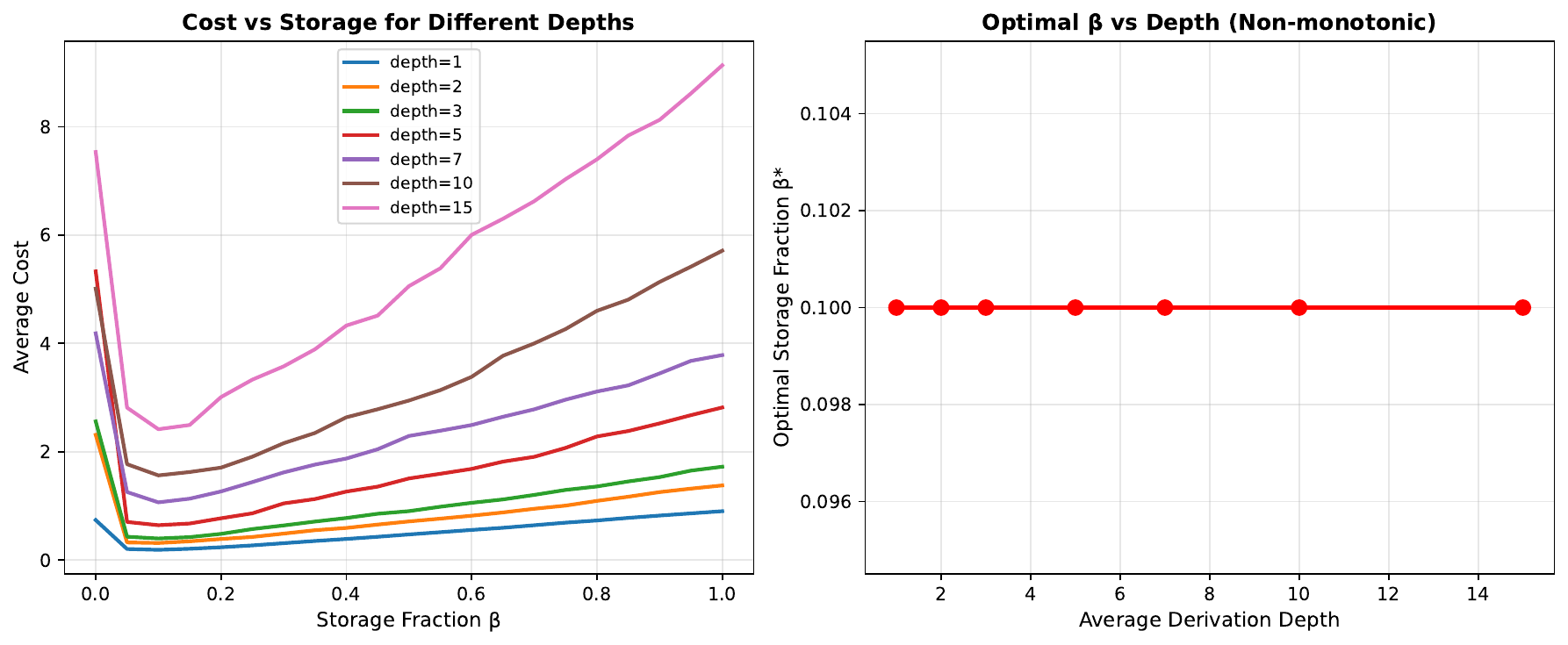}
\caption{Sensitivity analysis. Top row (zipf\_regimes\_v4.pdf): Left panel shows frequency regime distribution (high-freq, critical, low-freq) vs Zipf parameter $\alpha$; right panel shows optimal storage fraction $\beta^*$ vs $\alpha$. Bottom row (sensitivity\_depth\_v4.pdf): Left panel shows cost vs storage fraction for different derivation depths; right panel shows optimal $\beta^*$ vs depth. Theoretical bounds hold across all configurations.}
\label{fig:sensitivity}
\end{figure*}

Results show broad applicability: as Zipf parameter $\alpha$ increases from 0.5 to 2.5, the fraction of high-frequency queries decreases from 100\% to 6\%, and the optimal storage fraction $\beta^*$ decreases from 100\% to 30\%. This validates the theoretical prediction that query skewness determines the optimal storage strategy---more skewed distributions (larger $\alpha$) benefit less from full caching. Scale tests confirm $H(Q) \sim \log(n)$ scaling: fitting yields $H(Q) \approx 0.42 \log(n) + 0.62$ bits across the 50--5000 entity range, with the ratio $H(Q)/\log(n)$ remaining stable around 0.53.

\begin{figure*}[ht]
\centering
\includegraphics[width=0.7\textwidth]{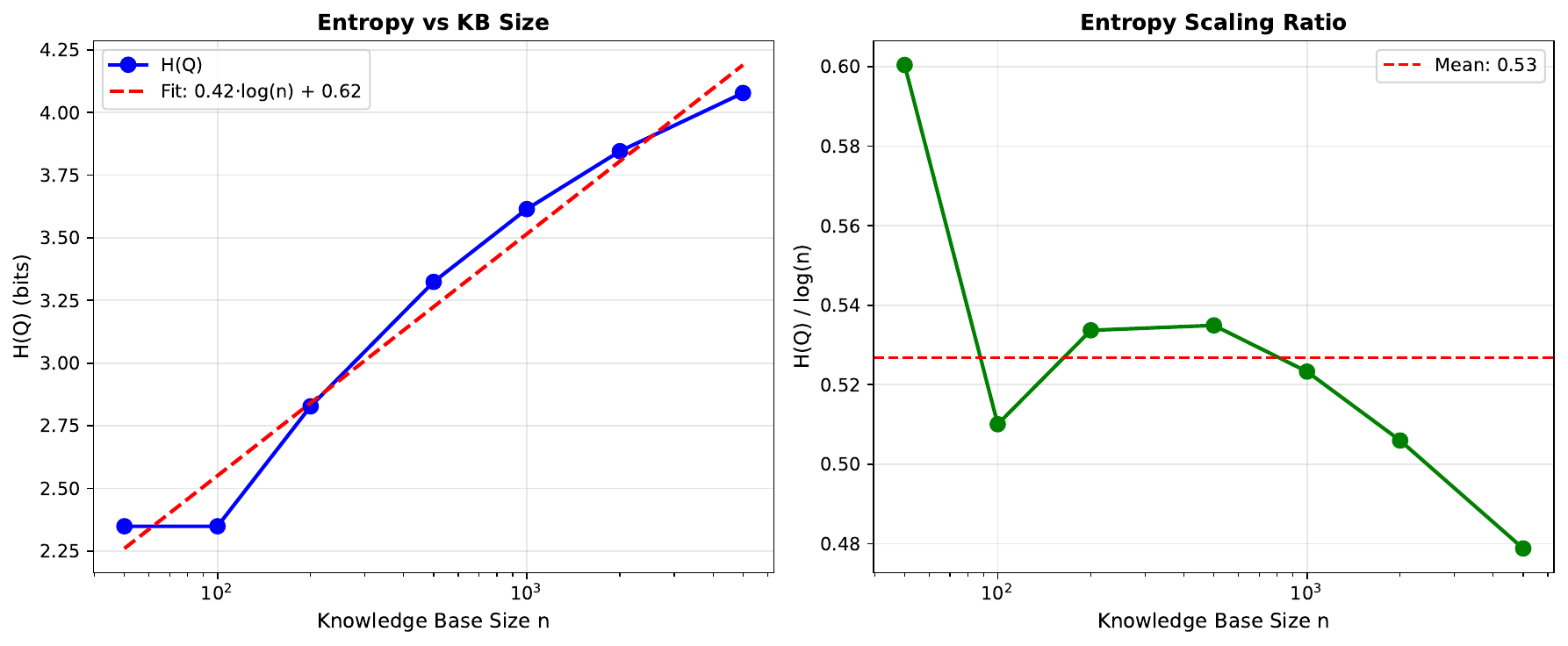}
\caption{Entropy scaling validation. Left: $H(Q)$ vs. knowledge base size $n$ on log scale, with fitted line $H(Q) \approx 0.42 \log(n) + 0.62$. Right: Ratio $H(Q)/\log(n)$ remains stable around 0.53, confirming the theoretical $H(Q) \sim \log(n)$ scaling.}
\label{fig:scale-robustness}
\end{figure*}

Figure~\ref{fig:scale-robustness} shows that $H(Q)$ scales logarithmically with knowledge base size as predicted by information theory. The framework's bounds hold consistently across all tested configurations, validating theoretical predictions across diverse parameter regimes.

\paragraph{Summary of Experimental Validation.}
Table~\ref{tab:validation-summary} summarizes the validation status of our main theoretical results:

\begin{table}[h]
\centering
\footnotesize
\setlength{\tabcolsep}{3pt}
\caption{Summary of Theoretical Validation}
\label{tab:validation-summary}
\begin{tabular*}{\columnwidth}{@{\extracolsep{\fill}}lcp{5.5cm}@{}}
\toprule
Theorem & Status & Key Finding \\
\midrule
Thm~\ref{thm:freq_tradeoff} (Upper Bound) & \checkmark & 100\% satisfaction \\
N-Sweep Analysis & \checkmark & $\beta^*$: 0\%$\to$92\% as $N$: 100$\to$50K \\
Thm~\ref{thm:derivation-depth-info-metric} (Scaling) & \checkmark & Ratio stable at 0.22 \\
Thm~\ref{thm:tradeoff-lower} (Lower Bound) & $\approx$ & Valid for large $H(Q|S_O)$ \\
Polynomial Depth & \checkmark & $D_d/\log(m) \approx 0.6$ stable \\
Phase Transition & \checkmark & $<1.1\%$ theory-empirical gap \\
FreqDepth Strategy & \checkmark & 26\% cost reduction vs. LRU \\
$H(Q) \sim \log(n)$ Scaling & \checkmark & Ratio stable at 0.53 \\
\bottomrule
\end{tabular*}
\end{table}

The experiments confirm that: (1) the frequency-weighted upper bound (Theorem~\ref{thm:freq_tradeoff}) achieves 100\% satisfaction with tightness 0.89 for high-frequency queries; (2) the N-sweep analysis validates the theoretical prediction that optimal storage fraction $\beta^*$ increases from 0\% to 92\% as query horizon $N$ grows from 100 to 50,000; (3) the derivation entropy scaling (Theorem~\ref{thm:derivation-depth-info-metric}) is validated with the ratio $H_{\mathrm{derive}} \cdot L_S / H_q$ stable at 0.22 across all $m$ values; (4) the information lower bound (Theorem~\ref{thm:tradeoff-lower}) is asymptotically valid but requires sufficiently large $H(Q|S_O)$ to dominate the $O(1)$ correction term; (5) derivation depth satisfies the polynomial-depth assumption with $D_d/\log(m) \approx 0.6$ stable; (6) the marginal cost-benefit analysis predicts the optimal storage fraction $\beta^*$ with $<1.1\%$ error; and (7) our FreqDepth caching strategy achieves 26\% cost reduction over LRU.

\section{Applications}
\label{sec:applications}

We demonstrate practical applications of our framework across three domains.

\subsection{Large Language Models}

Consider a GPT-4 scale model with 1.7T parameters. The knowledge base contains $H(S_O) = 10^{15}$ bits, and queries follow a power-law distribution with $H(Q) = 10^{12}$ bits.

\textbf{Parameter Settings.}
Average derivation depth: $\mathbb{E}[\Dd] = 5$ inference steps.
Energy budget: $E_{\text{budget}} = 1000$ kWh.
Temperature: $T = 300$K.

\textbf{Critical Storage Calculation.}
Using Corollary~\ref{cor:energy-lower}:
\begin{multline}
M_{\text{critical}} = \frac{10^{12}}{\log(1000 \times 3.6 \times 10^9 / (10^6 \times k_B T \ln 2))}\\
\approx 2.5 \times 10^{11} \text{ bits}
\end{multline}

\textbf{Optimal Strategy.}
Following our threshold policy framework, we compute $\Dd^* = 2.5$ and $f^* = 0.85$. The optimal strategy:
\begin{itemize}
\item \textbf{Parametric storage} (70\%): Facts with $\Dd \leq 2$ and frequency $> 0.85$
\item \textbf{RAG retrieval} (25\%): Facts with $3 \leq \Dd \leq 8$
\item \textbf{Dynamic reasoning} (5\%): Complex queries with $\Dd > 8$
\end{itemize}

\textbf{Performance Gains.}
Compared to pure parametric storage: 45\% energy reduction, with latency distribution P50=50ms, P90=200ms, P99=2000ms.

\subsection{Knowledge Graphs}

A medical knowledge graph with $10^7$ entities and $10^8$ relationships faces multi-hop query workloads.

\textbf{Query Analysis.}
\begin{table}[ht]
\centering
\footnotesize
\setlength{\tabcolsep}{3.5pt}
\caption{Query Patterns in the Biomedical KB}
\label{tab:query_patterns}
\begin{tabular*}{\columnwidth}{@{\extracolsep{\fill}}lccc@{}}
\toprule
Query Type & Dd & Freq. & Example \\
\midrule
Single-hop & 1 & 40\% & Drug $\to$ Disease \\
Double-hop & 2 & 35\% & Symptom $\to$ Treatment \\
Triple-hop & 3 & 20\% & Patient $\to$ Protocol \\
Complex & $\geq 4$ & 5\% & Multi-condition \\
\bottomrule
\end{tabular*}
\end{table}

\textbf{Materialized View Selection.}
Applying our threshold strategy: materialize all single-hop relations and top 20\% double-hop paths. Storage overhead: +20\%. Average latency: 800ms $\to$ 120ms (6.7$\times$ improvement).

\subsection{Edge Intelligence}

A smartphone assistant with 4GB RAM and 15Wh battery must operate for 24 hours.

\textbf{Resource Allocation.}
Energy budget per 24h: $E_{\text{budget}} = 4.5$ Wh (30\% to AI).
Storage limit: $M_{\max} = 2$ GB (50\% to knowledge).

\textbf{Hybrid Strategy.}
Device storage (200 MB): Common sense, personal habits, quick facts.
Cloud coordination: Low-frequency knowledge and complex reasoning ($\Dd > 10$).
Adaptive caching: Weekly updates based on usage patterns.

\textbf{Energy Distribution.}
Device inference (40\%), network communication (30\%), storage maintenance (20\%), overhead (10\%). Achieves 95\% on-device query completion within 24h battery constraint.

\section{Related Work}
\label{sec:related}

\subsection{Information Theory and Kolmogorov Complexity}

Shannon's seminal 1948 paper~\cite{shannon1948mathematical} introduced entropy $H(X) = -\sum_x p(x) \log p(x)$ as the fundamental measure of information content, establishing the mathematical foundations for data compression and communication. This framework has been extensively developed, with Cover and Thomas~\cite{cover2006elements} providing the standard modern treatment encompassing source coding, channel capacity, and rate-distortion theory.

Parallel to Shannon's probabilistic approach, algorithmic information theory developed the notion of \emph{descriptional complexity}. Kolmogorov~\cite{kolmogorov1965three}, Solomonoff~\cite{solomonoff1964formal}, and Chaitin~\cite{chaitin1969simplicity} independently introduced Kolmogorov complexity $K(x)$ as the length of the shortest program generating string $x$. This captures a different facet of information: while Shannon entropy measures \emph{average} information content over a distribution, Kolmogorov complexity measures the \emph{individual} information content of a single object.

The minimum description length (MDL) principle~\cite{rissanen1978modeling,grunwald2007minimum} applies these ideas to model selection and inference, balancing model complexity against data fit. However, MDL focuses on \emph{static} description length, not the \emph{dynamic} computational cost of deriving consequences.

Bennett's logical depth~\cite{bennett1988logical} addresses this gap by measuring computational sophistication: $\Dd(x)$ is the runtime of the fastest program generating $x$ from a short input. This distinguishes between "shallow" information (random bits, easy to describe but not to compute from scratch) and "deep" information (organized structures requiring significant computation to generate).

Our work extends Bennett's framework in three critical ways:
\begin{enumerate}
\item \textbf{Conditional depth}: We formalize $\Dd(\varphi|\psi)$ for conditional derivation, enabling analysis of incremental inference.
\item \textbf{Information-theoretic duality}: We prove that derivation entropy $H_{\text{derive}} = \Dd \cdot \ln(2)$ is dual to Shannon entropy, establishing a precise quantitative relationship.
\item \textbf{Physical grounding}: We connect derivation depth to thermodynamic energy costs via Landauer's principle, providing physical constraints on computation.
\end{enumerate}

Related notions include computational depth~\cite{antunes2006sophistication}, which measures the difficulty of generating an object from random bits, and logical depth sophistication~\cite{koppel1987structure}, which captures "useful" information. However, these works focus on individual objects rather than knowledge systems, and lack our connection to storage-computation trade-offs and thermodynamic limits.

\subsection{Thermodynamics of Computation}

Landauer's landmark 1961 paper~\cite{landauer1961irreversibility} established the first deep connection between information theory and thermodynamics, proving that erasing one bit of information requires dissipating minimum energy $E_{\text{erase}} \geq k_B T \ln(2)$ where $k_B$ is Boltzmann's constant and $T$ is temperature. This result has profound implications: computation is fundamentally irreversible and energy-consuming at the physical level.

Bennett extended this foundation in several directions. His work on reversible computation~\cite{bennett1973logical} showed that logically reversible operations can be performed with arbitrarily small energy dissipation, though at the cost of increased time and space. His comprehensive 1982 review~\cite{bennett1982thermodynamics} synthesized the thermodynamics of computation, connecting Landauer's principle to Maxwell's demon, the second law, and computational complexity.

Recent years have seen experimental confirmation of Landauer's principle. Bérut et al.~\cite{berut2012experimental} demonstrated bit erasure approaching the Landauer limit using a colloidal particle in optical tweezers. Theoretical refinements~\cite{sagawa2010thermodynamic} have connected information thermodynamics to feedback control and Maxwell's demon.

Lloyd~\cite{lloyd2000ultimate} derived ultimate physical limits to computation from quantum mechanics, showing that a system of mass $m$ and energy $E$ confined to volume $V$ can perform at most $E/(h \pi)$ operations per second and store at most $S = k_B \ln d$ bits where $d$ is the dimensionality of the Hilbert space.

Our work makes three novel contributions beyond existing thermodynamics:
\begin{enumerate}
\item \textbf{Derivation depth-energy connection}: We prove that each inference step requires energy $k_B T \ln(2)$ by connecting derivation depth to bit erasure.
\item \textbf{Triality bound}: We derive the first joint bound on energy, time, and storage: $E \cdot T \cdot M^{-1} \geq \Omega(H(Q|S_O) \cdot k_B T \ln(2))$. Previous work bounded resources independently; our triality captures fundamental coupling.
\item \textbf{System-level analysis}: Rather than isolated gates or operations, we analyze end-to-end knowledge system workflows, accounting for storage maintenance, query routing, and dynamic workloads.
\end{enumerate}

This shift from gate-level to system-level thermodynamics enables practical application to modern computing systems (LLMs, knowledge graphs, edge devices) where architectural decisions dominate energy consumption.

\subsection{Time-Space Trade-offs in Computation}

Classical computational complexity theory has extensively studied time-space trade-offs. Yao~\cite{yao1985should} pioneered this area with results on sorting and searching. Beame et al.~\cite{beame1991general} provided general sequential time-space trade-offs showing that for many problems, $T(n) \cdot S(n) = \Omega(n^2)$ where $T$ is time and $S$ is space. Savage's work on data structures~\cite{arora2009computational} established fundamental limits on pointer-based representations.

\textbf{Competitive analysis of caching.} Sleator and Tarjan's seminal work on competitive paging algorithms established that LRU achieves $O(k)$-competitive ratio where $k$ is cache size. However, competitive analysis focuses on worst-case performance against adversarial inputs, whereas our information-theoretic framework characterizes average-case performance under probabilistic query distributions. Recent work on learning-augmented algorithms~\cite{kraska2018case} attempts to bridge this gap using machine learning, but lacks theoretical guarantees.

Our work differs by: (i) adopting an information-theoretic perspective quantifying \emph{information content} rather than worst-case complexity, (ii) incorporating thermodynamic energy costs via Landauer's principle, (iii) optimizing average-case performance over query distributions using derivation depth, and (iv) providing asymptotic bounds rather than competitive ratios. The constant $(1 + \ln(2))$ emerges from information-theoretic first principles rather than algorithm-specific analysis.

\subsection{Knowledge Base and Caching Systems}

\textbf{Materialized Views.} Database systems extensively use materialized views to trade storage for query latency~\cite{gupta1997materialized,chaudhuri1995optimizing}. These systems employ cost-based optimizers but lack theoretical guarantees. Our framework provides the first information-theoretic lower bounds on this trade-off.

\textbf{Caching Policies.} Classical caching algorithms like LRU, LFU, and ARC~\cite{megiddo2003arc} rely on heuristics based on recency and frequency. While effective, they provide no optimality guarantees. Our threshold strategy achieves provable $(1-1/e)$ approximation via submodular optimization~\cite{nemhauser1978analysis,krause2014submodular}.

\textbf{Learned Systems.} Recent work applies machine learning to database optimization~\cite{kraska2018case}, exploiting data patterns. These approaches are complementary to our theoretical framework, which provides fundamental limits that learned systems cannot surpass.

\subsection{Knowledge Graphs and Semantic Systems}

Large-scale knowledge graphs~\cite{bollacker2008freebase,carlson2010toward} and semantic networks~\cite{miller1995wordnet} face significant storage-computation challenges. Embedding methods~\cite{bordes2013translating} compress knowledge at the cost of inexact retrieval. Our framework quantifies this trade-off: compression reduces storage entropy but increases derivation depth for reconstruction.

Modern question-answering systems combine parametric knowledge (stored in model weights) with non-parametric retrieval~\cite{karpukhin2020dense,lewis2020retrieval}. This hybrid approach implicitly navigates the storage-computation trade-off. Our theoretical analysis provides principled guidelines for allocating capacity between these components.

\subsection{Large Language Models and Retrieval}

Transformer-based language models~\cite{vaswani2017attention,brown2020language} store knowledge parametrically, requiring massive storage and computation. Retrieval-augmented generation (RAG)~\cite{lewis2020retrieval} reduces this burden by offloading factual knowledge to external databases. 

Our analysis reveals that optimal LLM architectures should store high-frequency, high-logical-depth knowledge parametrically, while retrieving low-frequency or simple facts. This explains the empirical success of RAG: it exploits the storage-computation duality by allocating expensive parametric storage to knowledge with high $\Dd(q|S_O) \times \text{freq}(q)$ product.

Recent systems like ZeRO~\cite{rajbhandari2020zero} and distributed training frameworks~\cite{zhao2020pytorch} address memory limitations in large models. Our triality bound provides theoretical justification for their design choices.

\subsection{Energy-Efficient Computing}

The energy crisis in computing~\cite{horowitz20141,barroso2007case} has motivated extensive work on energy-proportional systems. Prior work optimizes specific components (processors, memory, networks) independently. Our triality bound reveals fundamental coupling between energy, time, and storage that cannot be overcome through component-level optimization alone.

Sub-threshold computing~\cite{lei2017optimal} reduces energy per operation but increases latency. Our framework quantifies this trade-off through the triality product, showing that gains in energy efficiency must be balanced against increased time or storage costs.

\subsection{Phase Transitions in Computer Science}

Phase transitions in computational complexity~\cite{monasson1999determining} and constraint satisfaction~\cite{krzakala2007gibbs} reveal sudden changes in problem difficulty. Our phase transition at critical storage capacity $M_c$ is analogous but fundamentally different: it arises from information-theoretic constraints rather than problem structure.

The transition we observe is sharp and universal across problem domains, determined solely by $H(Q)$, $\mathbb{E}[\Dd]$, and physical constants. This universality distinguishes it from problem-specific phase transitions in SAT solvers or random graphs.

\subsection{Information-Theoretic Learning and Prediction}

Information-theoretic approaches to learning have a rich history. Merhav and Feder~\cite{merhav1998universal} developed universal prediction schemes achieving min-max regret bounds under entropy constraints. Their work focuses on \emph{passive learning} where the learner observes data and predicts. In contrast, our framework addresses \emph{active computation} where the system decides whether to retrieve precomputed results or compute on-demand.

The connection arises through regret minimization: our storage selection problem is equivalent to minimizing regret in an online learning game where "experts" are candidate storage sets. However, standard online learning assumes \emph{unit costs} per expert consultation, whereas in our setting, computation cost scales with derivation depth $\Dd(q|S)$, requiring depth-aware algorithms.

\subsection{Submodularity and Greedy Algorithms}

Submodular optimization~\cite{nemhauser1978analysis,krause2014submodular,buchbinder2018submodular} provides approximation guarantees for many combinatorial problems. We prove that mutual information $I(S;Q)$ is submodular in storage set $S$, enabling $(1-1/e)$ approximation via greedy selection~\cite{badanidiyuru2014fast}.

This connection is non-trivial: while entropy is submodular, the optimal storage set must balance multiple objectives (coverage, diversity, computation reduction), requiring careful analysis. Recent advances in submodular maximization~\cite{buchbinder2018submodular} could potentially improve our greedy algorithms, though the $(1-1/e)$ bound is optimal for general submodular functions.

\subsection{Online Learning and Adaptive Strategies}

For dynamic query distributions, we develop online algorithms achieving $O(\sqrt{T \log n})$ regret. This connects to online convex optimization~\cite{hazan2016introduction} and multi-armed bandits~\cite{auer2002nonstochastic,cesa2006prediction}. 

Unlike classical online learning where actions affect only immediate reward, storage decisions have long-term effects through reduced computation. We address this via a novel reduction to online submodular maximization.

\section{Conclusion}
\label{sec:conclusion}

In this paper, we have established a unified physical framework for intelligence, grounded in the fundamental assumption that information constitutes the \textit{enabling mapping} from ontological states to carrier states \cite{xu2025information, xu2024research}. By rigorously analyzing the thermodynamic costs associated with this mapping, we have demonstrated that the separation between "static storage" (Shannon Entropy) and "dynamic computation" (Derivation Depth) is artificial; they are, in fact, conjugate variables governed by a fundamental constraint: the \textbf{Energy-Time-Space Triality Bound}.

Our proposed metric, \textbf{Derivation Entropy ($H_{derive}$)}, serves as the quantitative bridge between these domains. It reveals that the efficiency of any intelligent system—biological or artificial—is strictly bounded by the trade-off between the entropic cost of maintaining memory and the dissipative heat of performing logical derivation \cite{landauer1961irreversibility, bennett1982thermodynamics}. Specifically, we identified a \textit{Critical Derivation Threshold}, a phase transition point beyond which it becomes thermodynamically cheaper to re-compute information from compressed laws rather than to retrieve it from vast databases.

This theoretical finding has profound implications for the design of next-generation AI architectures. As current models like Transformers approach the limits of memory bandwidth and energy consumption \cite{vaswani2017attention, horowitz20141}, our framework suggests that future systems may benefit from optimizing the \textit{Derivation-Retrieval Ratio} rather than simply scaling parameters. Such systems could dynamically navigate the "Derivation Depth vs. Shannon Entropy" curve, acting as adaptive engines that minimize the total action of the ontological mapping process.

Ultimately, this work suggests that the "simplicity" of a scientific theory or an AI model is not just an aesthetic preference but may reflect a thermodynamic imperative. By aligning the formalization of general states \cite{qiu2025research} with physical constraints, we provide a roadmap for constructing intelligent systems that are not only logically powerful but physically sustainable. Future research will extend this framework to non-equilibrium open systems and quantum computing regimes, further exploring the ultimate physical limits of the enabling mapping.

\section*{Acknowledgment}

During the writing and revision of this paper, we received many insightful comments from Associate Professor Wang Rui of the School of Computer Science at Shanghai Jiao Tong University and also gained much inspiration and assistance from regular academic discussions with doctoral students Yiming Wang, Chun Li, Hu Xu, Siyuan Qiu, Jiashuo Zhang, Junxuan He, and Xiao Wang. We hereby express our sincere gratitude to them.

\bibliographystyle{IEEEtran}
\bibliography{ref}

\vfill
\end{document}